\documentclass{aa}  

\usepackage{graphicx}
\usepackage{txfonts}
\usepackage[colorlinks=true,citecolor=blue,linkcolor=blue,urlcolor=blue,breaklinks=true]{hyperref} %
\usepackage{amsmath}	
\usepackage{subcaption}
\usepackage{xspace}
\usepackage{xcolor}
\usepackage{natbib}
\usepackage[]{placeins}

\newcommand{\PSRi}{PSR~J1842+0358\xspace}  
\newcommand{\PSRii}{PSR~J1926+0737\xspace} 
\newcommand{\orcidlink}[1]{\protect\href{https://orcid.org/#1}{\protect\includegraphics[width=8pt]{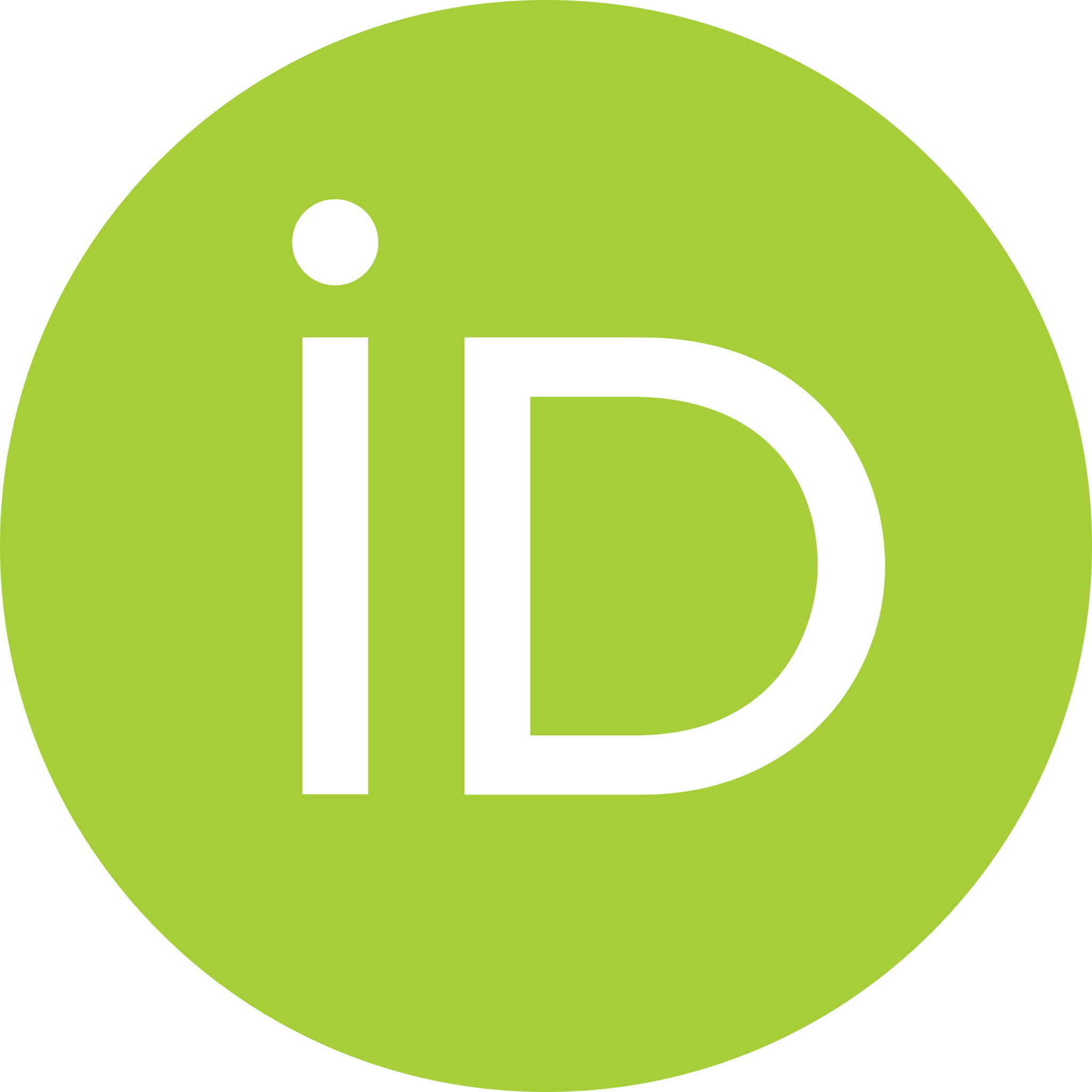}}}
\newcommand{\orcid}[1]{\unskip\textsuperscript{\orcidlink{#1}}}

\begin{document} 

\title{Main and interpulse interaction in PSRs J1842+0358 and J1926+0737: evidence for interpole communication
}

\author{X.~Song   \orcid{0000-0003-4924-0550}\ \inst{\ref{astron}}
\and P.~Weltevrede \orcid{0000-0003-2122-4540}\ \inst{\ref{jbca}}
\and J.~van~Leeuwen \orcid{0000-0001-8503-6958}\
\inst{\ref{astron}}
\and G.~Wright \orcid{0000-0001-8193-0557}\
\inst{\ref{jbca}}
\and M.~Keith \orcid{0000-0001-5567-5492}\
\inst{\ref{jbca}}
}
\authorrunning{X.~Song~et~al.}
\titlerunning{MP-IP interaction in PSRs~J1842+0358 and J1926+0737}  

   \date{Received; accepted}


\institute{
ASTRON, the Netherlands Institute for Radio Astronomy, Oude Hoogeveensedijk 4,7991 PD Dwingeloo, The Netherlands\label{astron} \\ e-mail: \url{song@astron.nl}
  \and
Jodrell Bank Centre for Astrophysics, Department of Physics and Astronomy, The University of Manchester, Manchester, M13 9PL, UK\label{jbca}
}

 
\abstract{
  Our understanding of the elusive radio-pulsar emission mechanism would be deepened by determining the
  locality of the emission. Pulsars in which the two poles interact can potentially help solve this challenge. We here report the discovery of interaction of emission between the main and the interpulse in two pulsars -- J1842+0358 and J1926+0737, based on FAST and MeerKAT data. When emission is bright in one pulse, it is dim in the other. Even when split in just 2 groups (strong versus weak) the anti-correlated brightness can change by a factor $\gtrsim$2. 
  Both sources furthermore show the same quasi-periodic modulation from the main and interpulse, at timescales exceeding 100 pulse periods. The longitude stationary modulation from at least one pulse suggests that it is a key signature for interpulse pulsars showing main and interpulse interaction. If the interaction happens within an isolated magnetosphere, without external influences, either communication between the opposite poles is required, or global changes drive both. This detailed study of these two sources was only made possible by improved sensitivity. The fact that both show two-pole modulation strongly suggests this is a general phenomenon in interpulse pulsars. In regular pulsars only one pole is visible, and a number of these show correlated changes between the profile and spin-down rate, that are also thought to be caused by global magnetospheric changes. Our results strengthen the case that such interactive magnetospheres are common to all pulsars. 
}


   \maketitle
%

\section{Introduction}
\label{sec:intro}

The individual pulses emitted by radio pulsars are highly variable. The diversity in the single pulse emission is stochastic, and some pulsars additionally show more systematic patterns. These are formed for example when subpulses (pulses of narrower widths than the average pulse profile) drift from one pulse longitude to another, or when periodic intensity modulations occur without significant subpulse phase variations. Both types of behaviour are relatively common in normal pulsars \citep{Weltevrede2006,Weltevrede2007,Basu2016,Basu2019,Basu2020,Song2023}, and are related to the fundamental radio emission from the pulsar magnetosphere. It has been theorised by \citet{RS} that the radio emission originates from localised and discrete subbeams rotating around the magnetic axis, forming the emission beams. Studying the single pulse properties provides useful constraints in the magnetospheric emission mechanisms. 

A subset of pulsars exists that is specifically interesting to study in this context: the interpulse pulsars, whose two pulses are both seen in one pulse period and usually half a stellar rotation apart. These pulses are dubbed the main pulse (hereafter MP) and the interpulse (hereafter IP). Only about 3\% of the total population are interpulse pulsars \citep{Maciesiak2011}, but these offer valuable insights into the magnetospheric structure. This is because the phenomenon is generally interpreted as emissions from the two opposite poles. These pulsars therefore  possess a specific geometry, with the  magnetic axis nearly perpendicular to the rotation axis: an orthogonal rotator. 

As one of the necessary simplifications in the \citet{RS} model consists of assuming  an aligned rotator (whose magnetic axis coincides with its rotation axis), it is interesting to study how emission is affected when the rotator is orthogonal, given its very different geometrical setup. Furthermore, as many models predict emission from both poles one may wonder if and how their emission are related or not -- an interesting aspect that can only be investigated in an orthogonal rotator, where both poles are seen. 

A variety of evidence suggests that the emission seen half a period apart in interpulse pulsars does in fact originate from both poles (and not from two emission regions on a single pole). The polarisation position angles (PAs), for example, can be measured for both the MP and the IP, and to be used to fit for the pulsar geometry according to the rotation vector model \citep[RVM;][]{RVM}, with the geometric parameters interlinked for both poles. One confirmed double-pole rotator is PSR~B0906$-$49, where a study of the detailed polarisation showed the  emission was from two magnetic poles \citep{Kramer2008}. Another line of evidence concerns the frequency evolution of the pulse profile: the separation between the MP and the IP is unchanged at different observing frequencies. If emission is from a single pole, components are generally seen to separate in phase at lower frequencies. The reasoning was offered by, e.g.,~\citet{Biggs1988} for PSR~B1702$-$19, as its MP peak to IP peak separation remained consistent with 180\degr\ from 408 to 1420\,MHz. Additional clues may show in the average pulse profile, where the MP and the IP profiles are usually very narrow in comparison to their separation, without `bridge' emission connecting the pulses even at a very low level. In contrast, broad pulse widths and `bridge' emissions are more likely to occur in single-pole interpulse pulsars, where the magnetic and rotational axes are  aligned. One  example is PSR~B1929+10 \citep{Rankin1997,Kou2021}, where the presence of bridge emission indicates a single-pole geometry.  

Some interpulse pulsars have been found to exhibit an interesting and surprising phenomenon, whereby the emission from the MP and the IP are correlated -- evidence of communication between the two poles. This was first discovered in PSR~B1822$-$09 (\citealt{Fowler1982}) but detailed analysis of this and other pulsars was not followed up until later decades. In PSRs~B1702$-$19 \citep{Weltevrede2007}, B1055$-$52 \citep{Biggs1990,Weltevrede2012}, B1822$-$09 \citep{Gil1994,Yan2019}, B1929+10 \citep{Kou2021} and B0823+26 \citep{Chen2023}, the correlation is furthermore associated with an intensity modulation of the same periodicity in both the MP and the IP. Even more intriguingly, these periodic variations are phase-locked, causing the correlations to be maintained over many thousands of pulse periods. In some sources, a phase delay between the MP and the IP emission is also present. These newly observed properties are not easily explained by pre-existing standard pulsar models. If the MP and the IP emission originate from different poles, even the \citet{RS} order-of-magnitude model assumptions would not predict such strictly identical configurations of the subbeams and periodicities. In addition, \citet{lt12} have shown that when derived exactly, the modulation in the \citet{RS} framework depends strongly on the local configuration of each pole. 

This makes  even more interesting  the fact that the two poles must be  connected in some way; and this is not possible through the traditionally defined open-field line regions where  radio emission is generated \citep[e.g.,~see discussions in][]{Weltevrede2009}. Although a single-pole model would explain the phase-locked correlation and the same periodicities from both pulses,  the geometric evidence described above (either from polarisation or from the pulse profiles) favours at least some pulsars to have emission from both poles. 

One problem for our understanding of the interpulse pulsar population, is that we do not know if the sources showing phase-locked emissions between the MP and the IP are somehow special, or the standard. 
And while these sources share many similarities, they exhibit significant differences in their detailed observed properties. Studying the interpulse pulsar sample as a whole would thus help establish whether the correlation always exist or not, and whether there are counter examples. In the Thousand-Pulsar-Array (TPA) project \citep{Johnston2020}, a sample of more than 30 interpulse pulsars was covered. Their subpulse modulation properties were studied in \citet{Song2023}. The TPA dataset provided uniform observations at L-band using the MeerKAT full-array of 64-dishes, with more than 1000 pulses for at least one single observation. However, even with the MeerKAT sensitivity, one of the pulses of these interpulse pulsars was usually much weaker than the other, sometimes preventing a clear detection of both the modulation and correlation. We thus further carried out a campaign using the FAST radio telescope \citep{Nan2011} on a number of sources, with higher sensitivity and longer single pulse sequences. 

In this paper, we report that two more interpulse pulsars, PSRs~J1842+0358 and J1926+0737, show correlated, quasi-periodic emission from the MP and the IP, based on data from both MeerKAT and FAST. Both pulsars were discovered in the Parkes Multibeam Pulsar Survey \citep{Lorimer2006,Eatough2010}. The two sources inhabit the middle yet relatively young part of the $P$-$\dot{P}$ diagram. PSRs~J1842+0358 and J1926+0737 have short spin periods of around 0.23\,s and 0.32\,s, with a spin-down rate $\dot{P}$ of 8.1E$-$16\,s/s and 3.8E$-$16\,s/s, respectively. The inferred spin-down luminosities are 2.5E33\,erg/s and 4.6E32\,erg/s, with characteristic ages of 4.6E6\,yr and 1.3E7\,yr. Both pulsars were claimed to be of double-pole kind in \citet[][and references therein]{Maciesiak2011}, but without supporting polarisation data. 

The structure of our paper is as follows: the MeerKAT and FAST observations are described in Sect.~\ref{sec:obs}, the single pulse analysis of PSRs~J1842+0358 and J1926+0737 are shown in Sect.~\ref{sec:j1842} and \ref{sec:j1926}, where the correlations in the MP and the IP emission are established, with the identified quasi-periodic modulations. Finally, we present our discussions and conclusions.

\section{Observations}
\label{sec:obs}

\subsection{MeerKAT}
\label{sec:MK}
The MeerKAT pulse stacks were taken from the TPA single pulse pipeline. A full description of the production of such single pulse stacks is in \citet{Song2023}. There was a single observation for each of PSRs~J1842+0358 and J1926+0737, covering 2343 and 4612 single pulses, respectively. The observations were centred at about 1284\,MHz, with a bandwidth of 856\,MHz. The data were polarisation calibrated but not flux calibrated. The onpulse regions were automatically identified and a straight-line-fit to the off-pulse region was subtracted from each single pulse to produce the baseline-subtracted single pulse stack. 

\subsection{FAST}
\label{sec:fastobs}
The FAST observations were taken in the Swift Calibration mode, with a 49.152\,$\mu$s sampling time and 4096 frequency channels. The observations were centred at 1250\,MHz with a bandwidth of 500\,MHz \citep{Jiang2019}. Polarisation calibration data using a noise diode was taken at the beginning of the observations for 60\,sec. Also these observations were  polarisation calibrated but not flux calibrated. The raw data were reduced with standard pulsar software. They were first passed to \textsc{dspsr}, to produce single pulses with 1024 phase bins. The sources were folded using ephemerides the MeerKAT observations. The original frequency resolution of 4098 channels was reduced to 512 for further processing. By adding all single pulses together, a single frequency versus pulse longitude spectrum was used for radio frequency interference (RFI) removal, therefore each single pulse has the same number of frequency channels removed. Polarisation calibration was applied to the data using the standard noise diode calibrator file, and the rotation measure (RM) was applied to the polarisation calibrated data. For PSR~J1842+0358 the RM value was derived from the MeerKAT observation while the value for PSR~J1926+0737 was fitted using \textsc{rmsynth} \citep{Weltevrede2016}. In the current paper, we focus on single pulse properties and do not study the polarisation properties of the sources in detail. Those are presented in \citet{sun+25}.

The individual files were next added to a single pulse stack using \textsc{psradd}. The onpulse regions were identified by eye and a linear slope was subtracted from each single pulse to remove the baseline. The observations of PSRs~J1842+0358 and J1926+0737 contained 4849 and 4476 single pulses after processing. For presentation purposes, the single pulses of the FAST data are normalised by the peak intensity of the corresponding average profile. 

\section{PSR~J1842+0358}
\label{sec:j1842}

In this Section, we discuss the MeerKAT and FAST data and analysis for PSR~J1842+0358. 

\subsection{MeerKAT pulse stack}
\label{sec:j1842tpa}

\begin{figure}
    \centering    \includegraphics[width=\linewidth]{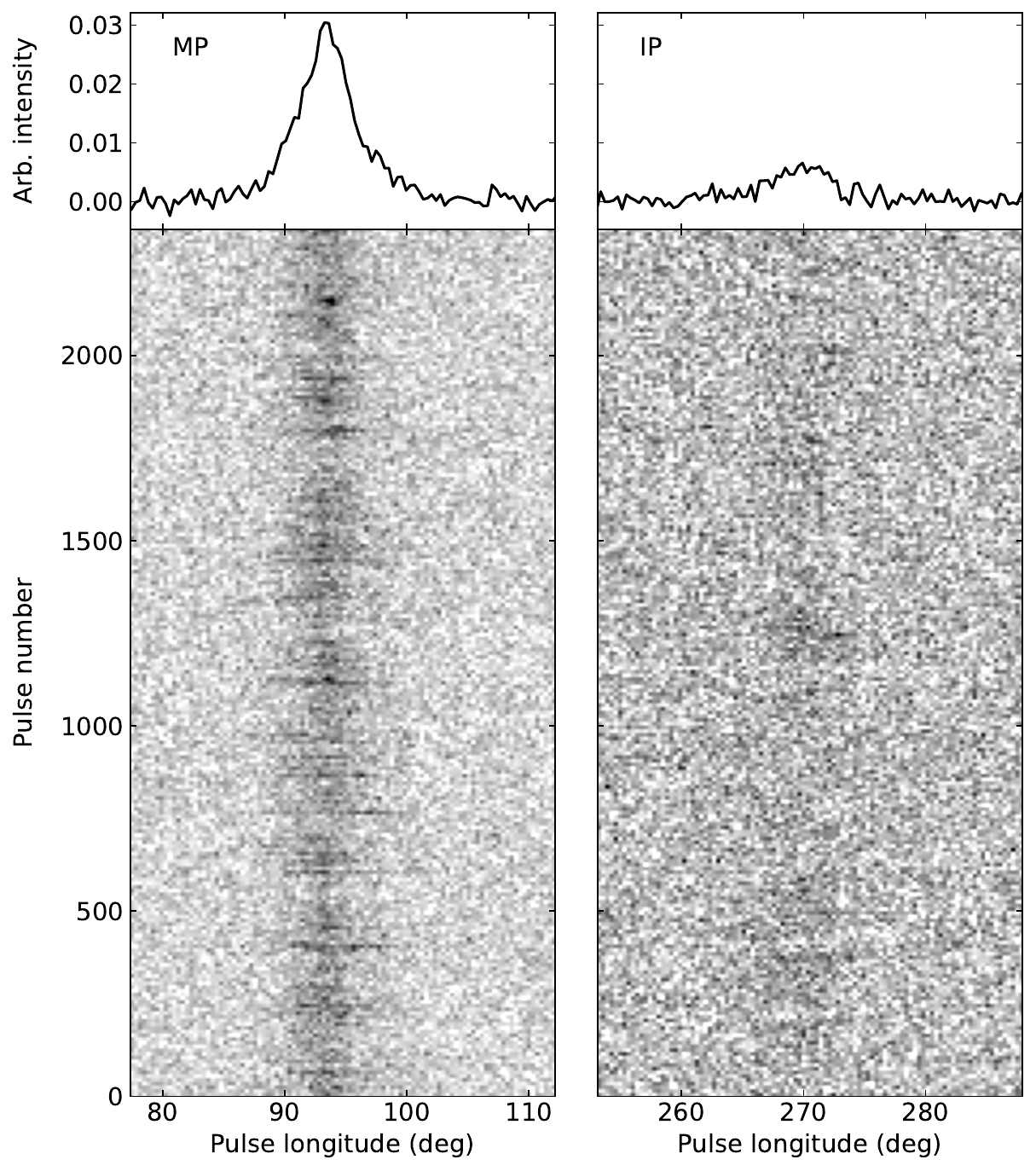}
    \caption{MP (left) and IP (right) pulse stacks of PSR~J1842+0358 from the MeerKAT observation. The pulse sequences are averaged over 10 single pulses but the pulse number on the the y-axis still denotes the single pulse number. The corresponding average profiles of all single pulses are shown on the top panels.
}
    \label{fig:J1842mk}
\end{figure}

We first present the only observation for this pulsar from the TPA project. No clear periodicity was identified for these data in \citet{Song2023}. The two-dimensional fluctuation spectrum (2DFS; \citealt{ES2002})  in that paper does show a slight increase in power towards the very lowest frequency of the MP, while no clear structure in the spectra of the IP could be identified. The pulse stacks are presented in Fig.~\ref{fig:J1842mk} in the bottom panel, with the corresponding average pulse profile (combining all single pulses) in the top panel. Given the low signal-to-noise ratio (S/N) in the IP, the pulse stacks are averaged over 10 single pulses. The MP (left) shows occasionally brighter and wider pulses (e.g.,~pulse number 400, 1100), but no clear periodicity is identified from the single pulses or the averaged sequence. The IP single pulses are generally hardly detected, although some pulses appear brighter (e.g.,~pulse number 1200 to 1300). The intensity modulation in both the MP and the IP does motivate further study of their behaviour. For this we exploit the higher S/N of the FAST observation, that is presented in the remainder of this section.

\subsection{Average pulse profile}
\label{sec:j1842prof}

\begin{figure}[b]
    \centering
    \includegraphics[width=\linewidth]{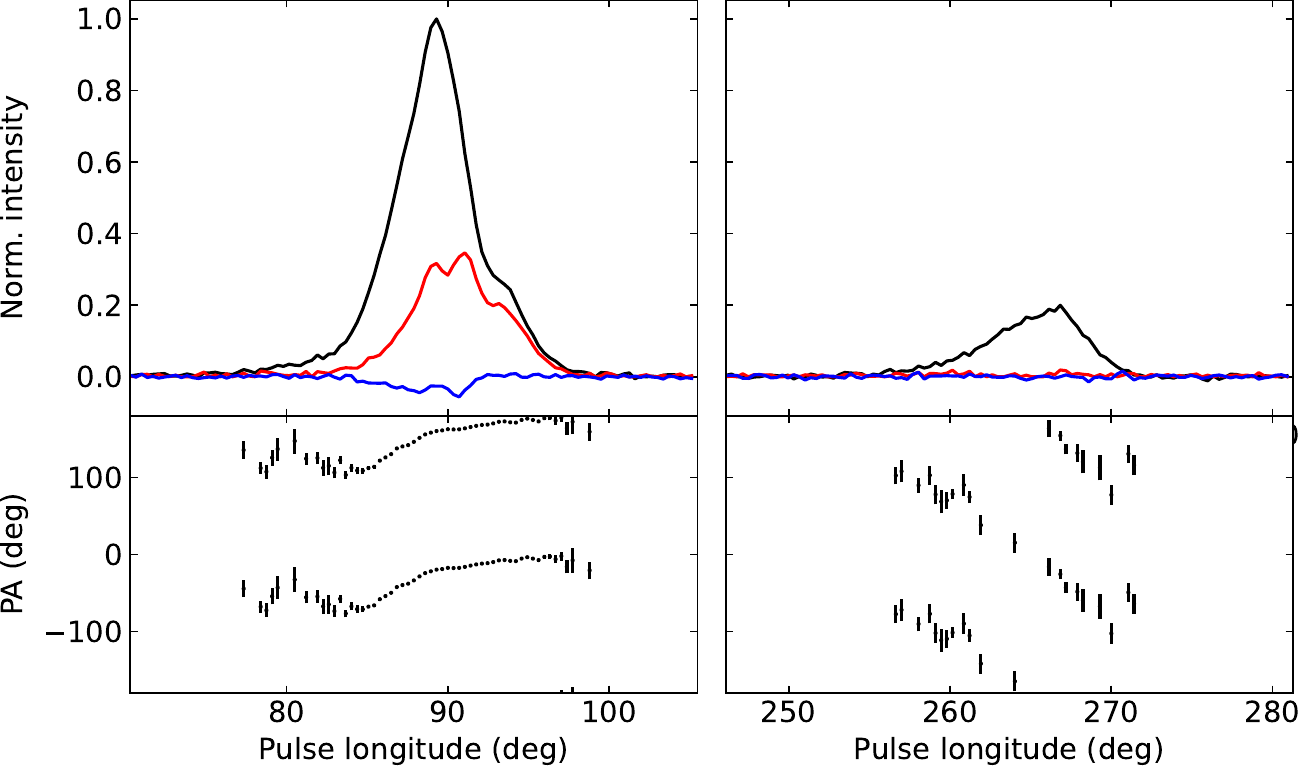}
    \caption{Average pulse profile of PSR~J1842+0358 from the FAST data, for the MP (left) and the IP (right). The total intensity profile (normalised by the MP peak) is in black, and the corresponding linear and circular polarisation are shown in red and blue, respectively. The bottom plots show the computed polarisation position angles of the onpulse regions for all points with a significance exceeding one sigma.}
    \label{fig:J1842prof}
\end{figure}

\begin{figure*}
    \centering
    \includegraphics[width=0.96\linewidth]{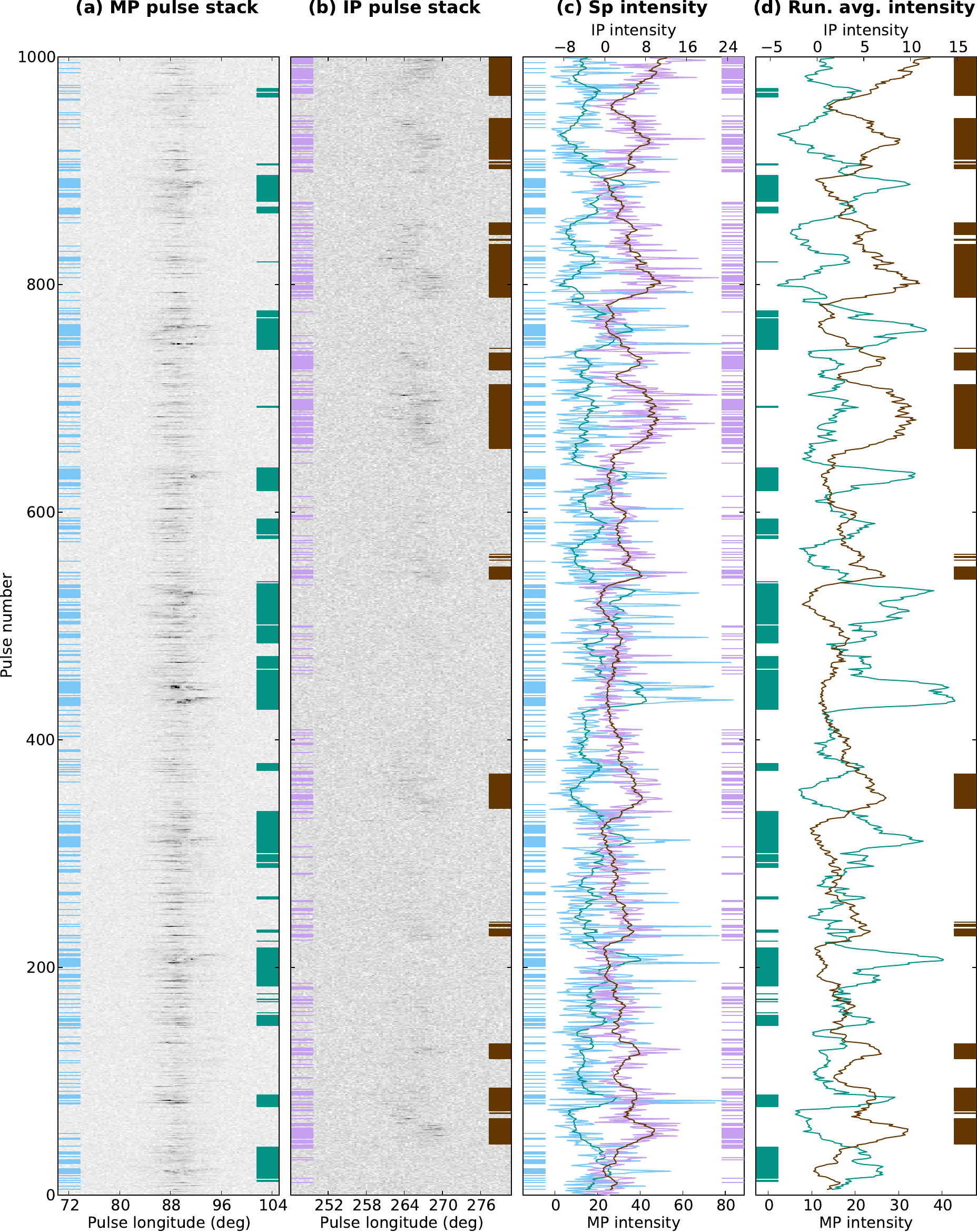}
    \caption{The first 1000 single pulses for PSR~J1842+0358:  pulse stacks (MP, IP) and intensities (single-pulse, running-averaged). (a)~Pulse stack of the MP. The blue and green margin bars mark the bright pulses, based on the intensities of the single pulses (blue, left) or their running averages (green, right). (b) As (a) but for the IP, now with violet and brown for single-pulse or running-averaged intensities, respectively. The single pulse intensities of the MP and the IP are in (c), with the running averaged intensities plotted on top. Margin bars identical to the left of (a) and (b). The running averaged intensities of the MP and the IP plotted alone are in (d), along with the corresponding selected bars (the same as those on the right of (a) and (b)). The rest of the \PSRi pulse stacks are in Appendix \ref{app:stacks1842} (Online only).}
    \label{fig:J1842sum}
\end{figure*}

The pulsar average pulse usually consists of multiple components. The average pulse profile of PSR~J1842+0358 detected by FAST is shown in Fig.~\ref{fig:J1842prof}. The MP (left) consists at least two components, with a small component appearing at the trailing edge of the pulse profile (pulse longitude about 92$^\circ$). Inspecting the single pulses, some  have bright and wide emission, especially towards the trailing edge of the pulse profile (e.g.,~pulse number around 200 in Fig.~\ref{fig:J1842sum}a), which also form the additional component in the average profile. The same wider pulses are seen in the MeerKAT observation in Fig.~\ref{fig:J1842mk}. The IP (right) is much weaker than the MP, and has a asymmetric profile. The widths of the MP and the IP at 10\% of their peaks, are around 11\degr\ and 14\degr, respectively. 

\begin{figure}
    \centering  \includegraphics[width=\linewidth]{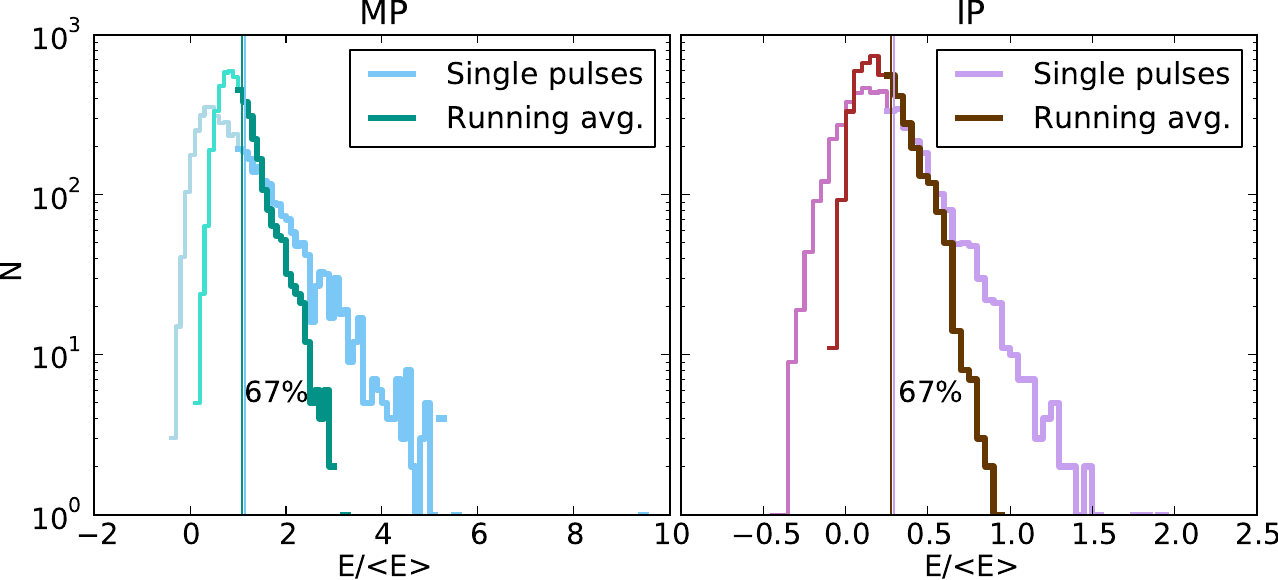}
    \caption{Pulse energy distributions of PSR~J1842+0358 for the MP (left) and the IP (right), with single pulse energies (blue and violet) and running averaged pulse energies (green and brown). The same colours of the pulse energies corresponding to the bright groups as in Fig.~\ref{fig:J1842sum} apply. The weak pulse energy distributions use the same but lighter set of colours.  Both pulse energies are normalised by the mean MP energy. The vertical line indicates the cut of the two groups, of  1/3 bright and 2/3 weak  pulse intensities (as described in Sect.~\ref{sec:j1842profmod}).}
    \label{fig:J1842dist}
\end{figure}

\subsubsection{Evidence for an orthogonal rotator}
\label{sec:j1842ortho}

Although \citet{Maciesiak2011} claimed the pulsar is an orthogonal rotator, there was no quantitative measurement of its geometry. The detailed polarisation study of the interpulse-pulsars data set is presented in \citet{sun+25};  we here focus only on the simple geometric arguments that suggest this pulsar is likely an orthogonal rotator.

The average profiles of the FAST observation are presented in Fig.~\ref{fig:J1842prof} top panels. The separation between the peak of the MP and the peak of the IP is 177.5\degr, close to 180\degr\ -- which is expected for an orthogonal rotator. Further evidence of a double-pole geometry is the lack of emission in between the MP and the IP. The pulse widths at 10\% of the peak intensity are of the order of 10\degr\ for both the MP and the IP, which are both much smaller than the MP-IP separation. Such behaviour in the pulse profile is less likely to be produced by a single-pole geometry, whose pulse profiles are relatively wide and tend to include bridge emission.

There is no significant evolution across the 500\,MHz bandwidths for the FAST observation and the L-band MeerKAT observation. We are not aware of data to check whether the MP-IP separation evolves between observing frequencies even farther apart, nor in X-ray or $\gamma$-ray to put more constraints on the geometry. 

Polarisation profiles were computed using \textsc{ppol} in \textsc{psrsalsa}\footnote{\url{https://github.com/weltevrede/psrsalsa}} \citep{Weltevrede2016} and are shown in Fig.~\ref{fig:J1842prof}. We obtained consist linear/circular polarisation profiles as presented in \citet{sun+25}. The position angles (PAs) were computed from the linear polarisation. According to the RVM model \citep{RVM}, the PAs form an S-shaped curve as the magnetic fields pass the line of sight (LoS). The PA points for the MP and the IP are shown in the bottom panels of Fig.~\ref{fig:J1842prof}. Although it includes less accurate PA points, the overall shape of the curve in the IP is clear enough. The positive slope of the MP PAs and the negative slope of the IP PAs suggest that the sign of the impact parameter $\beta$ is different for the MP and IP, with $\alpha$, the angle between the rotation and the magnetic axis, assumed to be close to 90\degr. Further discussion of how the slopes of the PA give rise to different LoS cut of the emission beam is in Sect.~\ref{sec:magneto}. In \citet{sun+25}, a polarisation fit indeed suggests that this pulsar is an orthogonal rotator. We note that their fitted PA direction of the IP is the opposite to ours here. However, the authors themselves point out that their result is based on only two PA points and needs clarification through longer observations.

From the multiple lines of evidence described above, we conclude PSR~J1842+0358 is an orthogonal rotator.

\subsection{Single pulse stack}
\label{sec:j1842pulse}

The most significant feature
in the single pulse stack (the first 1000 pulses shown in panels (a) and (b) of Fig.~\ref{fig:J1842sum}) is that both IP and MP display  intensity variability, and occasionally produce bright pulses. Especially the IP (panel b) shows significant variations in  single-pulse brightness, together with changes in emission phase. One example is found around pulse number 800. After a sequence of weak emission, the pulses start from a late pulse longitude of 270$^\circ$ at pulse number 785, but around 35 pulses the pulse is centred around 263$^\circ$ again. The pulses next shift towards later pulse longitude for another 30 pulses at around 267$^\circ$ until the emission is too weak to be seen. In the majority of the IP emission blocks, the trailing components appear first and the subpulses drift to earlier pulse longitudes. This suggests some form of subpulse modulation in the IP, though without a strict periodicity (for a detailed analysis see Sect.~\ref{sec:j1842spec}). The timescales of these IP periodicities are long, of the order of 100 pulses. 

In terms of the MP, the bright and narrow subpulses in the single pulses form a band-like structure (visible, e.g., in Fig.~\ref{fig:J1842sum}a, around pulse number 410), where pulses are brighter and wider than elsewhere. Such structure is even more visible in the 10-single pulse averaged MeerKAT observation in Fig.~\ref{fig:J1842mk}. There are some quasi-periodic intensity changes, but without a clear phase modulation. Some pulses show a wider pulse width with additional emission at pulse longitude 92 to 95$^\circ$, which also results in higher intensities in the single pulses (e.g.,~around pulse number 200, 420 and 620). The timescales of intensity changes seem to be shorter than that presented in the IP. More detailed analysis of the MP single pulse behaviour and its modulation is presented in Sect.~\ref{sec:j1842spec}.

From the single pulse stack, there is already a hint of anti-correlation between the MP and the IP, e.g.,~around pulse number 380 and 600 to 700, the MP is weak while the IP is bright. For pulse number 750, the opposite is seen -- the MP becomes bright while the IP fades. This anti-correlation is investigated in the remaining of this section, where we start a more quantitative way of selecting the bright and weak pulses.

\subsubsection{Separating bright and weak pulses}
\label{sec:j1842profmod}

Pulses can be  separated  based on  the summed onpulse intensities, which define bright and weak pulses. The two different resulting average pulse profiles can be used to investigate different behaviour. More interestingly, average profiles of the MP (or IP) can be constructed when the other component is bright or weak. This is useful to confirm whether  anti-correlations between the different intensity groups of the pulses exist (at zero phase delay), as hinted at by the pulse stack. For PSR~J1842+0358, the brightest 1/3 pulses of the MP and the IP are regarded as the bright group, respectively; the remainder are labelled weak. This selection is based on the fact that, by eye, the brightest 1/3 pulses best match the bright pulses in the IP. The selected bright pulses are indicated by horizontal bars in the left margin of Fig.~\ref{fig:J1842sum} (a, b), in blue for the MP and in violet for the IP, respectively. The single pulse intensities line graph is  shown in Fig.~\ref{fig:J1842sum}c, again using blue (MP) and violet (IP). The bright intensity blocks in the IP are more organised than the MP, and they match with the bright pulses. The variability in the MP seems to be more stochastic: the selected bright pulses hardly form coherent  blocks.

The pulse energy distributions help to identify whether there are modes (e.g.,~showing a bi-modal distribution). Such distributions are computed by the single pulse energies normalised by the mean of the MP intensities, and are shown in Fig.~\ref{fig:J1842dist}. Both the MP and the IP have a continuous distribution, without a significant bi-modality. This thus offers little evidence of mode changing. We suggest, rather, that (quasi-)periodic modulations produce the intensity variations in the single pulses (see also Sect.~\ref{sec:j1842spec}). The cut-off between the bright/weak groups of the MP and the IP intensities are shown in blue and violet lines, respectively.

Above, bright and weak single pulses were separated through  the individual intensities. We next perform this separation through the running-averaged intensities. That smooths out the stochastic pulse variability, while the long-term variability becomes more significant -- as the single pulse selected groups tend to have short transitions within a few single pulses (see the horizontal bars in Fig.~\ref{fig:J1842sum}c). Running-averaged intensities are comparable to  doing a smoothing by eye. The running averaged intensities are calculated as the following: for each single pulse, the intensity is given as the mean of 11 pulses around it. These are shown in Fig.~\ref{fig:J1842sum}d in green (MP) and brown (IP)\footnote{The meaning of the blue, violet, green and brown colour coding is identical throughout Figs.~\ref{fig:J1842sum}$-$\ref{fig:J1842selavgprof} and Figs.~\ref{fig:J1926sum}$-$\ref{fig:J1926selavgprof}.}. Now the quasi-periodic modulations are more pronounced. The resulted pulse energy distributions are in green lines in Fig.~\ref{fig:J1842dist}. The cut-off of the bright/weak pulses is similar to that of the single pulses, even though the distribution becomes narrower for both pulses because of the averaging. The same (1/3, 2/3) selection criterion is applied and the result is shown in the green and brown bars in the right margin of panels Fig.~\ref{fig:J1842sum}a, b, and in d. This method forms blocks with more continuity. The anti-correlation between the MP and the IP also becomes more visible -- both in the graph lines, and the block bars.

\subsubsection{(Anti-)correlation between the MP/IP brightness}
\label{sec:j1842profcorr}

From these bright/weak groups we now construct average pulse profiles, to show whether these profiles have different shapes/intensities, and study their possible (anti-)correlation.

\begin{figure}
    \centering
    \includegraphics[width=\linewidth]{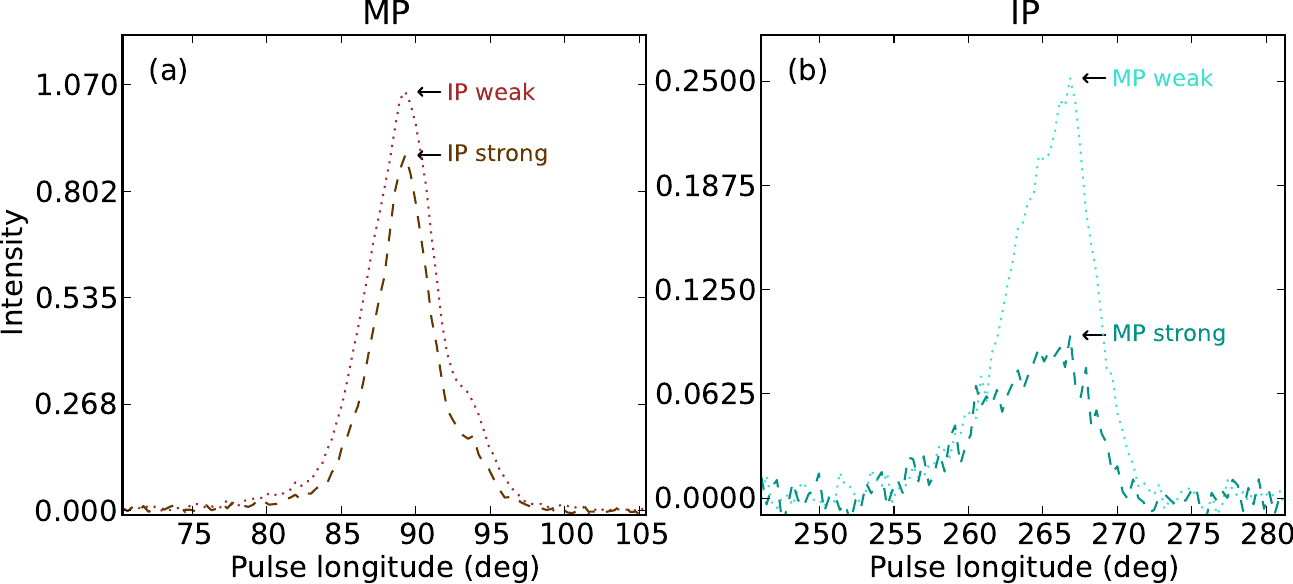}
    \caption{Average pulse profiles of the MP and the IP of PSR~J1842+0358, based on the weakest and strongest groups of running-averaged pulses from the other pulse. (a) MP profiles, based on, from top to bottom: weakest IP in running averaged (dark red), then brightest IP in running averaged (brown) marked in Fig.~\ref{fig:J1842dist}b. (b) IP average pulse profiles of the weakest and brightest groups of the MP, with colours in cyan and green, marked in Fig.~\ref{fig:J1842dist}a.
    }
    \label{fig:J1842selavgprof}
\end{figure}

These profiles are shown in Fig.~\ref{fig:J1842selavgprof}. They indeed demonstrate an anti-correlation between the MP and the IP. For simplicity, only the average profiles constructed using the bright and weak groups based on the running-averaged intensities are shown. In panel (a), the MP average pulse profile constructed using the weak group of IP intensities (dark red) has higher intensity than that of the bright group of IP intensity (brown). Similarly in the IP (panel b), the average profile from the MP weak pulses (cyan) is brighter than those constructed from the MP bright pulses (green). 

The same anti-correlation is revealed by comparing the profiles based on the bright and weak groups of single pulse intensities (see Fig.~\ref{fig:J1842selavgprofall}c, d in Appendix~\ref{app:addit}, and a description of the normalised profiles too). In addition, the two IP profiles based on the running averaged intensities of the MP show a larger difference in intensity than those based on the single pulse intensities. This indicates that the stochastic variability in the MP single pulses does not correlate with the changes in the IP intensities -- rather, the long-term variability (which is more pronounced in the running averaged intensities) is the reason for the anti-correlation between the MP and the IP. Profiles constructed using only the first and second half of the pulse sequence confirm the anti-correlation.

Another way of visualising the anti-correlation is to compute the longitude-resolved cross-correlation between the MP and the IP, at zero (or non-zero) phase delay. This is done by calculating the cross-correlation between each pair of intensities in the pulse profile for the single pulse sequence. At zero delay, strong positive correlation is expected between the onpulse of the MP and the IP themselves (as auto-correlations). On the other hand, negative correlations are seen between the MP and the IP, clearly confirming the MP and the IP anti-correlation. The longitude-resolved correlation maps for the FAST observation and the MeerKAT observation are presented in Figs.~\ref{fig:J1842phasecorr0} and \ref{fig:j1842mklrcc0} in Appendix~\ref{app:corrmap}, both reveal an anti-correlation between the MP and the IP. The result from both observations suggests that the anti-correlation is maintained over a long time on timescales of years. Such behaviour is intrinsic to the pulsar and is not caused by instrumental effects.

In summary, the averaged pulse profiles constructed using the bright and weak pulses of the other pulse indicate the anti-correlation between the MP and the IP, and such correlation is associated with the long-term variability. In addition, the longitude-resolved correlation map further demonstrates this anti-correlation appears for all onpulse longitudes between the MP and the IP at zero phase delay, and such correlation is maintained over years. The timescales of the long-term variability in the MP and the IP are investigated next.

\subsection{Long-term quasi-periodic modulations} \label{sec:j1842spec} 

Based on the analysis of the average profiles of the bright and weak pulses, we demonstrated that long-term single-pulse variability  is responsible for the anti-correlation. We investigate such proposed MP/IP (quasi-)periodic modulation in this Section. In the pulse stack (Fig.~\ref{fig:J1842sum}), both the MP and the IP show intensity modulations. Especially in the IP, the single pulses can change from nearly no emission to clearly detected pulses. Are these changes periodic?

\begin{figure}
\centering
\begin{subfigure}{0.5\linewidth}
  \centering
  \includegraphics[width=0.95\linewidth]{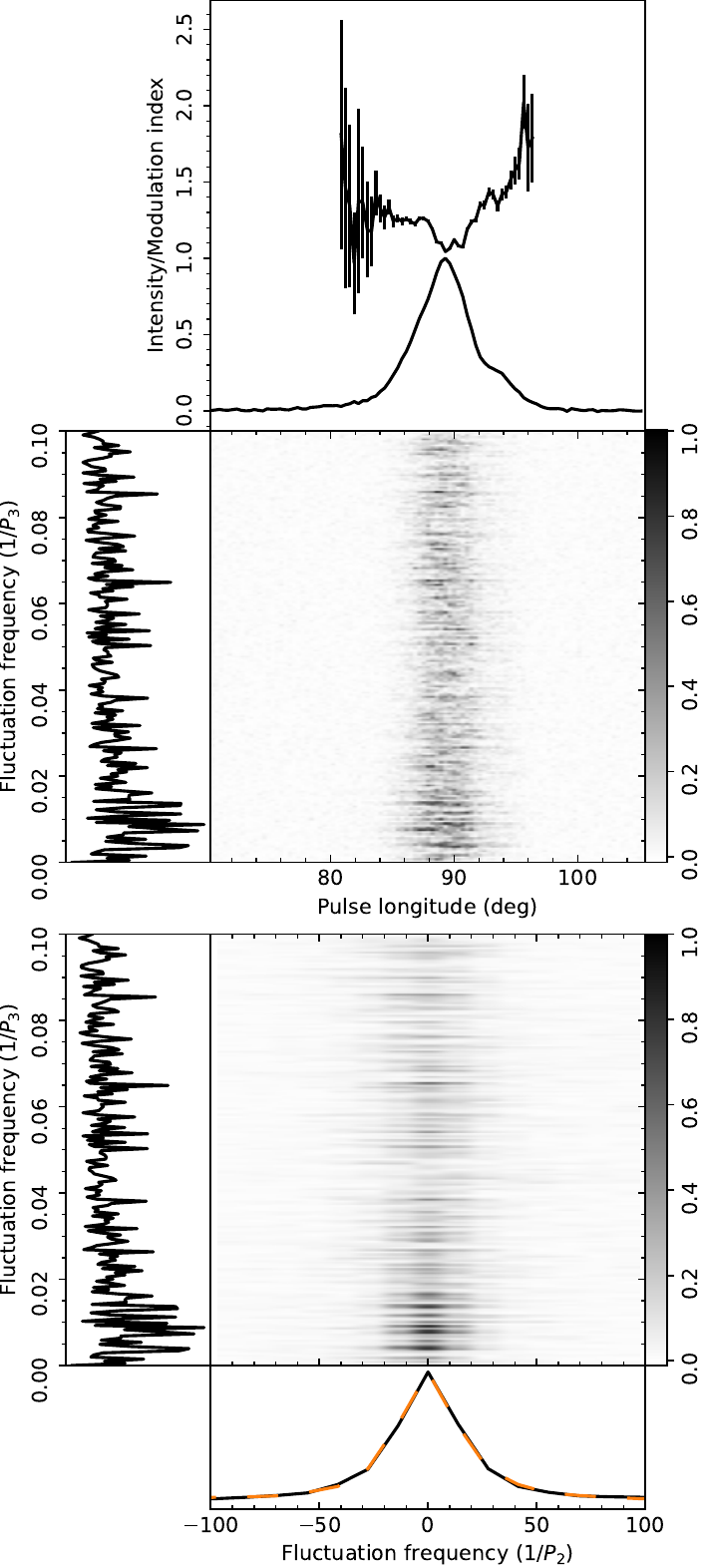}
  \label{fig:J1842specmp}
\end{subfigure}%
\begin{subfigure}{0.5\linewidth}
  \centering
  \includegraphics[width=0.95\linewidth]{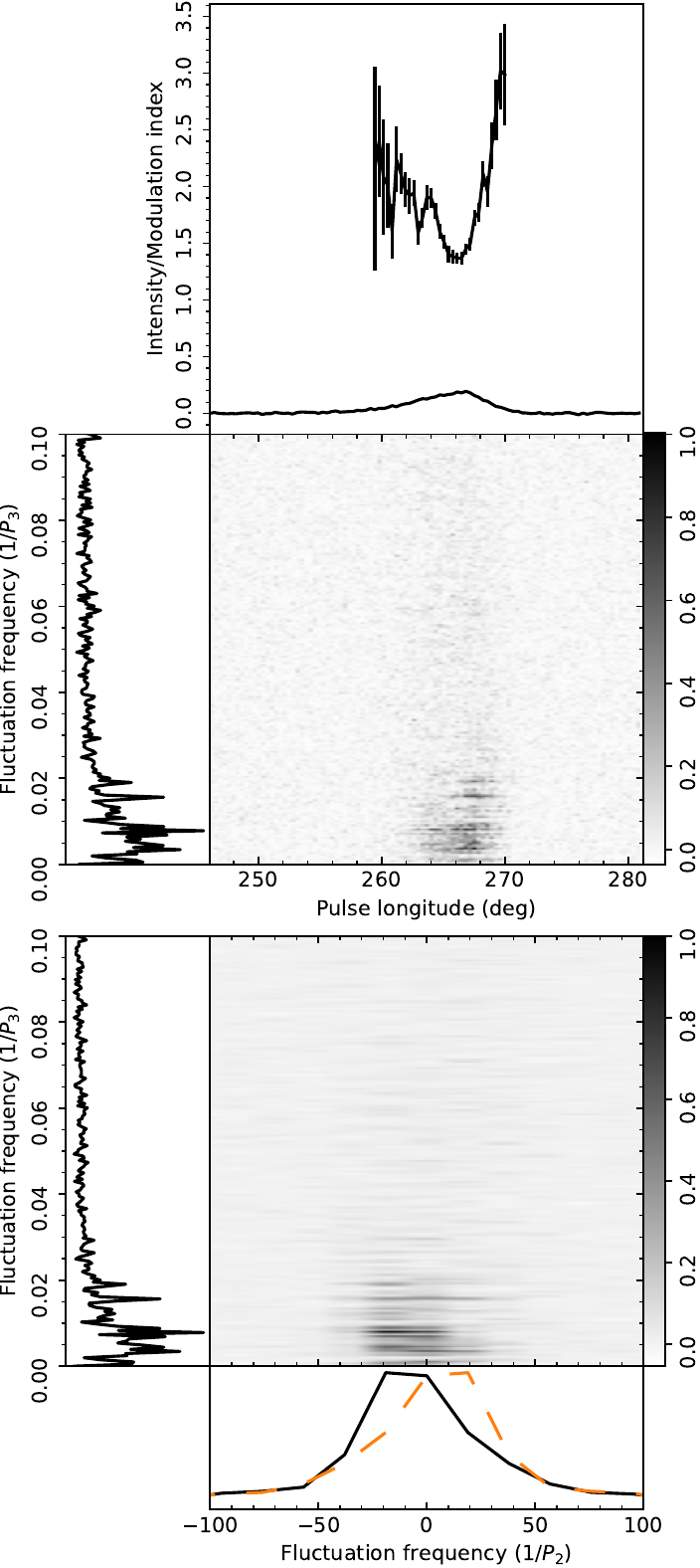}
  \label{fig:J1842specip}
\end{subfigure}
\caption{Fluctuation spectra of PSR~J1842+0358 for the MP and the IP (left and right panels). The top panel shows the average pulse profile, along with the modulation indices at each pulse longitude (black dots with errorbars). The middle panel shows the LRFS as fluctuation frequency ($1/P_3$, in cycles per period, cpp) versus pulse longitude (in degrees). The side panel is the integrated power along the horizontal direction. The bottom panel is the 2DFS, with fluctuation frequencies on both axes in cpp, with $1/P_3$ versus $1/P_2$. Both the left and bottom panels are integration of powers along the horizontal and vertical directions, respectively. The orange curve is the horizontal mirror image of the black curve in the bottom panel, indicating whether there are any possible drift in the pulse stack. The colour bar associated with the LRFS/2DFS indicates the scale in the plots. If the maximum is not 1, larger values have been clipped.}
\label{fig:J1842spec}
\end{figure}

Single pulse periodicities can be identified from the longitude-resolved fluctuation spectrum (LRFS; \citealt{Backer1970}), while the 2DFS (introduced earlier)  can next identify whether the periodicity shows drift or not. They are both calculated through fast Fourier transforms (FFTs) on single-pulse blocks of a certain length. Given the long-term intensity variabilities we choose a relatively long block size of 2048 pulses. The overall LRFS/2DFS is an averaged version of those computed from all blocks (two in this case). The LRFS computes FFTs at each pulse longitude, such that the periodicities ($P_3$) are determined per pulse number for each pulse longitude. The LRFS can also be used to compute the modulation indices, representing the fractional change of the single pulse intensity for each pulse longitude (black dots in the top panel of Fig.~\ref{fig:J1842spec}, and see \citealt{Weltevrede2006} for details on how these are computed). 

The modulation indices often form a U-shaped curve with the minimum occurring at the intensity peaks, and they increase towards the edge of the pulse profile. Both the MP and the IP show such a U-shaped curve. The modulation indices of the IP are relatively large, with values around 2 -- because the single pulses can change from nearly zero emission to clearly detected single pulses. Unlike the IP, the MP single pulses are always detected, and the corresponding modulation indices are between 1 to 1.5, with minimum values around the peak of the pulse profile. This behaviour in the FAST data is consistent with the TPA data. 

In the LRFS (second row in Fig.~\ref{fig:J1842spec}, zoomed in between 0 and 0.1~cpp in 1/$P_3$ frequency), both the MP and the IP display  a broad spectral feature at the lowest $1/P_3$ frequencies, of $\sim$0.02\,cpp and below. This broad spectral feature results in a periodicity centred around 112 but with a large errorbar ($P_3 \simeq 114 \pm 40$ for the MP and $P_3 \simeq 112 \pm 76$ for the IP, respectively). This periodicity is more pronounced in the IP, with a more significant power towards the lowest frequency. The MP LRFS reveals power along all $1/P_3$ frequencies for the onpulse regions, with the integrated $1/P_3$ curve clearly offset from zero -- indicating that the MP exhibits large stochastic variability. Having a closer look of the IP LRFS, the trailing component around 268$-$270\degr\ revealed a shorter periodicity, around 65\,$P$. This mostly relates to the bright trailing edge emission. This is also hinted by a peak in the integrated $1/P_3$ frequency side panel. Given the quasi-periodicity, such short periodicity is nevertheless blended with the long-term modulation. The frequency features are equally present when halving the sequence. 

From the 2DFS of
the MP (third left panel of Fig.~\ref{fig:J1842spec}), the lowest frequency feature is centred around $1/P_2=0$ frequency, without a clear offset, therefore ruling out significant drift. In the 2DFS of the IP (third right panel of Fig.~\ref{fig:J1842spec}), a negative drift is observed in the bottom integrated $1/P_2$ panel, indicating that subpulses drifting from the trailing to leading part of the IP pulse profile is more significant. Such drift is more significant in the trailing part of the IP, for example the sequences in Fig.~\ref{fig:J1842sum}b between pulse number 788 and 820, and between pulse number 920 and 945. 

Another way of testing whether drifting is present or not, is by constructing the subpulse phase track at each pulse longitude. This can further help identify the phase difference between the MP and the IP. Instead of taking the square of the complex values in calculating the LRFS, the complex phase is regarded as the subpulse phase, computed from a specific spectral bin where the strongest periodicity occurred in the LRFS (see \citealt{ES2002} and \citealt{Weltevrede2016}). This produces the phase differences against  pulse longitude. 

\begin{figure}
\centering
\includegraphics[width=\linewidth]{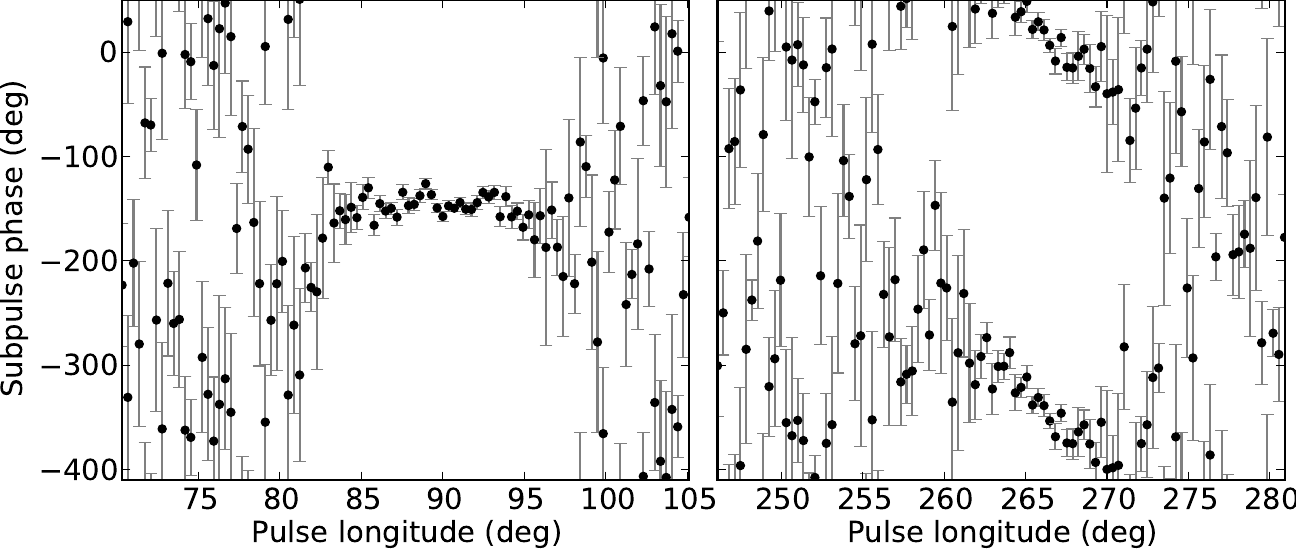}
  \label{fig:J1842trackmphigh}
\caption{Subpulse phase tracks of the MP and the IP (left and right column) of PSR~J1842+0358 computed at a $P_3$ periodicity around 112. The two panels presents the subpulse phases with errorbars, and with an arbitrary phase values in the vertical direction.}
\label{fig:J1842track}
\end{figure}

We focus on the $1/P_3$ frequency, for  $P_3 \simeq 112\,P$, to construct the subpulse phase relation (Fig.~\ref{fig:J1842track}). The MP periodicity is roughly independent of pulse longitude (left panel). In the IP, however, the subpulse phase decreases with pulse longitude (right panel). This means the subpulses drift from the trailing to the leading part of the IP profile. Such subpulse behaviour is similarly concluded from the 2DFS. In addition, the phase difference between the mid-point of the IP and the MP is roughly 180\degr. Because of the drifting nature in the IP, this 180\degr\ is a rough estimation. This indicates that there is an anti-correlation between the MP and the IP.

The spectral analysis shows that both the MP and the IP possess long-term quasi-periodicities, which have a broad spectral feature without a sharply defined periodicity. The MP reveals short-term quasi-periodicities as well towards higher $1/P_3$ frequencies. The MP modulation is longitude stationary without a clear drift feature. The IP has a more clearly defined lowest frequency feature and shows a negative drift across the pulse profile. The subpulse phase relation associated with the longest periodicity reveals a roughly 180\degr\ difference between the MP and the IP, further supporting the anti-correlation between the two. 

\subsection{Timescales of the correlated emission}
\label{sec:j1842corr}

\begin{figure}
    \centering
    \includegraphics[width=\linewidth]{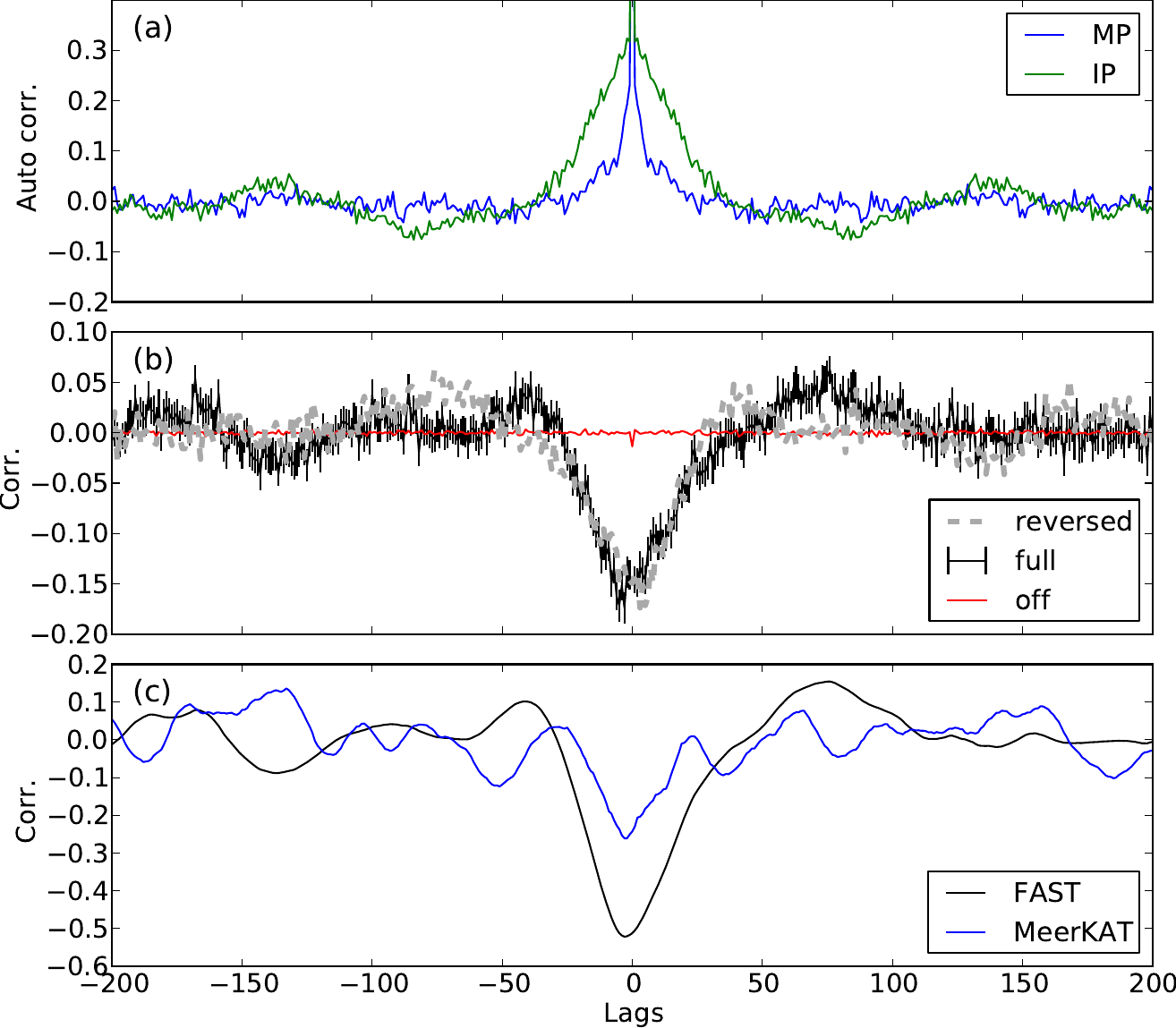}
    \caption{Correlation plots of PSR~J1842+0358. (a) shows the auto-correlation curves of the MP and the IP intensities, which reveal the periodicities in the pulse intensities. The plots are zoomed in to the y-range of $-$0.2 to 0.4. (b) shows the cross-correlation between the MP and the IP single pulse intensities and its errorbars in black line, and the correlation when reversing the order of the single pulses in grey dashed line. The corresponding correlation of the off-pulse regions is in red. (c) shows the correlation plots computed using the running averaged intensities of the MP and the IP, with the FAST sequence in black, and the MeerKAT sequence in blue, respectively.
      }
    \label{fig:J1842corr}
\end{figure}

As discussed in Sect.~\ref{sec:j1842pulse} and \ref{sec:j1842spec}, the MP and the IP have the same long-term quasi-periodic modulations, and these  variations are anti-correlated. The periodicity can also be tested using the auto-correlation of the summed pulse intensities. By definition the auto-correlation peaks at zero and is symmetric about zero. In Fig.~\ref{fig:J1842corr}a, the MP auto-correlation (in blue) shows a fast decrease for lags smaller than 10, with some evidence of an increasing at around lag 11/12. This indicates additional variations on top of the long-term variations, which are also revealed from the powers in the LRFS along all $1/P_3$ frequencies (left side panel of Fig.~\ref{fig:J1842spec}). Both the MP and the IP possess long-term variabilities, i.e.,~the first peak-to-peak periodicity for the MP and the IP are around 130 to 140$P$, but the peaks around this periodicity are wide therefore are not stable. This long-term periodicity is consistent with the lowest frequency feature in the LRFS. In the MP, the periodicities are less clear than in the IP, a result of the large stochastic MP variability.

To inspect the MP-IP anti-correlation on the single-pulse level, we compute the cross-correlation between the summed intensities of the pulses. The resulted curve is shown in Fig.~\ref{fig:J1842corr}b. The corresponding lags are computed such that a positive lag means the MP lags the IP, and vice versa. The maximum correlation is $-0.17\pm0.01$ and appears to occur at a negative lag (but as discussed in Appendix~\ref{app:lags}, this lag is insignificant and consistent with zero). The errors on the cross-correlation are determined by random shuffling the pulse order 100 times, determining each shuffled cross-correlation, and taking the standard deviation of those as the error.

Comparing the cross-correlation in panel b and the auto-correlations in panel a in Fig.~\ref{fig:J1842corr}, the long-term variabilities in the MP and especially the IP highly resemble the periodicities in the cross-correlation, although with a different sign at the same lag. This further supports the idea that the long-term variabilities are responsible for the anti-correlation between the MP and the IP.

To demonstrate the significance of the anti-correlation, we compared it to  the  cross-correlation of the off-pulse region (in red in panel b). All aspects of the method  other than the pulse-phase range  were unchanged. The on-pulse anti-correlation is much larger than that of the off-pulse, indicating it is significant and not caused by instrumental effects or RFI. We also checked that the anti-correlation presents in the two halves of the sequence. 

From the cross-correlation curve, the most significant feature is that the curve is asymmetric around zero -- the first peak with positive correlation appears around 80 lags for positive lags, which is longer than the first positive peak at negative lags around $-$40 lags. Such asymmetry is best seen in Fig.~\ref{fig:J1842corr}b, by comparing against the dashed curve, computed by reversing the order of the pulses. The asymmetry is also visible in the cross-correlation curve of the two halves of the pulse sequence. The behaviour of the correlation curve is further investigated in Sect.~\ref{sec:J1842simcurve}, which is likely to be caused by the different emission patterns in the MP and the IP. 

As followed from Fig.~\ref{fig:J1842selavgprof}, the IP intensity difference based on weak and bright MP pulses increases if the running average is used, compared to using the individual pulses. This suggests the long-term variability is responsible for the anti-correlation. To show this, we now also determine cross-correlation using the running averaged method, applied to both the FAST and MeerKAT data. In Fig.~\ref{fig:J1842corr}c, the cross-correlation curve of the FAST running-averaged intensities (black) reveals an anti-correlation of increased significance, with a maximum of $-0.52\pm0.01$. The curve is similar to that computed with the single pulse intensities, showing an asymmetry around zero lag, but with a much larger correlation and a more significant quasi-periodicity. The maximum cross-correlation between the MP and the IP running-averaged intensities of the MeerKAT data (blue) is weaker than that of the FAST data. This is because the MeerKAT S/N is lower. 

\subsection{Summary}

Using the FAST data we have shown, from the average profiles constructed from different pulse intensities (Sect.~\ref{sec:j1842pulse}), and from the cross-correlation analysis (Sect.~\ref{sec:j1842corr}), that there is an anti-correlation between the MP and the IP intensities, and that this correlation is caused by the long-term intensity variability. From the spectral analysis (Sect.~\ref{sec:j1842spec}), both pulses reveal quasi-periodicity modulation at a long timescale $P_3$ over 100 pulses.

\section{PSR~J1926+0737}
\label{sec:j1926}

In this Section, we discuss the analysis of the MeerKAT and FAST data for PSR~J1926+0737. The approach is similar to that described in Sect.~\ref{sec:j1842} for PSR~J1842+0358, so please refer to that Section for analysis details if needed. 

\subsection{TPA pulse stack}
\label{sec:j1926tpa}

\begin{figure}
    \centering
    \includegraphics[width=\linewidth]{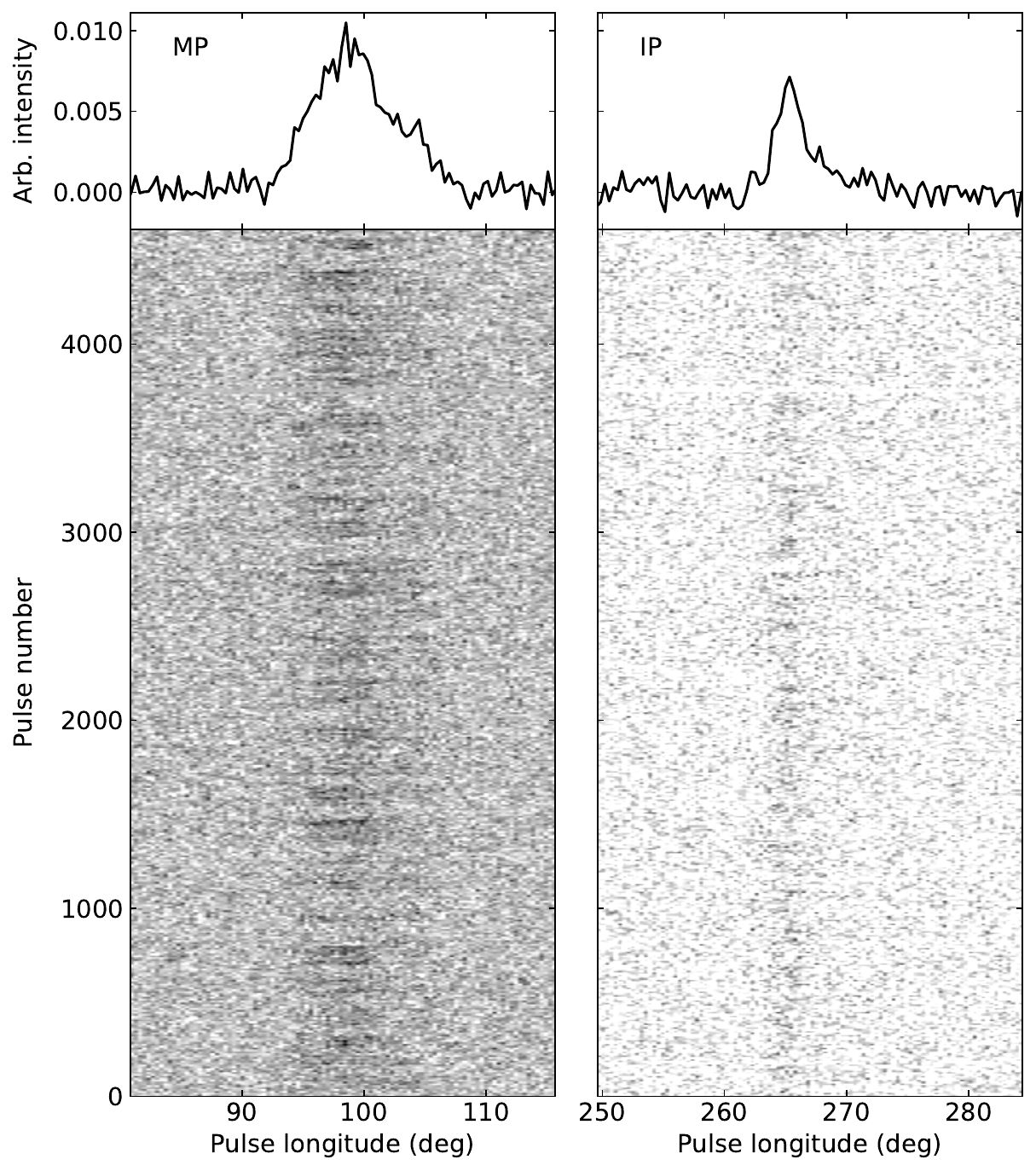}
    \caption{MP (left) and IP (right) pulse stacks of PSR~J1926+0737. The pulse sequences were averaged over 10 single pulses but the pulse number shows the number of single pulses. The corresponding average profiles of all single pulses are shown on the top panels.}
    \label{fig:J1926mk}
\end{figure}

The pulse stacks and  profiles of PSR~J1926+0737 from the MeerKAT TPA data are shown in Fig.~\ref{fig:J1926mk}. The MP shows some modulation, on very long timescales ($\sim$100\,$P$) and not strictly periodic. No clear periodicity is easily identified in the IP emission pattern. From the 2DFS in \citet{Song2023}, only the MP has a low-frequency feature, without a clear drift. To determine if the quasi-periodicity in the MP extends to the IP too, we  pursued higher S/N observations using
FAST. 

\subsection{Average pulse profile}
\label{sec:j1926prof}

\begin{figure}
    \centering
    \includegraphics[width=\linewidth]{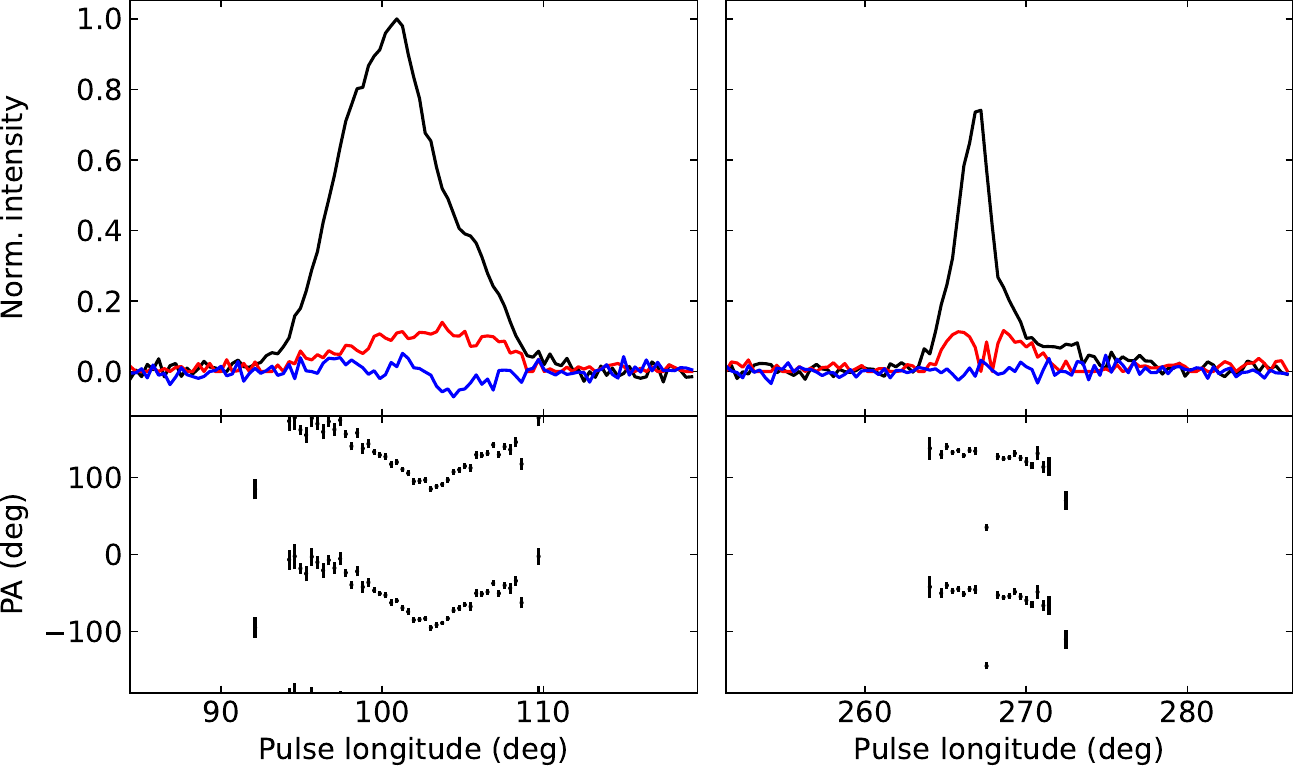}
    \caption{Average pulse profile of PSR~J1926+0737 (top plot), for the MP (left) and the IP (right). The total intensity profile (normalised by the MP peak) is in black, and the corresponding linear and circular polarisation are shown in red and blue, respectively. The bottom plots show the computed polarisation position angles.}
    \label{fig:J1926prof}
\end{figure}

Both the MP and IP of this pulsar have asymmetric profiles (top panels in Fig.~\ref{fig:J1926prof}). The 14\degr-wide MP consists of multiple components. In particular, one component in the trailing part is marked by bright emission in some single pulses. The IP is narrower  ($\simeq9$\degr) and consists of a Gaussian convolved with an exponential tail. There is an increase of emission around pulse longitude 270\degr, but it is not clear from the single pulses (further discussed in Sect.~\ref{sec:j1926pulse}) where this originates.

\subsubsection{Evidence of an orthogonal rotator}
\label{sec:j1926ortho}

The separation between the peak of the MP and the peak of the IP is about 166.3\degr, smaller than the two-pole expectation of 180{\degr}. But the narrowness of the MP/IP profiles and the lack of other, bridging emission, argue for a double-pole pulsar. Figure~\ref{fig:J1926prof} shows the polarisation profiles and PAs. The PA-swing of the MP forms a V-shape. The IP PAs are mostly flat across the pulse longitude. There is a trend of decreasing PAs towards the trailing part of the pulse profile, but with larger errorbars. The PA measured around the pulse profile peak (around 268\degr) shows a dip that is 100\degr\ lower than the other PA points. This point is likely to be insignificant due to the low linear polarisation (in red on the top panel). These are consistent with the findings in \citet{sun+25}.

From these PAs we cannot constrain the sign of $\beta$ as was done in Sect.~\ref{sec:j1842ortho}. As the PAs are similar at both the MP and IP peaks ($\sim$100{\degr}), the pulsar is consistent with an orthogonal rotator however. 

\subsection{Single pulse stack}
\label{sec:j1926pulse}

\begin{figure*}
    \centering
    \includegraphics[width=0.96\linewidth]{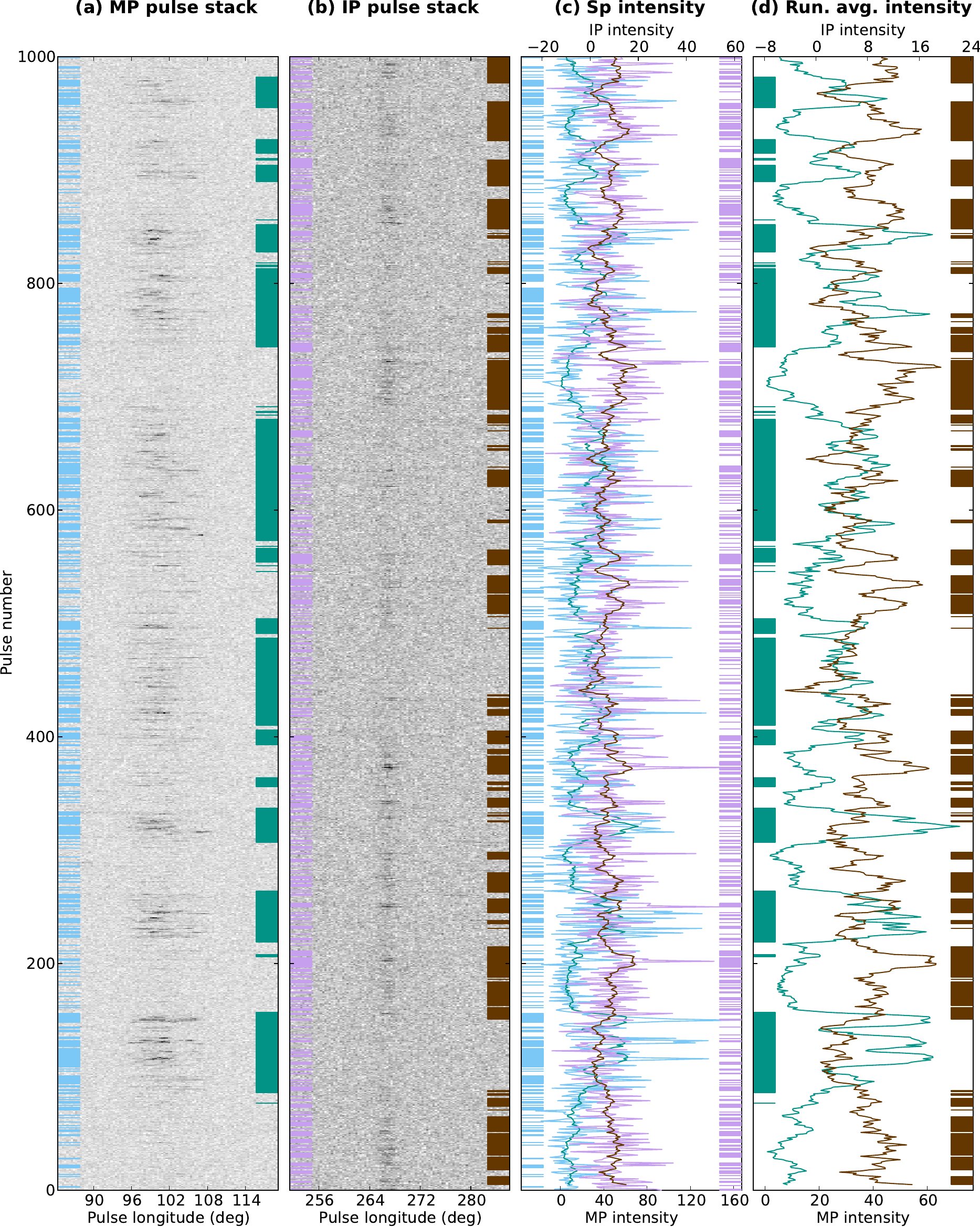}
    \caption{The first 1000 single pulses for PSR~J1926+0737:  pulse stacks (MP, IP) and intensities (single-pulse, running-averaged). (a)~Pulse stack of the MP. The blue and green margin bars mark the bright pulses, based on the intensities of the single pulses (blue, left) or their running averages (green, right). (b) As (a) but for the IP, now with violet and brown for single-pulse and running-averaged intensities, respectively. The single pulse intensities of the MP and the IP are in (c), with the running averaged intensities plotted on top. Margin bars identical to the left of (a) and (b). The running averaged intensities of the MP and the IP plotted alone are in (d), along with the corresponding selected bars (the same as those on the right of (a) and (b)). The rest of the \PSRii pulse stacks are in Appendix \ref{app:stacks1926} (Online only).}
    \label{fig:J1926sum}
\end{figure*}

From the pulse stack of the first 1000 pulses in Fig.~\ref{fig:J1926sum}, it is clear that both the MP (panel a) and the IP (panel b) emission show intensity modulation. Such modulation does not have strong pulse longitude variation (i.e.,~drifting subpulses). Whether there are any periodicities within the single pulses or not is not entirely clear from just looking at the pulses (but see Sect.~\ref{sec:j1926spec}). The MP seems to have an additional emission around pulse longitude 108${^\circ}$, which coincides with the extra component in the average pulse profile.

\subsubsection{Separating bright and weak pulses}
\label{sec:j1926profmod}

\begin{figure}[bt]
    \centering
    \includegraphics[width=\linewidth]{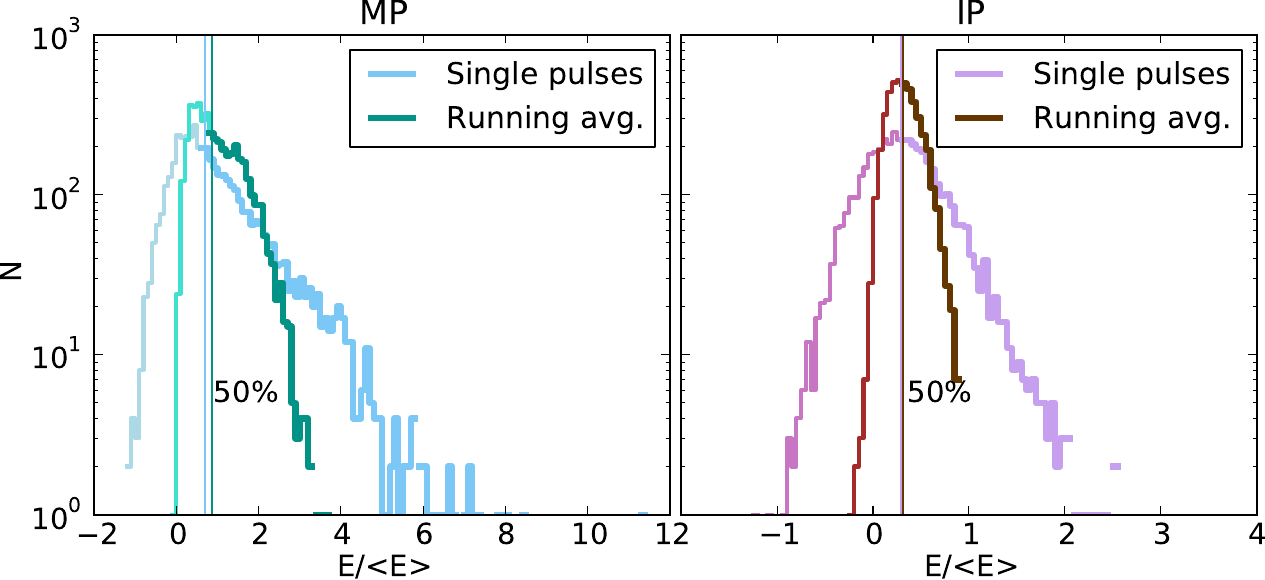}
    \caption{Pulse energy distributions of PSR~J1926+0737 for the MP (left) and the IP (right), with single pulse energies (blue and violet) and running averaged pulse energies (green and brown). Both pulse energies are normalised by the mean MP energy. The vertical line shows two groups that are split at the median pulse intensity (as described in Sect.~\ref{sec:j1926profmod}).}
    \label{fig:J1926dist}
\end{figure}

The single pulse stack (Fig.~\ref{fig:J1926sum}a, b) already visually suggests  an anti-correlation between the MP and the IP -- at around pulse number 100, for example, the IP is much brighter and the MP is nearly undetected. Around pulse number 150, on the other hand, the MP becomes more visible while the IP is not seen. 

To see if the  brightness in one pulse correlate oppositely of  the other pulse, we apply again the methodology from Sect.~\ref{sec:j1842profmod}. We construct average pulse profiles for different total intensities of the other component (IP or MP). From a pulse-stack identification by eye we determine the pulse-energy split fraction, and we put it at the 50th percentile. These brightest half are again marked in blue (MP) and violet (IP) bars in Fig.~\ref{fig:J1926sum}a, b with their intensities are in panel c in the same colour. Pulse blocks where the emission is stronger have higher density of bars (e.g.,~for the MP around pulse number 100, 210, for the IP around pulse number 0 to 50, around 200). The single pulse intensities also show some spiky structures. 

The single-pulse energy distributions (blue for MP and violet for IP in Fig.~\ref{fig:J1926dist}) show no significant bi-modality. The intensity modulation is therefore not due to mode changing. To reduce the stochastic variability inherent to the single pulses, we also again analyse the data using 11-single-pulse running-averaged intensities. Now the pulse energy distributions (green lines in Fig.~\ref{fig:J1926dist}) are indeed narrower. 

Using the running averaged intensities (Fig.~\ref{fig:J1926sum}d in green and brown for the MP and the IP), the long-term variations become much more significant. The selected 50-percentile bright/weak blocks of the MP and the IP are now clearly anti-correlated.

\subsubsection{(Anti-)correlation between the MP/IP brightness}
\label{sec:j1926profcorr}

\begin{figure}
    \centering
    \includegraphics[width=\linewidth]{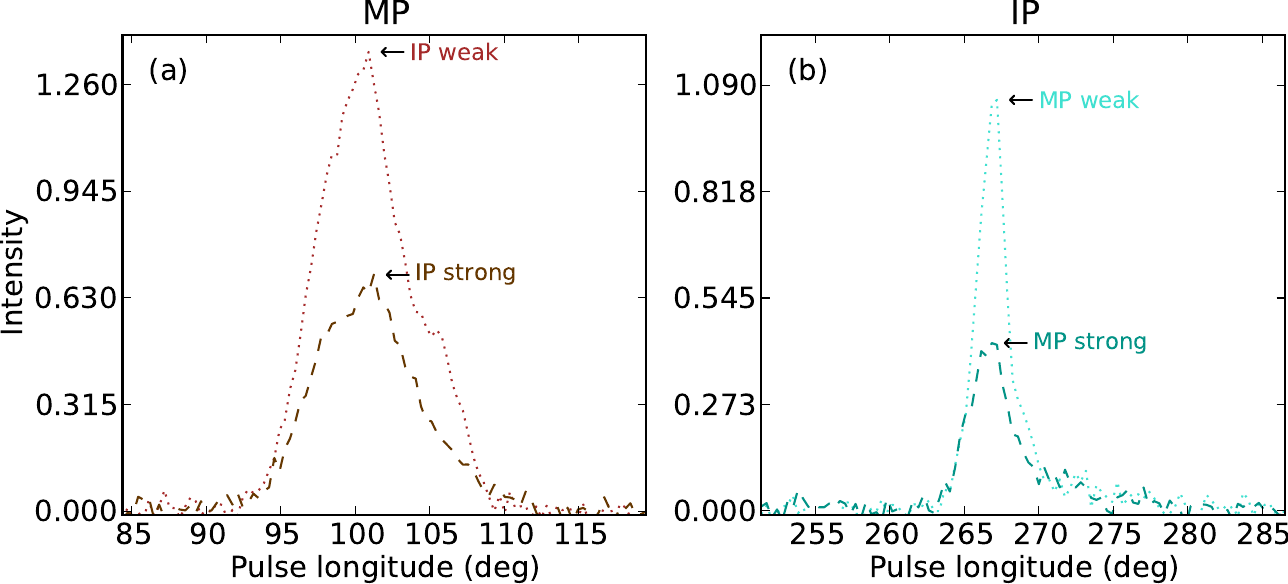}
    \caption{Average profiles of the MP and the IP of PSR~J1926+0737 based on grouping pulses by the brightest and weakest halves of the opposite pulse. (a) MP profiles, based on, from top to bottom: weakest IP in running averaged (dark red), then brightest IP in running averaged (brown) marked in Fig.~\ref{fig:J1926dist}b. (b) IP average pulse profiles, based on the MP halves, with colours in cyan and green, as marked in Fig.~\ref{fig:J1926dist}a.
      An anti-correlation is clearly present.}
    \label{fig:J1926selavgprof}
\end{figure}

This MP/IP anti-correlation
is further demonstrated when again averaging profiles when the opposite pulse is weak or bright (Fig.~\ref{fig:J1926selavgprof}a, b). The MP is brighter when the IP is in its weaker half from the running average (dark red). It is weaker when the IP is bright (brown) -- corroborating the MP-IP intensity anti-correlation. The difference in  running-averaged-based intensities is larger than in the single pulse constructed profiles (see the full version in Appendix Fig.~\ref{fig:J1926selavgprofall}c, d). This indicates that the long-term variability in the IP correlates strongly to the MP intensity variation, rather than the stochastic variability, which is not correlated between the MP and the IP. The IP average profiles in panel b lead to exactly the same conclusions. The shape of the profile does not significantly change (see further description in Appendix and Fig.~\ref{fig:J1926selavgprofall}). Our first/last-half sequence tests, finally, confirm our finding: the anti-correlation is real and always present. 

The longitude-resolved cross-correlation map is computed to identify which parts of the profile anti-correlate. We find that the anti-correlation is observed along all onpulse longitudes between the MP and the IP -- identical to PSR~J1842+0358 (Sect.~\ref{sec:j1842profcorr}). The plots are shown in Appendix~\ref{fig:J1926phasecorr0} and ~\ref{fig:j1926mklrcc0} for the FAST and MeerKAT observation, respectively. As similarly concluded for PSR~J1842+0358, such anti-correlation is maintained over years and is not caused by instrumental effects.

In this section, we have shown, 
through constructing the average profiles of the bright and weak groups of the pulses, that the brightness of the MP and IP are anti-correlated. This anti-correlation is present along all on-pulse longitudes and is maintained over a long time. Moreover, it is the long-term variability in the MP and the IP that is  responsible for the anti-correlation. The quasi-periodicity will be investigated in the next Section.

\subsection{Probing the quasi-periodic modulations}
\label{sec:j1926spec}

\begin{figure}
\centering
\begin{subfigure}{0.5\linewidth}
  \centering\includegraphics[width=0.95\linewidth]{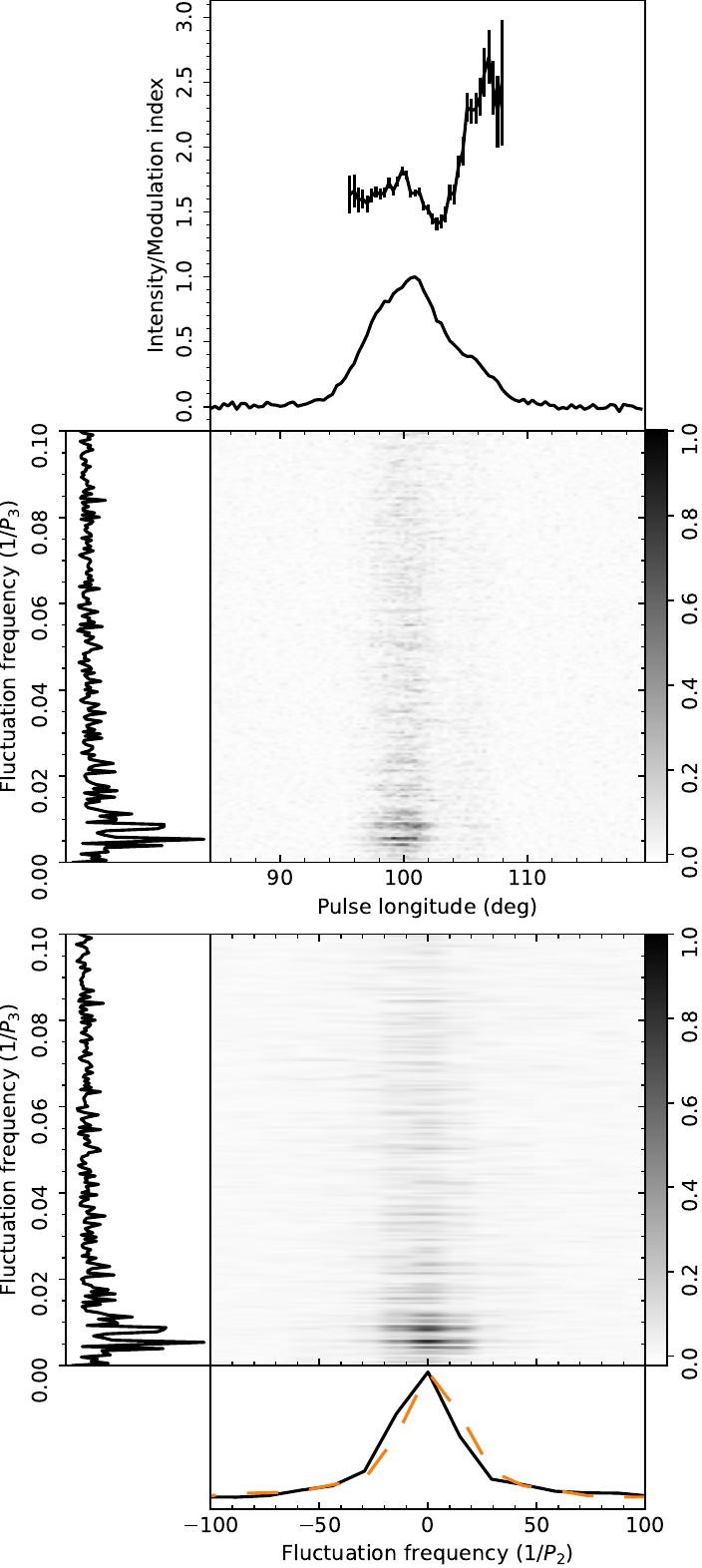}
  \label{fig:J1926specmp}
\end{subfigure}%
\begin{subfigure}{0.5\linewidth}
  \centering
  \includegraphics[width=0.95\linewidth]{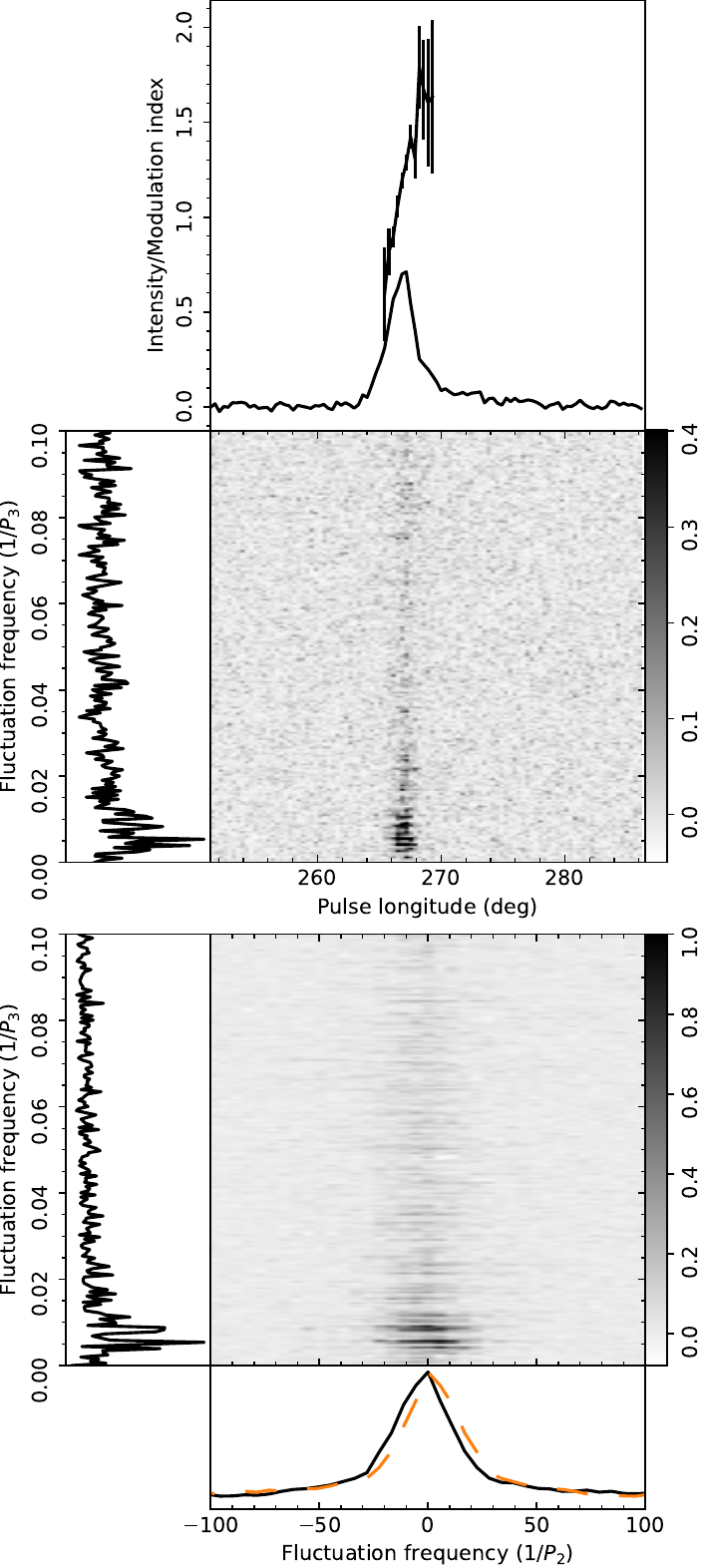}
  \label{fig:J1926specip}
\end{subfigure}
\caption{Fluctuation spectra of PSR~J1926+0737 for the MP and the IP (left and right panels). For full description, see Fig.~\ref{fig:J1842spec}.}
\label{fig:J1926spec}
\end{figure}

The running averaged intensities in Fig.~\ref{fig:J1926sum}d hint at the existence of (quasi-)periodic modulation in the single pulses. Such modulation can be revealed by constructing the LRFS and 2DFS (as described in Sect.~\ref{sec:j1842spec}) and these spectra are shown in Fig.~\ref{fig:J1926spec}. 

The modulation indices in the top panel (black lines with errorbars), show the MP (left panel) modulation is  mostly  flat in the leading part, at a modulation index of $\sim$1.7, followed by an increase towards the trailing edge. In the IP the single pulses are weaker, and the modulation indices are thus less constrained -- but still these form an increasing trend. Both the MP and IP  modulation indices are relatively high ($\gtrsim$1), indicating  intensity variability of the single pulses.
The FAST data, with its clearly detected and variable single pulses, produces larger MP modulation indices, with smaller errors, than the MeerKAT data.

The most significant feature in the LRFS (middle panels) is the lowest $1/P_3$ frequency feature below 0.02~cpp. Selecting a broad spectral feature resulted in a periodicity $\simeq 153 \pm 55$ from the MP, and $\simeq 147 \pm 63$ for the IP, respectively. The single pulse stack (Fig.~\ref{fig:J1926sum}) also suggests  shorter-term intensity variability on timescales of  a few tens of pulses, but there is no clearly constrained periodicity there in the LRFS. Stochastic variabilities towards higher $1/P_3$ frequencies (e.g.,~from 0.02 to 0.3 cpp) are seen as a red-noise spectrum, due to the intensity changes in the single pulses. Possibly, this short-term variability has an unstable timescale and is washed out by the longer-term (a few hundred $P$) modulation.

The 2DFS (bottom panels of Fig.~\ref{fig:J1926spec}) indicates  some overall negative drift for both MP and IP, but it  is not clearly associated with the lowest modulation frequency. Therefore, the dominant variability is longitude stationary. 

\begin{figure}
\includegraphics[width=\linewidth]{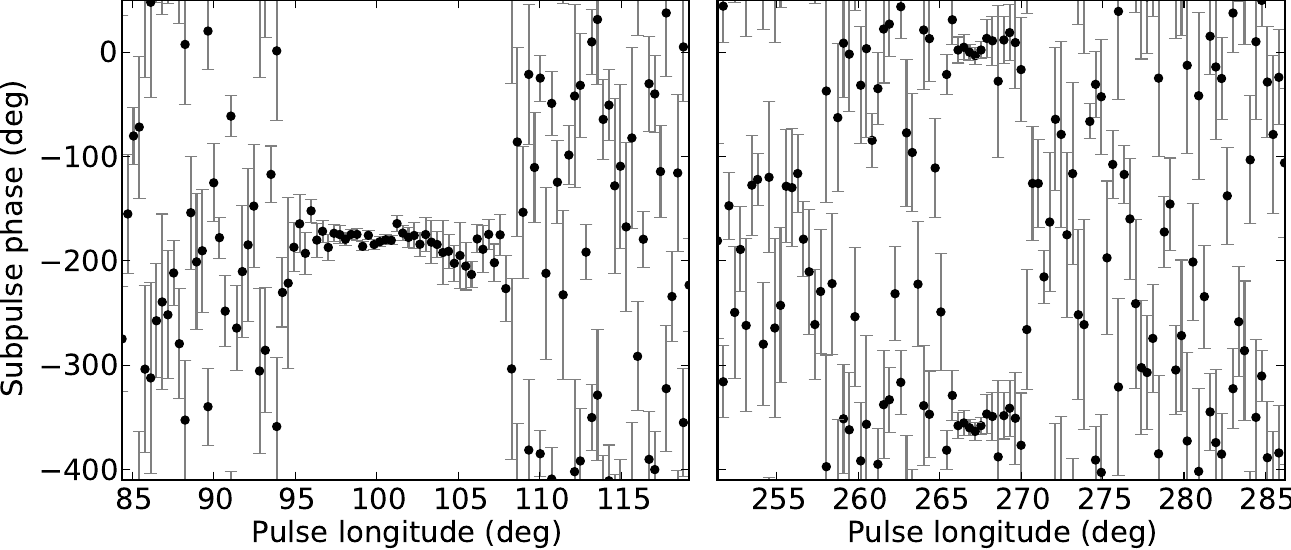}
\caption{Subpulse phase tracks of the MP and the IP (left and right columns) of PSR~J1926+0737 computed at a $P_3$ periodicity around 180. See caption of Fig.~\ref{fig:J1842track} for more details.
  }
\label{fig:J1926track}
\end{figure}

To visualise the phase variations associated with the main periodicity, we construct subpulse phase tracks (bottom panels of Fig.~\ref{fig:J1926track}). We use the bright MP/IP feature around $P_3 \simeq 185$. As the resulting phases are mostly flat, the periodicity lacks clear subpulse drift. The phase difference between the MP and the IP is close to 180\degr, albeit with high uncertainty given the weak IP. Nevertheless, this phase difference provides further evidence of the anti-correlation between the MP and the IP intensity modulation.

\subsection{Timescales of the correlated emission}
\label{sec:j1926corr}

\begin{figure}
    \centering
    \includegraphics[width=\linewidth]{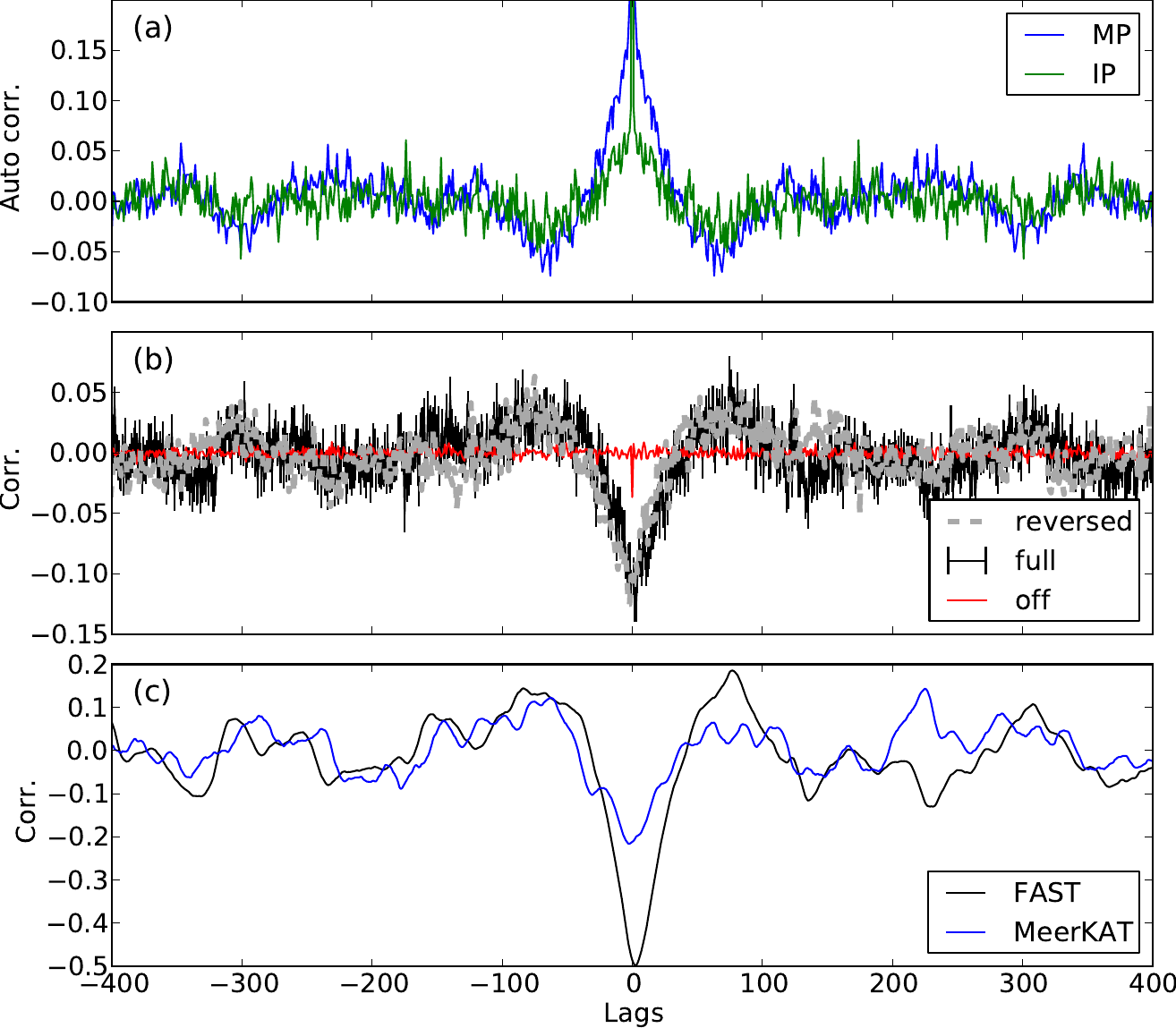}
    \caption{Correlation plots of PSR~J1926+0737. (a) shows the auto-correlation curves of the MP and the IP intensities, which reveal any periodicities in the pulse intensities. The plots are zoomed-in to the y-range of $-$0.1 to 0.2. (b) shows the cross-correlation between the MP and the IP single pulse intensities and its errorbars in black line, and the correlation when reversing the order of the single pulses in grey dashed line. The corresponding correlation of the off-pulse regions is in red. (c) shows the correlation plots computed using the running averaged intensities of the MP and the IP, with the Fast full sequence in black, and the MeerKAT full sequence in blue, respectively.}
    \label{fig:J1926corr}
\end{figure}

Correlation between the MP and IP summed intensities can determine the timescales  and delays of the quasi-periodicities, as determined in the previous Sections (e.g., Fig.~\ref{fig:J1926sum}c). Fig.~\ref{fig:J1926corr}a indicates a peak-to-peak periodicity of about 120$P$, consistent with the higher frequency spectral feature in the LRFS/2DFS. In the MP, the second minimum appears around lag 150, with further positive peaks around lag 240. This indicates that the periodicities are unstable.

To investigate any time delays in the MP-IP interaction, we determine the their intensity cross-correlation at different lags (Fig.~\ref{fig:J1926corr}b, black line). The maximum anti-correlation is $-0.13\pm0.02$, significantly larger than the reference off-pulse correlation (red), at a positive lag (but being consistent with zero lag, see Appendix~\ref{app:lags}). Such anti-correlation also presents in both halves of the sequence (not shown).

The cross-correlation based on the reversed pulse order
reveals no significant asymmetry (Fig.~\ref{fig:J1926corr}b). This  differs from PSR~J1842+0358 (Fig.~\ref{fig:J1842corr}), where the modulation in the IP featured a fast rise and slow fade. In addition, the periodicity in the cross-correlation is very similar to those in the auto-correlations of the MP and the IP, confirming the single-pulse quasi-periodicity. 

Applying these same methods  on the running averages in stead of the single pulses results in an even larger maximum correlation of $-0.50\pm0.02$ (Fig.~\ref{fig:J1926corr}d, black for FAST observation). The MeerKAT observation also reveals an anti-correlation around zero lag, with the cross-correlation curve in blue. This validates that the long-term variability (not the short-term changes) is the reason for the MP-IP anti-correlation. 

\subsection{Summary}

We established that PSR~J1926+0737 shows an anti-correlation between the MP and the IP. Both display quasi-periodic changes on timescales of $>$100 pulses. The quasi-periodicities do not display  subpulse phase modulations, in contrast to the subpulse drift in the IP of PSR~J1842+0358.

\section{Discussion}
\label{sec:discussion}
In the following we discuss the established emission correlations and comment on potential interpretations.

\subsection{Simulating the cross-correlation curve of J1842+0358}
\label{sec:J1842simcurve}

The cross-correlation curves of PSR~J1842+0358 (Fig.~\ref{fig:J1842corr}b, c) show a number of interesting features. The large anti-correlation near lag 0 is asymmetric, such that the negative correlation extends furthest to positive lags. A positive correlation appears  on  both sides of this central negative correlation, but the peak on the right-hand side is located about twice as far from lag 0 as that on the left.
Here we show that these asymmetries can be explained by the MP and IP having different intensity modulation patterns in their variability cycles.

To demonstrate this, we simulated a sequence of MP and IP intensities to mimic their behaviour within one $P_3$ period. Instead of assuming any analytic description of these sequences, we used the measurements directly. A complication is that the modulation cycle is variable. To deal with this, the pulse stacks for the MP and IP were classified by eye in bright and weak sequences. Because each sequence had a different length, interpolation was used to combine the individual sequences. The result is a single averaged bright and weak mode sequence for the MP and IP, with a length equal to their median length.

We established that the MP and IP share an identical $P_3$ periodicity (as shown in Sec.~\ref{sec:j1842spec}). Therefore, to explain the correlations, the brightening and weakening of the MP and IP during the $P_3$ cycle must follow a typical and repeated structure. Here we aim to simulate this repeated periodic structure, without considering any variability from cycle to cycle. The length of the MP and IP cycle are scaled to be the same to reflect the synchronisation of the $P_3$ periodicities. To reproduce the anti-correlation, the MP starts in the weak sequence, while the IP starts in the bright sequence.

\begin{figure}
    \centering
    \includegraphics[width=\linewidth]{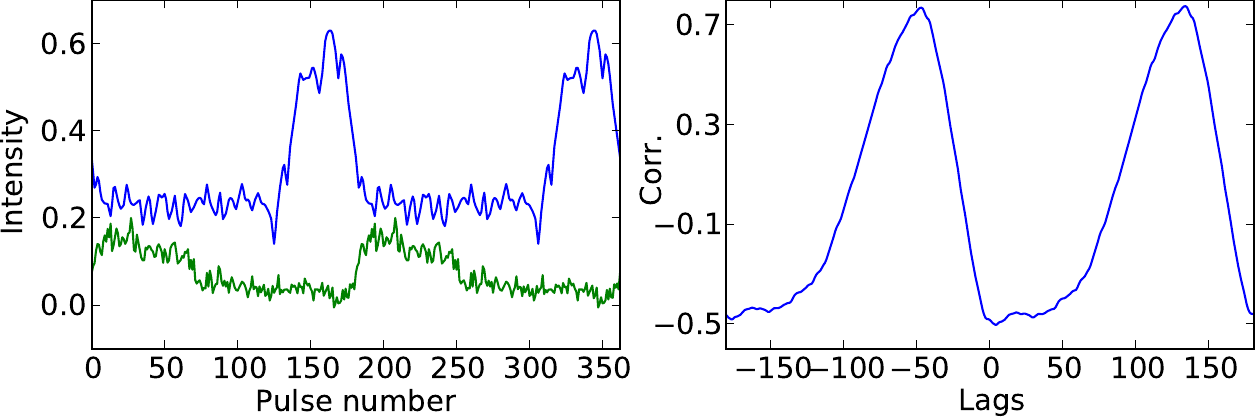}
    \caption{Left panel: simulated pulse intensity sequence within two $P_3$ cycles of the MP (blue) and the IP (green) obtained as the average of selected single pulses (see text for details). The right panel is the corresponding cross-correlation curve.}
    \label{fig:J1842simseq}
\end{figure}

Two cycles of the obtained sequence are shown in the left panel of Fig.~\ref{fig:J1842simseq}. This reveals that the MP (blue) shows a relatively sharp and symmetric brightening profile during the modulation cycle. On the other hand, the IP (green) shows a sharp rise in intensity followed by a gradual fall. Such behaviour is also visible in the average intensities shown in panel d of Fig.~\ref{fig:J1842sum}. For example, around pulse number 200, 300, 550 and 750, the MP average intensities are bright and the bright IP intensities follow. The simulated sequences agree well with the (2/3, 1/3) weak and bright duration split adopted in Sect.~\ref{sec:j1842profmod}.

The cross-correlation curve obtained from the simulated sequences is shown in the right panel. Important qualitative features of the observed cross-correlation curve are reproduced. The negative correlation around zero lag is asymmetric as observed. Also, a positive correlation appears at a negative lag of $-$49 pulses, and at a positive lag of 134 pulses -- nearly twice as large. This is because in the simulated intensity sequence the bright MP phase occurs just before the sharp rise in the IP intensity and well after the preceding bright IP phase. This simulation further demonstrates that the observed correlations in the MP and IP intensities arise because of slow phase-locked $P_3$ modulation.

\subsection{Longitude stationary modulation}
\label{sec:long}

The two pulsars discussed in this paper reveal that the main and interpulse emission possess the same subpulse modulation periodicity. In addition, at least the modulation in one pulse is longitude stationary. Longitude stationary modulation is observed in other interpulse pulsars that show main and interpulse communication. In particular for those with evidence of orthogonal rotators, PSRs~B1702$-$19 \citep{Weltevrede2007}, B1055$-$52 \citep{Weltevrede2012}, and B1822$-$09 quiet mode \citep{Yan2019} have longitude stationary features for both their MP and IP emission. PSR~B0823+26 \citep{Chen2023} shows drifting subpulses in the MP while the IP is longitude stationary. All six orthogonal rotators with main-interpulse communication (including the two pulsars studied here) have longitude stationary modulation. This suggests that having longitude stationary modulation in at least one of the two pulses is common, if not universal, for orthogonal rotators exhibiting subpulse modulation.

The presence of longitude stationary modulation in these interpulse pulsars can be contrasted with longitude stationary modulation found in the general pulsar population.Therefore, all pulsars with $\dot{E} \geq 10^{32}$~erg/s -- above which the six interpulse pulsars with orthogonal geometry and periodic subpulse modulation are found -- were selected. Of those pulsars for which also periodic subpulse modulation was detected, about 40 per-cent had at least one profile component with longitude stationary modulation (estimated from \citealt{Song2023}). Thus the probability of finding longitude stationary features in six out of six interpulse pulsars with periodic modulation by chance is small. This suggests that subpulse modulation in orthogonal rotators could well be different compared to the general population of pulsars.

The nearly orthogonal geometry could well be important, as such a geometry leads to a different current distribution and variation of the accelerating potential across the polar cap (see, e.g.,~Fig.~1 in \citealt{TA12}) compared to aligned rotators. Because the spark drift is governed by this potential variation \citep{sl24}, the resulting subpulse motion in orthogonal rotators might be affected, leading to longitude stationary modulation only.

\subsection{Model interpretation}
\label{sec:model}

We established that the subpulse modulations in the MP and IP of PSRs~J1842+0358 and J1926+0737 are synchronised. The MP and IP share the same $P_3$, and despite $P_3$ not being strictly constant, the modulation patterns of the MP and IP remain phase-locked. Therefore, the two pulsars studied here add to a fundamental problem with the text-book picture of pulsar emission. The MP and IP are normally assumed to come from open fieldlines anchored to opposite magnetic poles, and it is the local polar cap physics that is thought to be responsible for subpulse modulation. To maintain phase locking, the emission regions responsible for the MP and IP emission need to have some means of inter-communication.

In searching for a model, the simplest assumption would be to assign identical and completely synchronised emission regions to both MP and IP and attribute the observed apparent asymmetry to differing LoS cuts through the emission beams. However, the widespread observations of complex and often subtle asymmetry in such pulsars (including the longitude stationary modulation features discussed in section~\ref{sec:long}) seems to call this into question. 

An alternative view is to see the MP and IP together forming a single inter-connected system but with the MP and IP playing different roles so that the modulation arising from each driving, with delay, modulation in the other -- thus creating, in effect, a hysteresis loop. Evidence of such behaviour is suggested in PSR~J1842+0358 by brief but intense emission at the MP triggering a sudden enhancement of intensity and subpulse drift at the IP. In turn, the bright stationary emission is restarted in the MP as the drift process fades. This would result in the quasi-periodic periodicity at both poles demonstrated by the simulation in Fig.~\ref{fig:J1842simseq}.

In the following, we explore a range of potential mechanisms in light of these possibilities. The models considered include: firstly, emission from the main and interpulse are from the opposite poles, with the interaction through the large-scale magnetosphere. Secondly, there is an external driver responsible for the periodicities at both poles. Lastly, a single-pole mechanism may be able to avoid the need for interpole communication.

\subsubsection{Interpole communication}
\label{sec:magneto}

Over the years pulsar magnetosphere models have been theoretically examined and modelled. Generally it has been assumed that the magnetosphere is in a steady state with the implicit assumption that the observed complex radio emission patterns are a small-scale perturbation (e.g.,~sparks) to the model.

In this picture the magnetosphere has two distinct zones: open fieldlines extending to infinity and a closed region connecting the magnetic poles with zero potential difference and a trapped corotating plasma. In cases where the pulsar is near-aligned, \citet{RS} postulated that there is a plasma $\mathrm{\bf{E}} \times \mathrm{\bf{B}}$ drift across the open fieldlines so that it lags corotation with a periodicity $P_4 \sim 5.6B/P^2$ (where $B$ is the dipolar magnetic field in units of $10^{12}$~G). It is reasonable to expect a periodicity of this order even when the pulsar is highly inclined, although, as we point out above, the simple picture of the RS model has to be significantly adapted for such pulsars.

The lag increases as the pulsar ages until, in old pulsars, its theoretical value becomes comparable to the spin of the star. But for middle-aged pulsars such as J1842+0358 and J1926+0737, the expected lag is quite modest, being 45$P$ and 19$P$ respectively. These figures are significantly smaller than the modulation timescales observed here, but can be lengthened within the RS framework by adopting a smaller closed region and hence a polar cap radius wider by a factor $s$ giving $P_4\sim5.6s^2B/P^2$. This would suggest $s\sim$2 or 3. Such an extended emission region has been inferred observationally for PSR~B1055$-$52 \citep{Weltevrede2009}, where polarimetry suggests that the MP emission originated from regions between $s=1.5$ and $s=2$.

The challenge presented by these pulsars is therefore not so much the timescale of the modulations but the fact that they demonstrate strong communication between the poles, which are traditionally thought to be separated  by fieldlines enclosing an inactive corotating plasma. We are therefore compelled to abandon the assumption that the magnetospheres of such pulsars are in a steady state, even on the inter-pole travel timescale of one pulse period \citep{Biggs1990}. The important further implication is that if interpulse pulsars have proven interaction between their poles, then we may suppose that other (maybe all) pulsars have interactive magnetospheres (as argued by \citealt{Melrose2012,Melrose2016,Wright2022} and by \citealt{Basu2020} from observational evidence). In such a picture the observed emission patterns are created by the interactions themselves.

In highly inclined pulsars, both the MP and IP beam will have a central region (not necessarily defined by the last closed fieldline touching the light cylinder) whose fieldlines drive the angular momentum loss of the pulsar to infinity. But the fieldlines which generate the outer regions of the beam (i.e.~those which communicate between poles via closed fieldlines) will have a strongly differing physical environment, depending on whether they are equatorial with respect to the rotation axis and pass close to the light cylinder -- or perpendicular to  the equator and pass close to the rotation axis itself (well within the light cylinder). Even the equatorial regions themselves will likely be asymmetric since the trailing and leading fieldlines will carry particle flow with or counter to the pulsar rotation.

Thus any polar beam will have at least three/four distinct regions with separate emission characteristics, and the separate LoS at MP and IP will probably sample very different regions -- even if the beams are intrinsically identical at each pole. This is potentially a way to explain why longitude stationary modulation is observed at one pole and drift with the same modulation at the other, especially, as in the case of PSR~J1842+0358 (Section~\ref{sec:j1842prof}), we are probably viewing fieldlines on opposite sides of the beam axis.
 
However, the physics behind interpole interaction, although argued to be widespread on geometric and observational grounds \citep{Wright2022}, has been little studied. Global changes in the magnetosphere have been suggested to explain correlated changes in the pulsar spin-down rate and the pulse shape (e.g.,~\citealt{Kramer2006,Lyne2010}). And it should be pointed out that the simultaneous radio and X-ray mode change in PSR~B0823+26 \citep{Hermsen2018} -- a pulsar now known to have simultaneous modulations at both poles in radio on timescales of a few pulse periods \citep{Chen2023} -- is a clear indication of a magnetosphere which undergoes dramatic but not understood physical changes.

\subsubsection{Extrinsic drivers}
\label{sec:ext}

The fact that MP and IP emission show the same periodicity could point towards an external mechanism driving the modulation in both emission regions. This mechanism could be material external to the pulsar interacting with the magnetosphere. Such material could come
from regions connected to the neutron star magnetosphere. For example, \citet{Philippov2015,Lyubarsky2019,Philippov2019} have considered magnetic reconnection around a Y-shaped current sheet just outside the light cylinder and this may provide a viable way of influencing both poles via a return current.

Other external materials could be provided by a `fallback' disk formed when the star was born \citep{Michel1981}. As discussed in \citet{Cordes2008}, such fallback disk, circulating around the neutron star, could be in the form of asteroids (rocks). These materials need to be within the light cylinder to influence the emission. Given the modulation periodicity around 100 pulses for PSRs~J1842+0358 and J1926+0737, the corotation radii of the two pulsars are outside the light cylinder. Interaction can only be expected if asteroids come within the light cylinder, a scenario that could be possible, despite the long periodicities, if the asteroids are in a highly elliptical orbit. The existence of asteroids has been suggested to explain the emission changes in PSR~J0738$-$4042 \citep{Brook2014}. 

If indeed the synchronised modulation in the MP and IP are caused by a single external driving mechanism, some asymmetry between the poles will be required unless caused by difference in the way the LoS crosses the emission beams. Otherwise, asymmetries could arise from the orientation of the inflow of material (e.g.,~the orientation of the orbit of an asteroid or the return current). An extrinsic model gives a natural explanation of the synchronisation -- as the external materials interact with the magnetosphere similarly for different emission regions.

\subsubsection{Single-pole mechanism}
\label{sec:bi-dir}

In the standard picture, the MP and IP emission are produced at opposite poles, making interpole communication difficult to explain. However, as suggested by \citet{Dyks2005}, this issue can be avoided if the MP and IP emission come from the same pole. Then one pulse is composed of outward, and the other of inward directed emission. The inward  emission travels through the magnetosphere, past the neutron star, and is seen half a turn later. This model was proposed to explain the anti-correlation between the IP and the pre-cursor of PSR~B1822$-$09.

In a single-pole interpretation the MP and IP are not expected to be exactly 180\degr\ apart -- the inward emission has to travel further as it needs to pass the neutron star before reaching the observer. This could explain why for both pulsars discussed in this paper the MP-IP separation deviates from 180\degr. The inward emission is expected to be delayed relative to the outward emission \citep{Dyks2004,Dyks2005,Dyks2005caustic}. This leads us to conclude that for both our pulsars, since the separation between the MP and the IP is less than 180$\degr$, the MP is dominated by inward directed emission, and the IP by outward directed emission. The required emission heights \citep{Dyks2004,Dyks2005,Weltevrede2007} are of the order of a few hundred kilometres. Such emission heights are compatible with those estimated from polarisation properties of many pulsars (e.g.,~\citealt{Rookyard2015}).

Most importantly, the single-pole model avoids the need for communication between emission regions far apart. Both pulses possess synchronised modulation would be naturally explained, as both pulses originate from the same region in the magnetosphere. However, the model lacks of a physical motivation as why the inward emission should be produced, and how emission can propagate through the inner magnetosphere with strong magnetic fields \citep{Dyks2005}. In addition, it is not clear why inward emission is only seen in one pole but not the other. 

\section{Conclusions}
\label{sec:conclusions}

Due to its high sensitivity, FAST allowed for detailed observations of two new interpulse pulsars. Strikingly, each of these two pulsars show similar behaviours: the quasi-periodicity is the same for both the MP and the IP, and their modulation is anti-correlated. The periodicity is long, of the order of 100 pulses. The (anti-)correlation is clearly visible in the average pulse profiles of the bright and dim groups of pulses, whose intensity is brighter when selecting the weak pulses of the opposite pole. The correlation is confirmed from secondary analysis, and was maintained over a timespan of several years.

One interesting but poorly-understood phenomenon in all interpulse pulsars with interaction, is that modulation of one pulse is longitude stationary. Potentially this is a requirement for communication between two emission regions. This further requires that the model can provide some form of asymmetry to allow different behaviours from the two emission regions. 

Studying interpulse pulsars provides important evidence to the underlying emission mechanism -- the MP-IP interaction is not predicted by a simple rotating carousel model -- which is local and confined within the polar cap regions. The interaction requires some global changes, which could be fulfilled by an interactive magnetosphere. We conclude that an interactive magnetosphere is common to all pulsars, with both MP-IP communication and the correlation between spin-down rate and long-term pulse shape variations being observational manifestations.

\begin{acknowledgements}
We thank L.~Wang and F.~Kou for their support at the FAST proposal stage, and for assistance with the data transfer. We further thank the investigators of PT2023\_0052 for their useful data. This research was supported by Vici project `ARGO' with project number 639.043.815 and by `CORTEX' (NWA.1160.18.316), under the research programme NWA-ORC; both financed by the Dutch Research Council (NWO). 
This work made use of the data from FAST (Five-hundred-meter Aperture Spherical radio Telescope) (https://cstr.cn/31116.02.FAST). FAST is a Chinese national mega-science facility, operated by National Astronomical Observatories, Chinese Academy of Sciences.
The MeerKAT telescope is operated by the South African Radio Astronomy Observatory, which is a facility of the National Research Foundation, an agency of the Department of Science and Innovation.
\end{acknowledgements}

%
%

\bibliographystyle{aa}
\bibliography{ip}

\begin{thebibliography}{48}
\expandafter\ifx\csname natexlab\endcsname\relax\def\natexlab#1{#1}\fi

\bibitem[{{Backer}(1970)}]{Backer1970}
{Backer}, D.~C. 1970, \nat, 227, 692

\bibitem[{{Basu} {et~al.}(2020){Basu}, {Mitra}, \& {Melikidze}}]{Basu2020}
{Basu}, R., {Mitra}, D., \& {Melikidze}, G.~I. 2020, \apj, 889, 133

\bibitem[{{Basu} {et~al.}(2016){Basu}, {Mitra}, {Melikidze}, {Maciesiak}, {Skrzypczak}, \& {Szary}}]{Basu2016}
{Basu}, R., {Mitra}, D., {Melikidze}, G.~I., {et~al.} 2016, \apj, 833, 29

\bibitem[{{Basu} {et~al.}(2019){Basu}, {Mitra}, {Melikidze}, \& {Skrzypczak}}]{Basu2019}
{Basu}, R., {Mitra}, D., {Melikidze}, G.~I., \& {Skrzypczak}, A. 2019, \mnras, 482, 3757

\bibitem[{{Biggs}(1990)}]{Biggs1990}
{Biggs}, J.~D. 1990, \mnras, 246, 341

\bibitem[{{Biggs} {et~al.}(1988){Biggs}, {Lyne}, {Hamilton}, {McCulloch}, \& {Manchester}}]{Biggs1988}
{Biggs}, J.~D., {Lyne}, A.~G., {Hamilton}, P.~A., {McCulloch}, P.~M., \& {Manchester}, R.~N. 1988, \mnras, 235, 255

\bibitem[{{Brook} {et~al.}(2014){Brook}, {Karastergiou}, {Buchner}, {Roberts}, {Keith}, {Johnston}, \& {Shannon}}]{Brook2014}
{Brook}, P.~R., {Karastergiou}, A., {Buchner}, S., {et~al.} 2014, \apjl, 780, L31

\bibitem[{{Chen} {et~al.}(2023){Chen}, {Wen}, {Duan}, {He}, {Wang}, {Lyu}, {Wang}, {Li}, {Yuan}, {Han}, {Wang}, {Yuen}, {Yan}, {Xiang}, {Dang}, {Wang}, {Niu}, {Wang}, {Ergesh}, {Cui}, \& {Zhang}}]{Chen2023}
{Chen}, J.~L., {Wen}, Z.~G., {Duan}, X.~F., {et~al.} 2023, \apj, 946, 2

\bibitem[{{Cordes} \& {Shannon}(2008)}]{Cordes2008}
{Cordes}, J.~M. \& {Shannon}, R.~M. 2008, \apj, 682, 1152

\bibitem[{{Dyks} {et~al.}(2005{\natexlab{a}}){Dyks}, {Fr{\k{a}}ckowiak}, {S{\l}owikowska}, {Rudak}, \& {Zhang}}]{Dyks2005caustic}
{Dyks}, J., {Fr{\k{a}}ckowiak}, M., {S{\l}owikowska}, A., {Rudak}, B., \& {Zhang}, B. 2005{\natexlab{a}}, \apj, 633, 1101

\bibitem[{{Dyks} {et~al.}(2004){Dyks}, {Rudak}, \& {Harding}}]{Dyks2004}
{Dyks}, J., {Rudak}, B., \& {Harding}, A.~K. 2004, \apj, 607, 939

\bibitem[{{Dyks} {et~al.}(2005{\natexlab{b}}){Dyks}, {Zhang}, \& {Gil}}]{Dyks2005}
{Dyks}, J., {Zhang}, B., \& {Gil}, J. 2005{\natexlab{b}}, \apjl, 626, L45

\bibitem[{{Eatough} {et~al.}(2010){Eatough}, {Molkenthin}, {Kramer}, {Noutsos}, {Keith}, {Stappers}, \& {Lyne}}]{Eatough2010}
{Eatough}, R.~P., {Molkenthin}, N., {Kramer}, M., {et~al.} 2010, \mnras, 407, 2443

\bibitem[{{Edwards} \& {Stappers}(2002)}]{ES2002}
{Edwards}, R.~T. \& {Stappers}, B.~W. 2002, \aap, 393, 733

\bibitem[{{Fowler} \& {Wright}(1982)}]{Fowler1982}
{Fowler}, L.~A. \& {Wright}, G.~A.~E. 1982, \aap, 109, 279

\bibitem[{{Gil} {et~al.}(1994){Gil}, {Jessner}, {Kijak}, {Kramer}, {Malofeev}, {Malov}, {Seiradakis}, {Sieber}, \& {Wielebinski}}]{Gil1994}
{Gil}, J.~A., {Jessner}, A., {Kijak}, J., {et~al.} 1994, \aap, 282, 45

\bibitem[{{Hermsen} {et~al.}(2018){Hermsen}, {Kuiper}, {Basu}, {Hessels}, {Mitra}, {Rankin}, {Stappers}, {Wright}, {Grie{\ss}meier}, {Serylak}, {Horneffer}, {Tiburzi}, \& {Ho}}]{Hermsen2018}
{Hermsen}, W., {Kuiper}, L., {Basu}, R., {et~al.} 2018, \mnras, 480, 3655

\bibitem[{{Jiang} {et~al.}(2019){Jiang}, {Yue}, {Gan}, {Yao}, {Li}, {Pan}, {Sun}, {Yu}, {Liu}, {Tang}, {Qian}, {Lu}, {Yan}, {Peng}, {Zhang}, {Wang}, {Li}, {Li}, \& {FAST Collaboration}}]{Jiang2019}
{Jiang}, P., {Yue}, Y., {Gan}, H., {et~al.} 2019, Science China Physics, Mechanics, and Astronomy, 62, 959502

\bibitem[{{Johnston} {et~al.}(2020){Johnston}, {Karastergiou}, {Keith}, {Song}, {Weltevrede}, {Abbate}, {Bailes}, {Buchner}, {Camilo}, {Geyer}, {Hugo}, {Jameson}, {Kramer}, {Parthasarathy}, {Reardon}, {Ridolfi}, {Serylak}, {Shannon}, {Spiewak}, {van Straten}, {Venkatraman Krishnan}, {Jankowski}, {Meyers}, {Oswald}, {Posselt}, {Sobey}, {Szary}, \& {van Leeuwen}}]{Johnston2020}
{Johnston}, S., {Karastergiou}, A., {Keith}, M.~J., {et~al.} 2020, \mnras, 493, 3608

\bibitem[{{Kou} {et~al.}(2021){Kou}, {Yan}, {Peng}, {Lu}, {Liu}, {Zhang}, {Strom}, {Wang}, {Yuan}, {Yuen}, {Yu}, {Yao}, {Liu}, {Yan}, {Jiang}, {Jin}, {Li}, {Qian}, {Yue}, {Zhu}, \& {FAST Collaboration}}]{Kou2021}
{Kou}, F.~F., {Yan}, W.~M., {Peng}, B., {et~al.} 2021, \apj, 909, 170

\bibitem[{{Kramer} \& {Johnston}(2008)}]{Kramer2008}
{Kramer}, M. \& {Johnston}, S. 2008, \mnras, 390, 87

\bibitem[{{Kramer} {et~al.}(2006){Kramer}, {Lyne}, {O'Brien}, {Jordan}, \& {Lorimer}}]{Kramer2006}
{Kramer}, M., {Lyne}, A.~G., {O'Brien}, J.~T., {Jordan}, C.~A., \& {Lorimer}, D.~R. 2006, Science, 312, 549

\bibitem[{{Lorimer} {et~al.}(2006){Lorimer}, {Faulkner}, {Lyne}, {Manchester}, {Kramer}, {McLaughlin}, {Hobbs}, {Possenti}, {Stairs}, {Camilo}, {Burgay}, {D'Amico}, {Corongiu}, \& {Crawford}}]{Lorimer2006}
{Lorimer}, D.~R., {Faulkner}, A.~J., {Lyne}, A.~G., {et~al.} 2006, \mnras, 372, 777

\bibitem[{{Lyne} {et~al.}(2010){Lyne}, {Hobbs}, {Kramer}, {Stairs}, \& {Stappers}}]{Lyne2010}
{Lyne}, A., {Hobbs}, G., {Kramer}, M., {Stairs}, I., \& {Stappers}, B. 2010, Science, 329, 408

\bibitem[{{Lyubarsky}(2019)}]{Lyubarsky2019}
{Lyubarsky}, Y. 2019, \mnras, 483, 1731

\bibitem[{{Maciesiak} {et~al.}(2011){Maciesiak}, {Gil}, \& {Ribeiro}}]{Maciesiak2011}
{Maciesiak}, K., {Gil}, J., \& {Ribeiro}, V. A.~R.~M. 2011, \mnras, 414, 1314

\bibitem[{{Melrose} \& {Yuen}(2012)}]{Melrose2012}
{Melrose}, D.~B. \& {Yuen}, R. 2012, \apj, 745, 169

\bibitem[{{Melrose} \& {Yuen}(2016)}]{Melrose2016}
{Melrose}, D.~B. \& {Yuen}, R. 2016, Journal of Plasma Physics, 82, 635820202

\bibitem[{{Michel} \& {Dessler}(1981)}]{Michel1981}
{Michel}, F.~C. \& {Dessler}, A.~J. 1981, \apj, 251, 654

\bibitem[{{Nan} {et~al.}(2011){Nan}, {Li}, {Jin}, {Wang}, {Zhu}, {Zhu}, {Zhang}, {Yue}, \& {Qian}}]{Nan2011}
{Nan}, R., {Li}, D., {Jin}, C., {et~al.} 2011, International Journal of Modern Physics D, 20, 989

\bibitem[{{Philippov} {et~al.}(2019){Philippov}, {Uzdensky}, {Spitkovsky}, \& {Cerutti}}]{Philippov2019}
{Philippov}, A., {Uzdensky}, D.~A., {Spitkovsky}, A., \& {Cerutti}, B. 2019, \apjl, 876, L6

\bibitem[{{Philippov} {et~al.}(2015){Philippov}, {Spitkovsky}, \& {Cerutti}}]{Philippov2015}
{Philippov}, A.~A., {Spitkovsky}, A., \& {Cerutti}, B. 2015, \apjl, 801, L19

\bibitem[{{Radhakrishnan} \& {Cooke}(1969)}]{RVM}
{Radhakrishnan}, V. \& {Cooke}, D.~J. 1969, \aplett, 3, 225

\bibitem[{{Rankin} \& {Rathnasree}(1997)}]{Rankin1997}
{Rankin}, J.~M. \& {Rathnasree}, N. 1997, Journal of Astrophysics and Astronomy, 18, 91

\bibitem[{{Rookyard} {et~al.}(2015){Rookyard}, {Weltevrede}, \& {Johnston}}]{Rookyard2015}
{Rookyard}, S.~C., {Weltevrede}, P., \& {Johnston}, S. 2015, \mnras, 446, 3367

\bibitem[{{Ruderman} \& {Sutherland}(1975)}]{RS}
{Ruderman}, M.~A. \& {Sutherland}, P.~G. 1975, \apj, 196, 51

\bibitem[{{Song} {et~al.}(2023){Song}, {Weltevrede}, {Szary}, {Wright}, {Keith}, {Basu}, {Johnston}, {Karastergiou}, {Main}, {Oswald}, {Parthasarathy}, {Posselt}, {Bailes}, {Buchner}, {Hugo}, \& {Serylak}}]{Song2023}
{Song}, X., {Weltevrede}, P., {Szary}, A., {et~al.} 2023, \mnras, 520, 4562

\bibitem[{{Sun} {et~al.}(2025){Sun}, {Wang}, {Yan}, \& {Wang}}]{sun+25}
{Sun}, S.~N., {Wang}, N., {Yan}, W.~M., \& {Wang}, S.~Q. 2025, arXiv e-prints, arXiv:2503.13824

\bibitem[{{Szary} \& {van Leeuwen}(2024)}]{sl24}
{Szary}, A. \& {van Leeuwen}, J. 2024, \mnras, 532, 4075

\bibitem[{{Timokhin} \& {Arons}(2013)}]{TA12}
{Timokhin}, A.~N. \& {Arons}, J. 2013, \mnras, 429, 20

\bibitem[{{van Leeuwen} \& {Timokhin}(2012)}]{lt12}
{van Leeuwen}, J. \& {Timokhin}, A.~N. 2012, \apj, 752, 155

\bibitem[{{Weltevrede}(2016)}]{Weltevrede2016}
{Weltevrede}, P. 2016, \aap, 590, A109

\bibitem[{{Weltevrede} {et~al.}(2006){Weltevrede}, {Edwards}, \& {Stappers}}]{Weltevrede2006}
{Weltevrede}, P., {Edwards}, R.~T., \& {Stappers}, B.~W. 2006, \aap, 445, 243

\bibitem[{{Weltevrede} \& {Wright}(2009)}]{Weltevrede2009}
{Weltevrede}, P. \& {Wright}, G. 2009, \mnras, 395, 2117

\bibitem[{{Weltevrede} {et~al.}(2012){Weltevrede}, {Wright}, \& {Johnston}}]{Weltevrede2012}
{Weltevrede}, P., {Wright}, G., \& {Johnston}, S. 2012, \mnras, 424, 843

\bibitem[{{Weltevrede} {et~al.}(2007){Weltevrede}, {Wright}, \& {Stappers}}]{Weltevrede2007}
{Weltevrede}, P., {Wright}, G.~A.~E., \& {Stappers}, B.~W. 2007, \aap, 467, 1163

\bibitem[{{Wright}(2022)}]{Wright2022}
{Wright}, G. 2022, \mnras, 514, 4046

\bibitem[{{Yan} {et~al.}(2019){Yan}, {Manchester}, {Wang}, {Yuan}, {Wen}, \& {Lee}}]{Yan2019}
{Yan}, W.~M., {Manchester}, R.~N., {Wang}, N., {et~al.} 2019, \mnras, 485, 3241

\end{thebibliography}

\begin{appendix} 

\section{Additional Analysis Details}
\label{app:addit}

\subsection{Anti-correlation between the MP/IP brightness}
\label{app:modes}

As discussed in Sect.~\ref{sec:j1842profcorr} and \ref{sec:j1926profcorr}, anti-correlation is revealed by constructing the average pulse profiles based on the strong/weak running averaged intensity of the other pulse. For completeness, we describe the set of average profiles based on the single pulse intensity groups, and those of normalised profiles to see any possible shape changes.

\begin{figure}
    \centering
    \includegraphics[width=\linewidth]{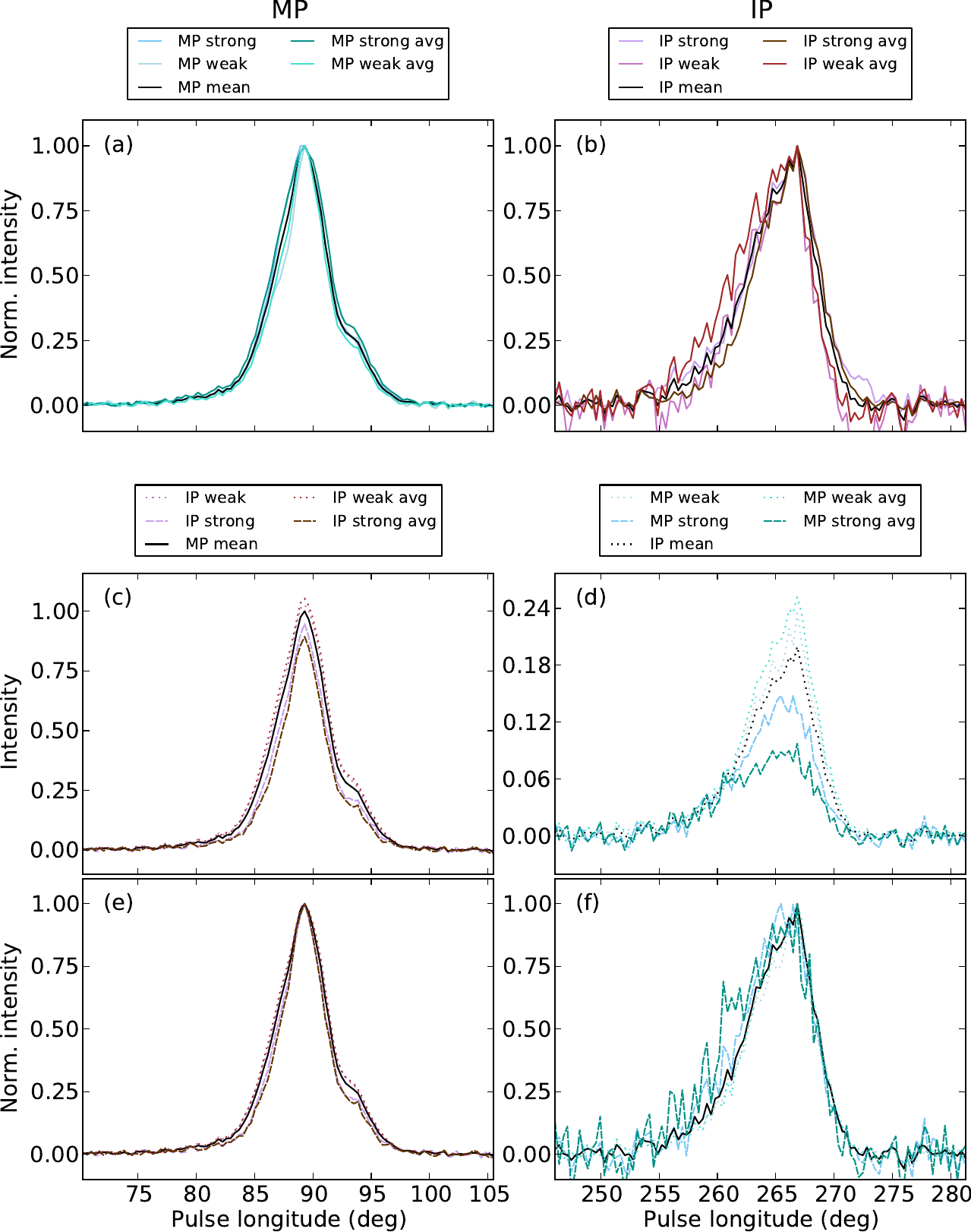}
    \caption{Average pulse profiles of the MP and the IP of PSR~J1842+0358 based on groups of pulses with different
      intensities. (a) MP peak-normalised average profiles of the bright and weak single pulse intensities (blue and light blue) and running-averaged intensities (green and cyan), respectively. The average profile in black is also shown; (b) same as (a) but IP peak-normalised average profiles, single pulse intensities in violet and pink, and running averaged intensities in brown and dark red, respectively; (c) the MP averaged profiles based on the weakest (dotted) and brightest (dashed) groups of single pulse intensities and the running averaged intensities of the IP. Colours corresponds to the IP colours in (b); (d) same as (c) but for the IP averaged profiles based on the two groups of the MP; the profiles in (e) and (f) are the same as in (c) and (d) but now are peak-normalised.}
    \label{fig:J1842selavgprofall}
\end{figure}

Fig.~\ref{fig:J1842selavgprofall}a, b show the MP and IP normalised pulse profiles using either the single-pulse or running-averaged intensities, respectively. The colours corresponds to those in Fig.~\ref{fig:J1842dist} for the weak and bright groups. The mean pulse profile is also shown for comparison. The MP average profiles based on bright pulses are wider than those based on weak pulses (in green and blue for single pulse intensities and running-averaged intensities, respectively). Bright MP pulses having larger pulse widths is a signature of the single pulses too (as in Fig.~\ref{fig:J1842sum}). In the IP (panel b), the trailing component of the brighter pulses is wider than that of the weaker pulses. This suggests that the bright pulses tend to be brighter at the trailing part of the IP profile. In contrast, the IP profiles constructed with the MP intensities do not show such difference in the trailing component. We speculate that because of the drifting subpulses in the IP (see the analysis in Sect.~\ref{sec:j1842spec}), bright pulses are more likely to start from the trailing part of the pulse profile (transitioning from weak to bright pulses). While when using the MP weak pulses, the corresponding IP pulses selected include all pulse phases, without a preference of the pulses with a bright trailing component.

Panel c and d give the same set of plots as in Fig.~\ref{fig:J1842selavgprof} but now include the profiles constructed by the single pulse intensities. These profiles are constructed using the other pulse, therefore the colours used correspond to the other pulse. The MP intensity difference between profiles constructed based on the running averaged intensities of the IP is larger than those based on the single pulse intensities (and is also true for the IP). This leads us to conclude that the long-term variability is responsible for the anti-correlation between the MP and the IP.

These sets of profiles are also normalised to see any shape changes (in panels e and f). Visible there is that when the IP is weak, the MP profiles are wider, in both the leading and trailing part (panel e). The IP profile shape does not significantly change with MP weak versus bright.

\begin{figure}
    \centering
    \includegraphics[width=\linewidth]{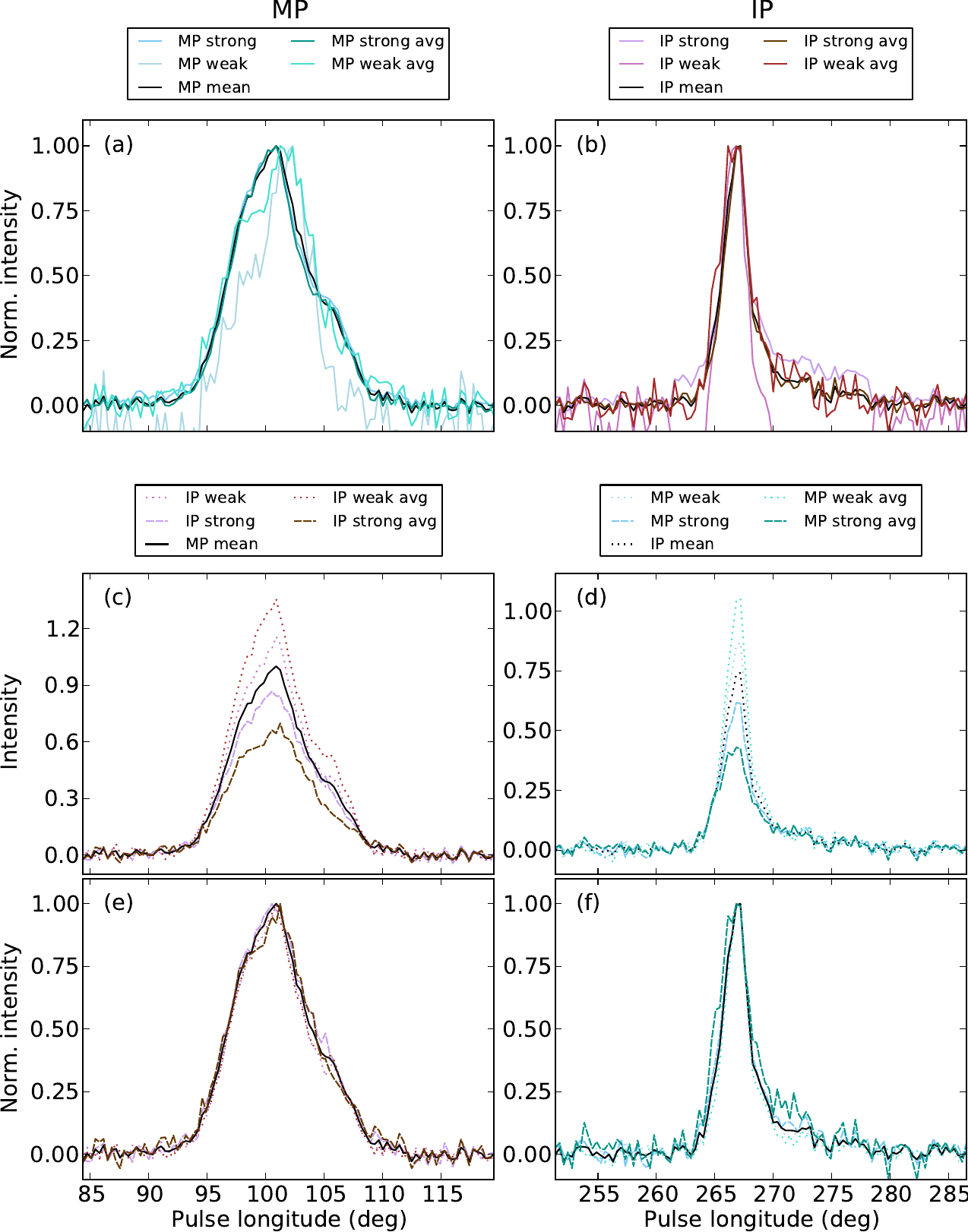}
    \caption{Average pulse profiles of the MP (first column) and the IP (second column) of PSR~J1926+0737 based on groups of pulses with different intensities. See the caption of Fig.~\ref{fig:J1842selavgprofall} for details of each panel.}
    \label{fig:J1926selavgprofall}
\end{figure}

The same sets of plots but for PSR~J1926+0737 are shown in Fig.~\ref{fig:J1926selavgprofall}. The MP/IP peak-normalised average profiles using the bright and weak intensities are in panels a and b, respectively. There is no significant change in the profile shape.

Panels c and d show the average profiles of the weak/bright intensities of the other pulse, the same as in Fig.~\ref{fig:J1926selavgprof} but adding the profiles based on the single pulse intensities. Again, the intensity difference is larger between the profiles based on the weak and bright running averaged intensities, comparing to those based on the single pulse intensities, for both the MP and the IP. This suggests that the long-term variability is strongly correlated, rather than the stochastic variability.

Lastly as in panels e and f, the normalised version of profiles are shown. These do not reveal significant shape changes, which could be because the longitude stationary modulation does not have significant phase variation.

\subsection{Longitude-resolved correlation map}
\label{app:corrmap}

\begin{figure}
    \centering
    \includegraphics[width=\linewidth]{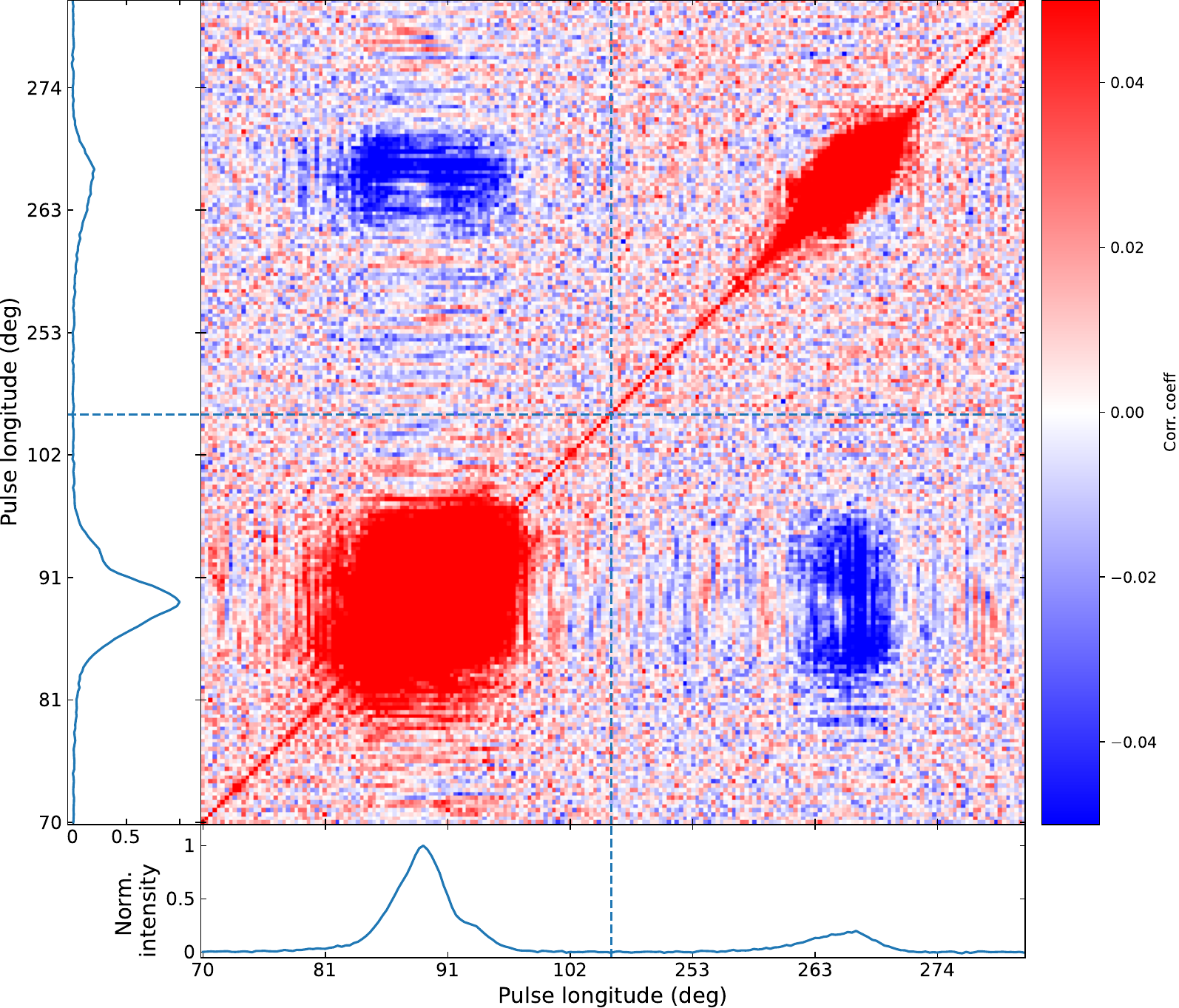}
    \caption{Longitude-resolved correlation map of PSR~J1842+0358 at zero phase delay, and only the selected onpulse regions are shown (note that the pulse longitudes are not consecutive, where the off-pulse regions are removed). The colour plot shows the corresponding cross correlations along pairs of pulse longitude. The side and bottom panels are the normalised integrated pulse profiles, with the dashed lines indicate boundary of the MP and the IP.}
    \label{fig:J1842phasecorr0}
\end{figure}

\begin{figure}
    \centering
    \includegraphics[width=\linewidth]{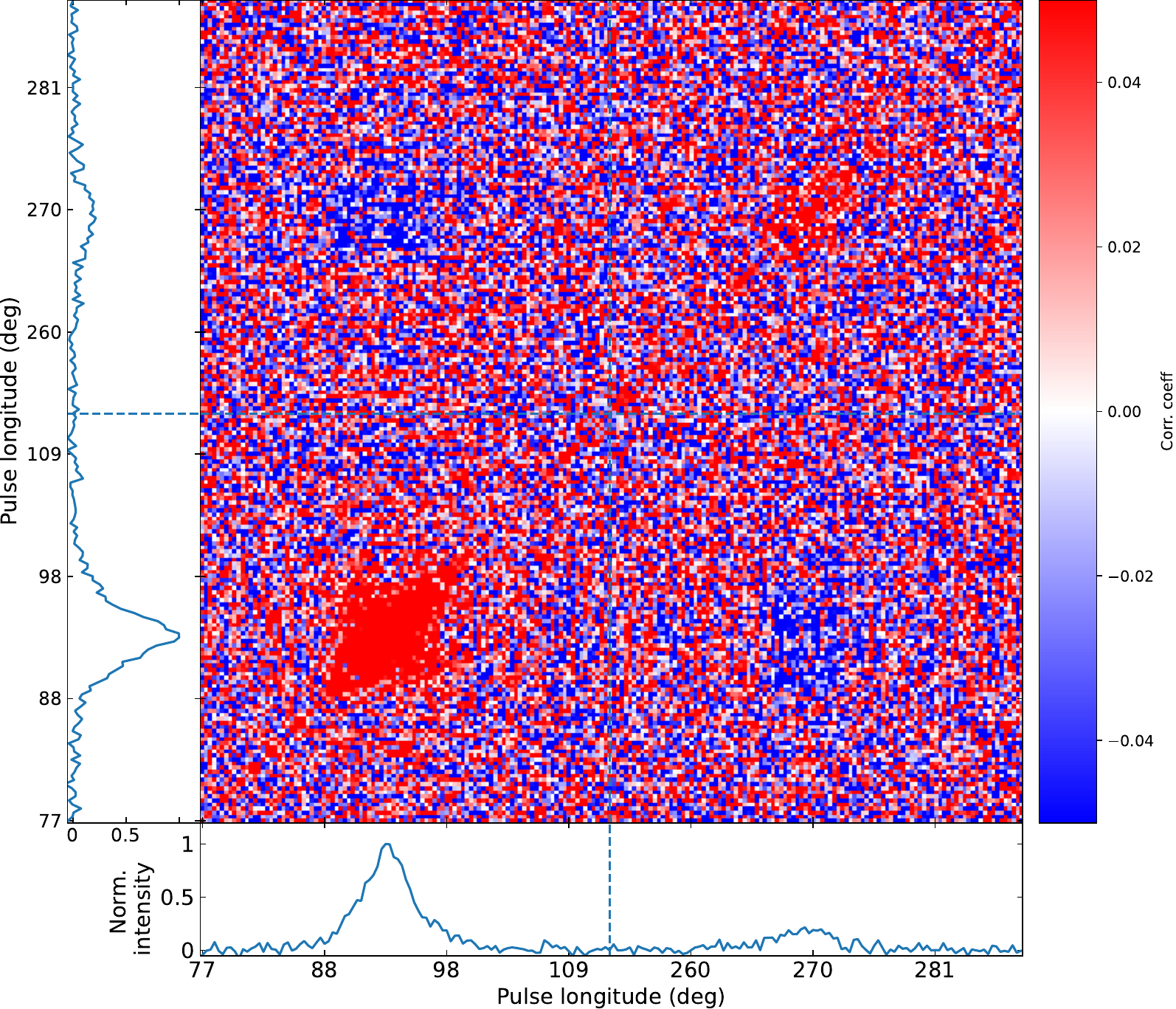}
    \caption{Longitude-resolved correlation map of PSR~J1842+0358 at zero phase delay using the MeerKAT 10-pulse averaged data. Layout otherwise as Fig.~\ref{fig:J1842phasecorr0}.}
    \label{fig:j1842mklrcc0}
\end{figure}

\begin{figure}
    \centering
    \includegraphics[width=\linewidth]{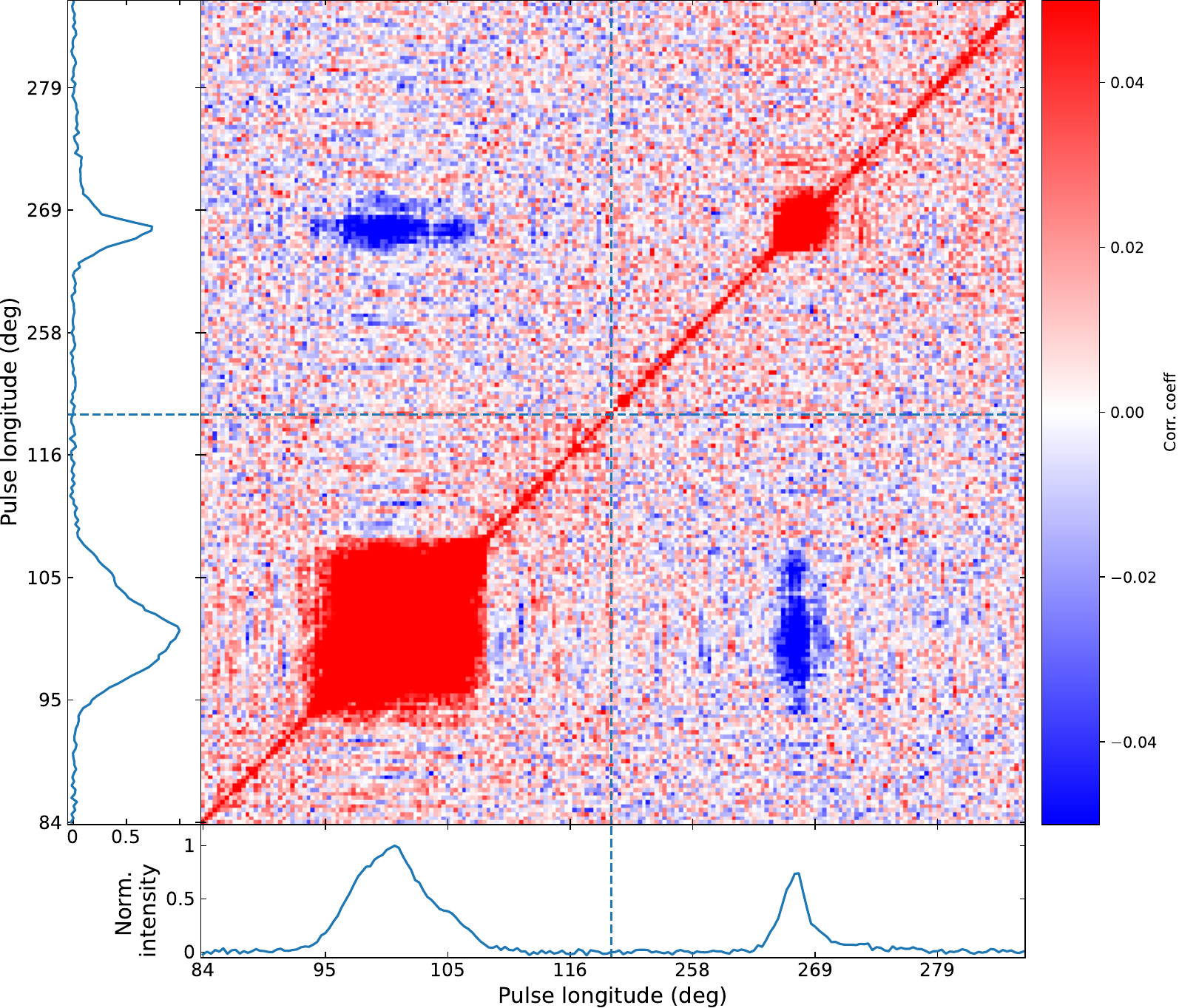}
    \caption{Longitude-resolved correlation map of PSR~J1926+0737 at zero phase delay.
      Layout otherwise as Fig.~\ref{fig:J1842phasecorr0}.}
    \label{fig:J1926phasecorr0}
\end{figure}

\begin{figure}
    \centering
    \includegraphics[width=\linewidth]{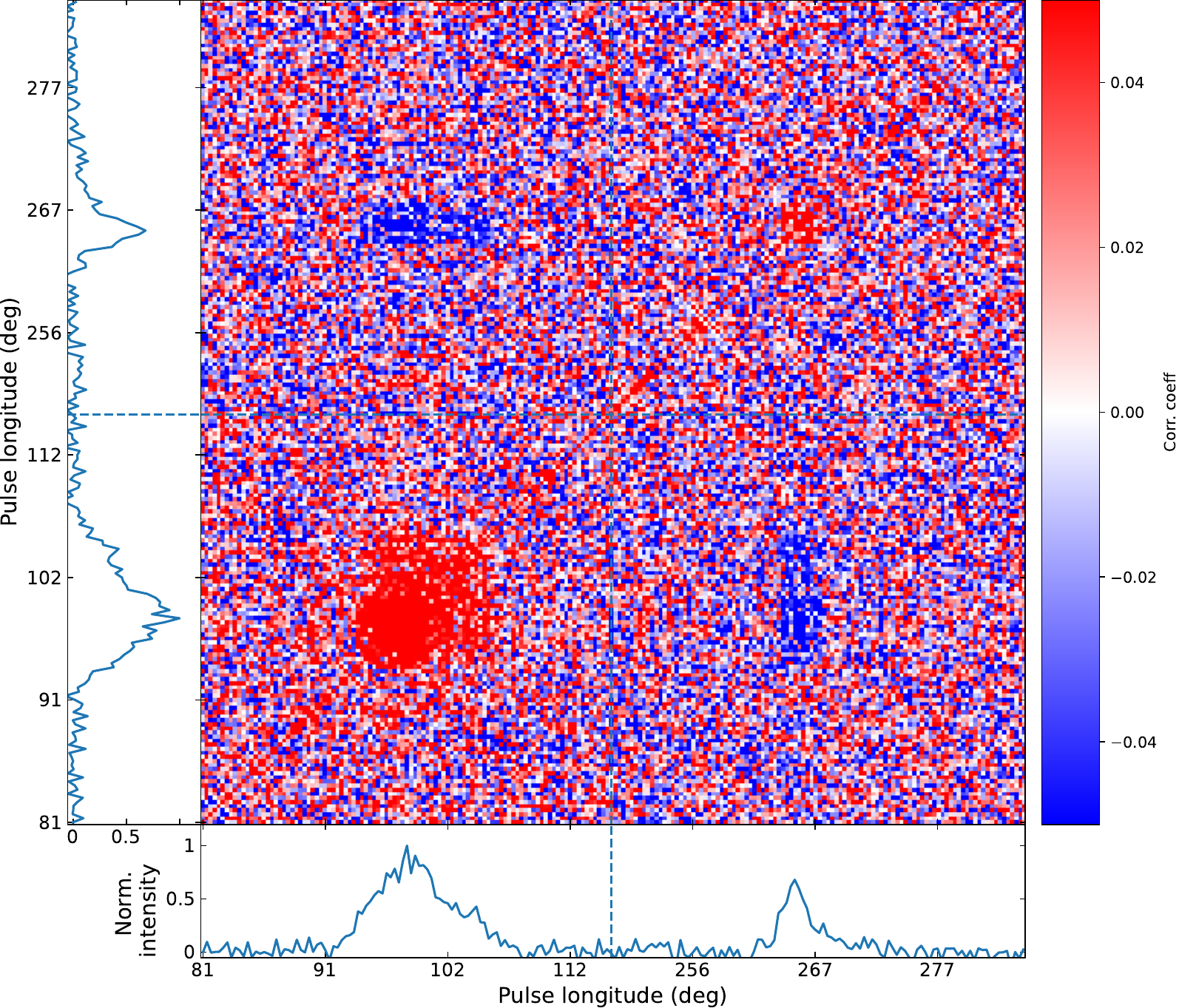}
    \caption{Longitude-resolved correlation map of  PSR~J1926+0737 at zero lag using the MeerKAT 10-pulse averaged data. See the caption of Fig.~\ref{fig:J1926selavgprofall} for further details.}
    \label{fig:j1926mklrcc0}
\end{figure}

The longitude-resolved correlation plots for the FAST data and the 10-single pulse averaged MeerKAT data are in Figs.~\ref{fig:J1842phasecorr0} and \ref{fig:j1842mklrcc0} for PSR~J1842+0358, and Figs.~\ref{fig:J1926phasecorr0} and \ref{fig:j1926mklrcc0} for PSR~J1926+0737, respectively. These are at lag zero.

In these plots, the bottom-left and top-right corners show the auto-correlations between the onpulse of the MP and the IP, respectively. Strong positive correlations (in red) are present there for all observations. Negative correlations are seen in the top-left corner (in blue), representing the relation between the IP and the MP (mirrored bottom-right, correlation between the MP and the IP). This clearly confirms that the MP and IP are anti-correlated for both pulsars. 

\newpage
\newpage

\subsection{Lag of maximum correlation}
\label{app:lags}

For both pulsars, we found that the lags where the maximum anti-correlation occurs are not at zero, but we argue here that they are likely consistent with zero. This is because, firstly, the lags at the correlation maximum are different for the two halves of the data. Secondly, we fit the correlation with a parabolic curve at small lags (between $-$10 to 10 lags) and find that the peak is consistent with zero. For PSR~J1926+0737, the parabolic fit to the cross-correlation coefficients favours a slight positive lag of 1.6, but it is consistent with 0 at 3.1$\sigma$. The longitude-resolved correlation map at the cross-correlation maximum reveals a similar pattern as those shown at zero lag (Figs.~\ref{fig:J1842phasecorr0} and \ref{fig:J1926phasecorr0}) -- there is always an anti-correlation among all on-pulse phases. 

\FloatBarrier
~\newpage
\section{\PSRi pulse stacks (online only)}
\label{app:stacks1842}

\begin{figure*}
    \centering
    \includegraphics[width=\linewidth]{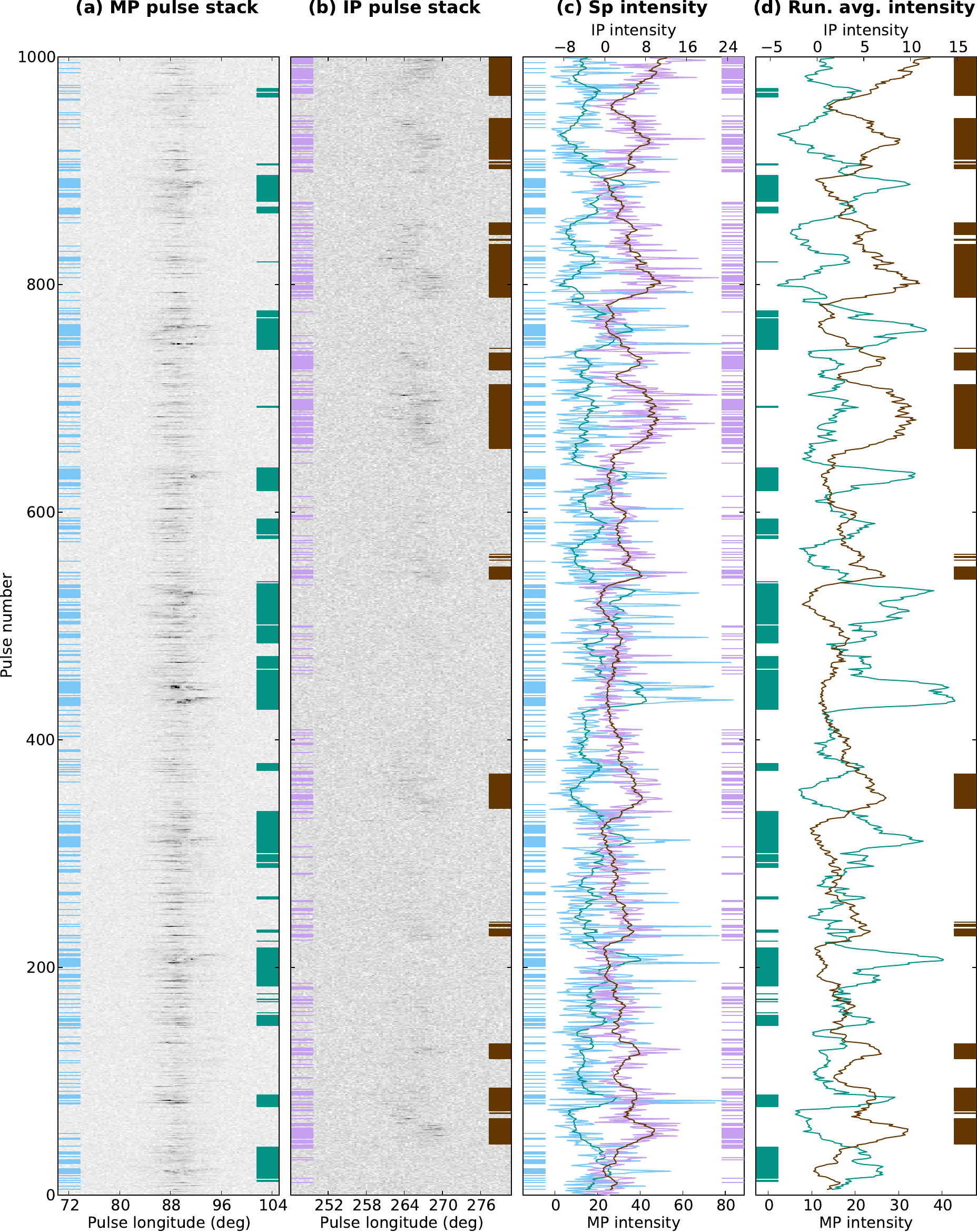}
    \caption{Pulse stacks of the MP (first panel) and the IP (second panel) for the first 1000 pulses for PSR~J1842+0358, and the corresponding single pulse (third panel) and running averaged intensities (fourth panel). See text for the explanation of the lines.}
    \label{fig:J1842sumall1}
\end{figure*}

\begin{figure*}
    \centering
    \includegraphics[width=\linewidth]{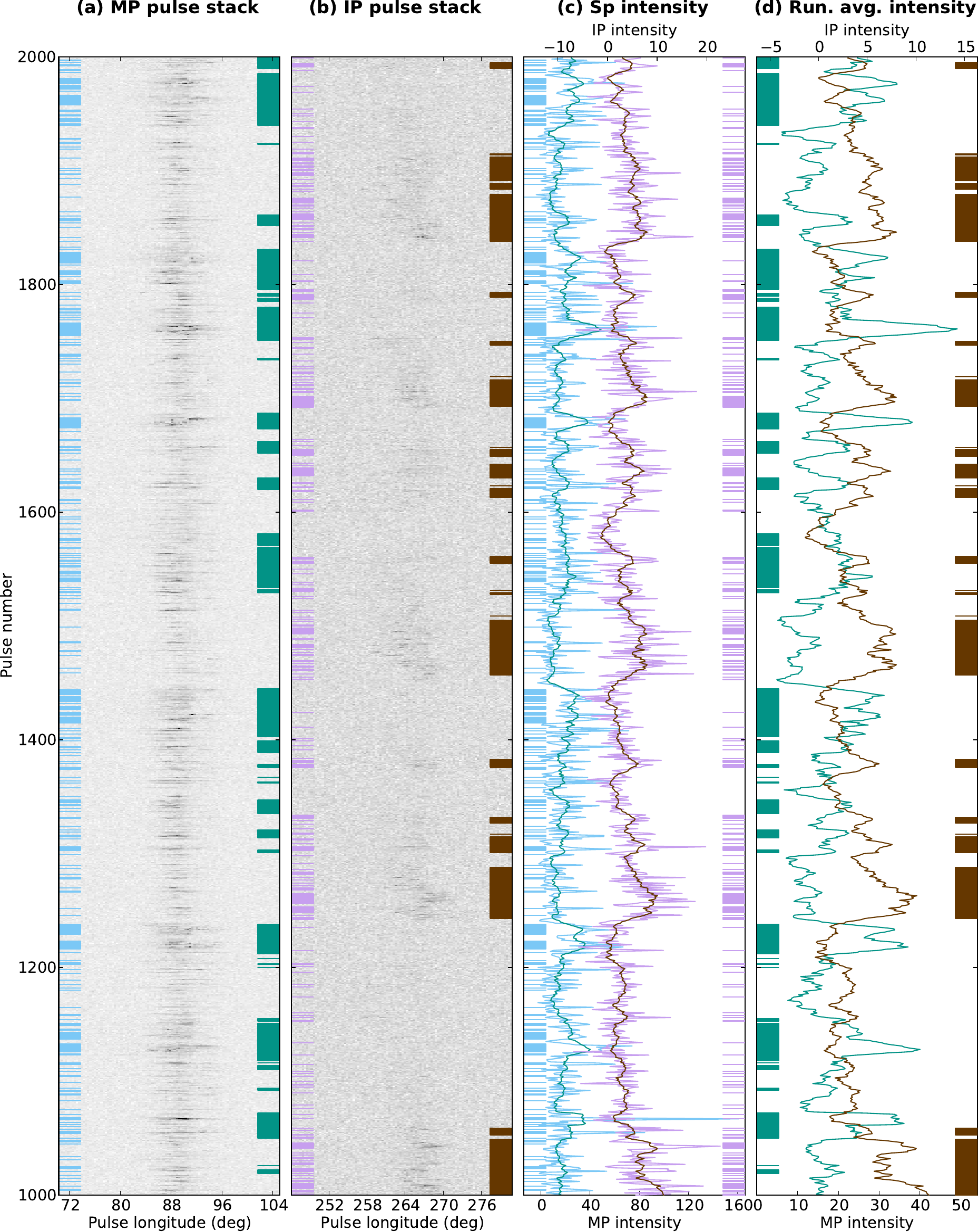}
    \caption{Pulse stacks of the MP (first panel) and the IP (second panel) for pulse number 1000 to 2000 for PSR~J1842+0358, and the corresponding single pulse (third panel) and running averaged intensities (fourth panel). See text for the explanation of the lines.}
    \label{fig:J1842sumall2}
\end{figure*}

\begin{figure*}
    \centering
    \includegraphics[width=\linewidth]{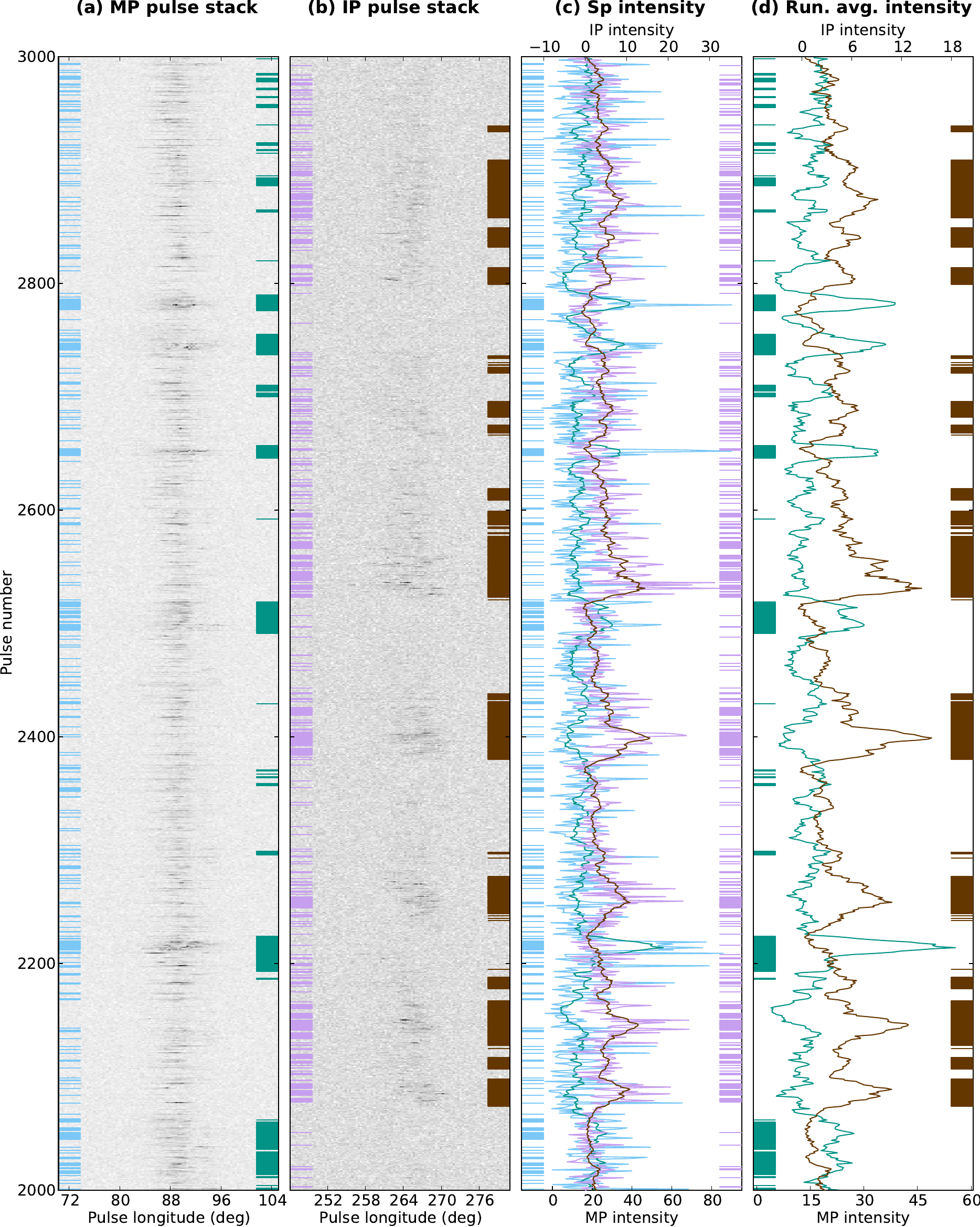}
    \caption{Pulse stacks of the MP (first panel) and the IP (second panel) for pulse number 2000 to 3000 for PSR~J1842+0358, and the corresponding single pulse (third panel) and running averaged intensities (fourth panel). See text for the explanation of the lines.}
    \label{fig:J1842sumall3}
\end{figure*}

\begin{figure*}
    \centering
    \includegraphics[width=\linewidth]{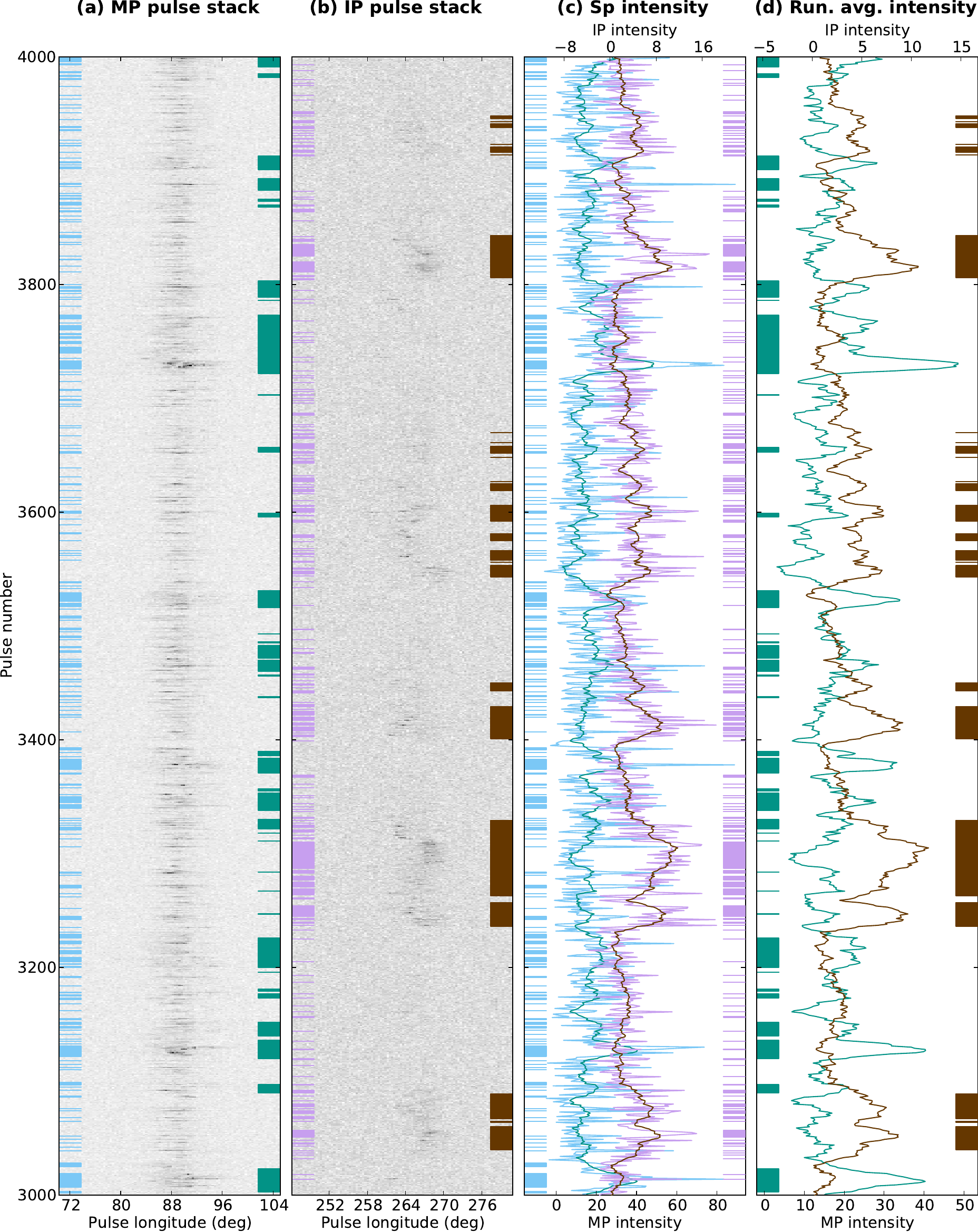}
    \caption{Pulse stacks of the MP (first panel) and the IP (second panel) for pulse number 3000 to 4000 for PSR~J1842+0358, and the corresponding single pulse (third panel) and running averaged intensities (fourth panel). See text for the explanation of the lines.}
    \label{fig:J1842sumall4}
\end{figure*}

\begin{figure*}
    \centering
    \includegraphics[width=\linewidth]{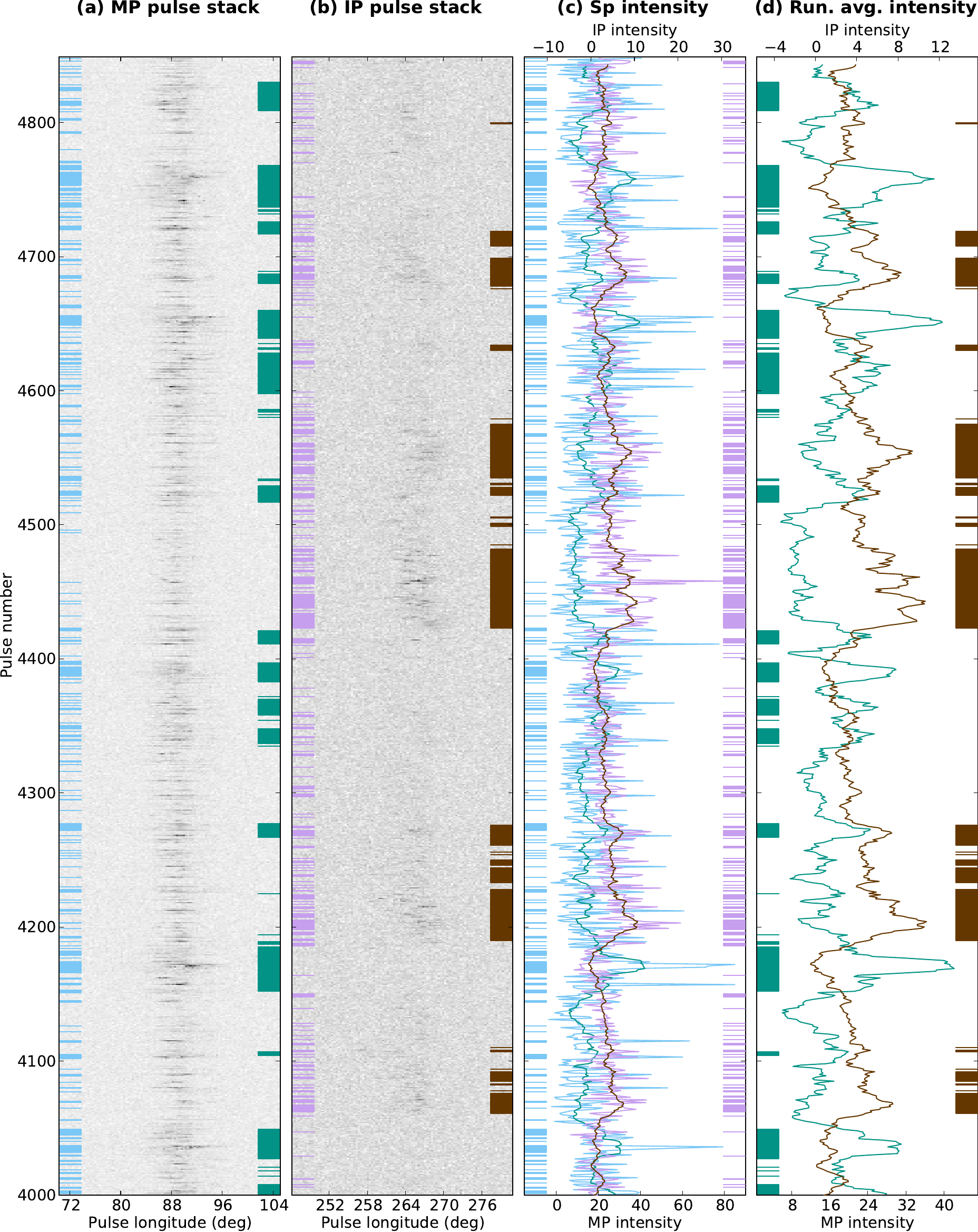}
    \caption{Pulse stacks of the MP (first panel) and the IP (second panel) for pulse number 4000 to 4849 for PSR~J1842+0358, and the corresponding single pulse (third panel) and running averaged intensities (fourth panel). See text for the explanation of the lines.}
    \label{fig:J1842sumall5}
\end{figure*}

\section{\PSRii pulse stacks (online only)}
\label{app:stacks1926}

\begin{figure*}
    \centering
    \includegraphics[width=\linewidth]{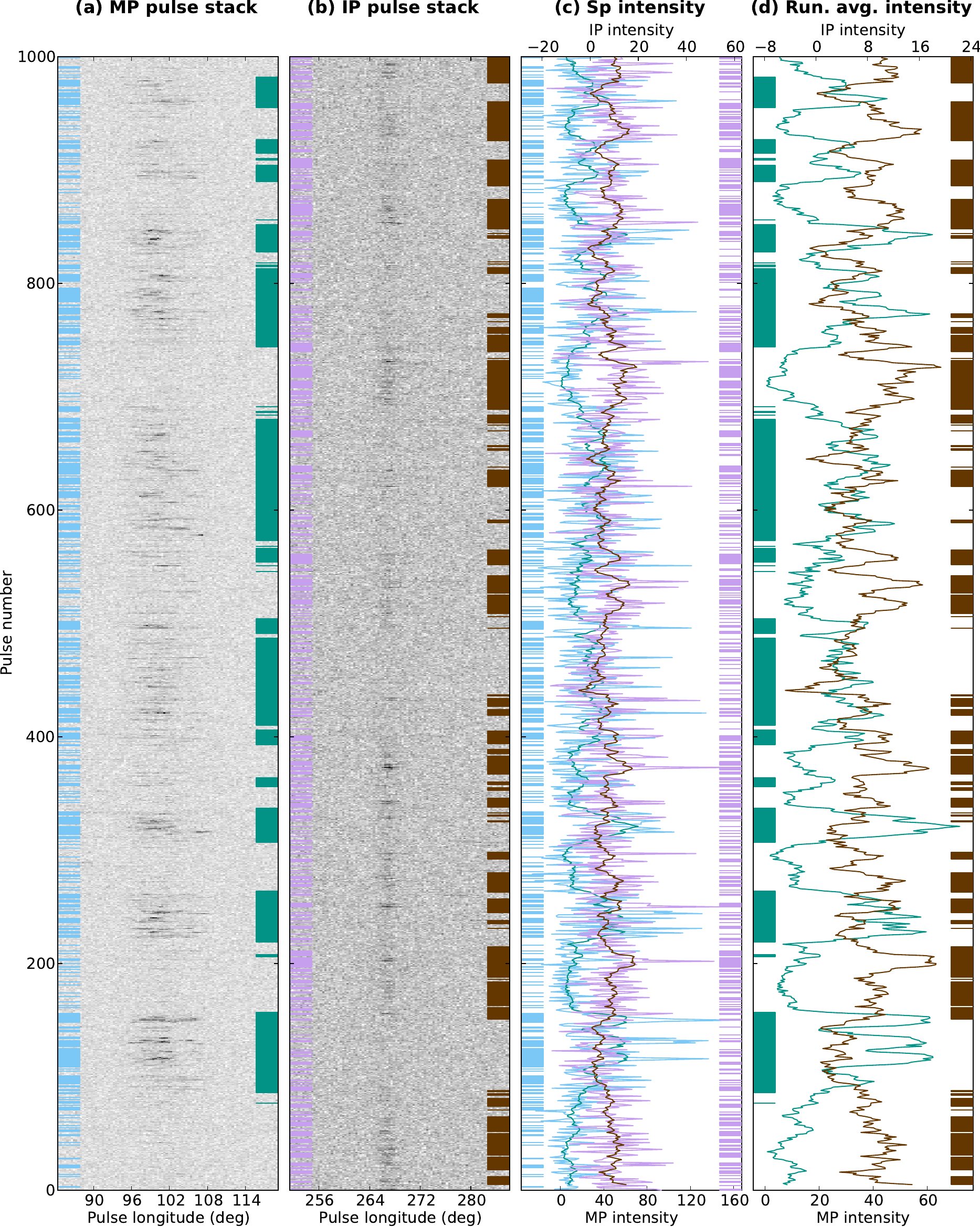}
    \caption{Pulse stacks of the MP (first panel) and the IP (second panel) for the first 1000 pulses for PSR~J1926+0737, and the corresponding single pulse (third panel) and running averaged intensities (fourth panel). See text for the explanation of the lines.}
    \label{fig:J1926sumall1}
\end{figure*}

\begin{figure*}
    \centering
    \includegraphics[width=\linewidth]{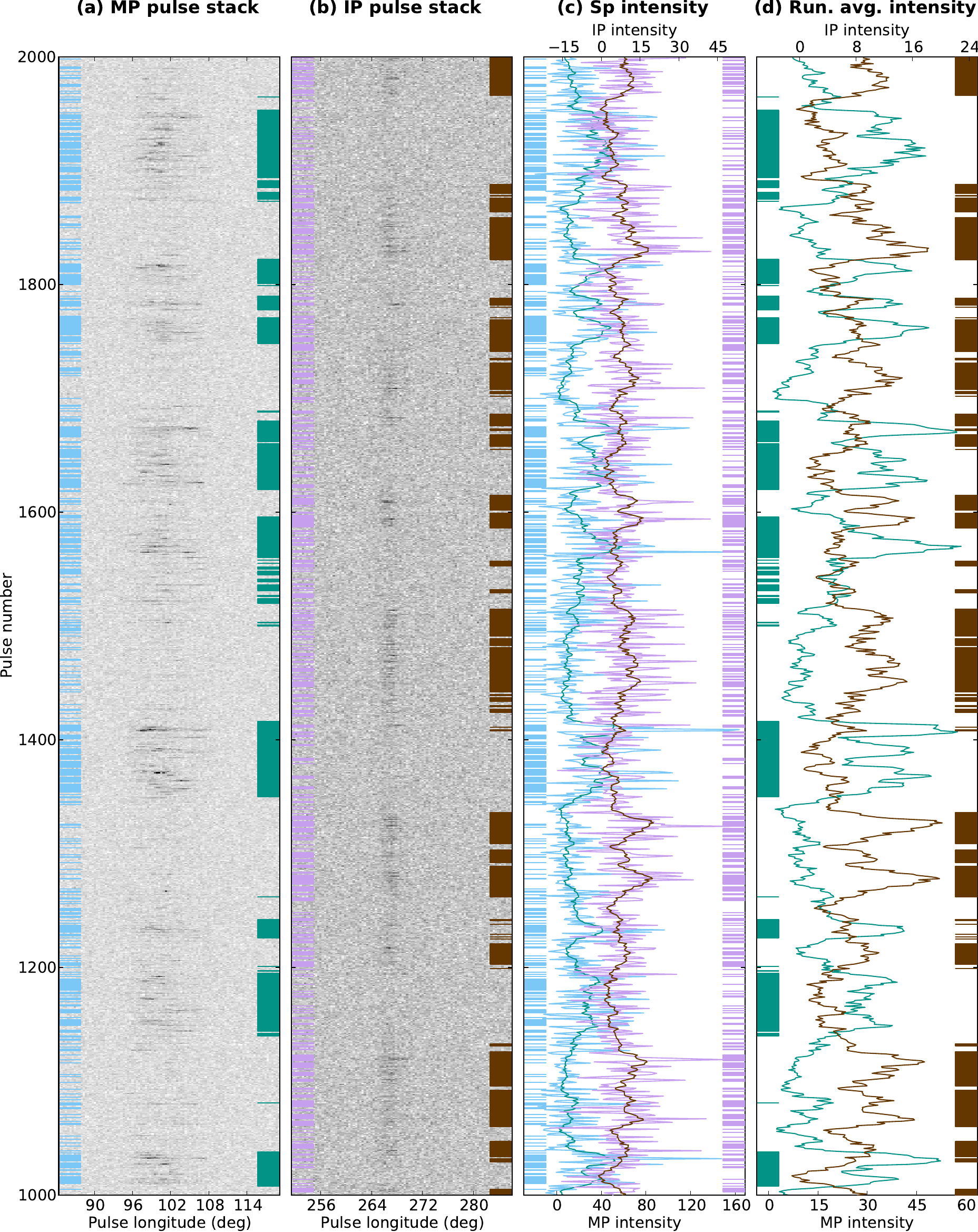}
    \caption{Pulse stacks of the MP (first panel) and the IP (second panel) for pulse number 1000 to 2000 for PSR~J1926+0737, and the corresponding single pulse (third panel) and running averaged intensities (fourth panel). See text for the explanation of the lines.}
    \label{fig:J1926sumall2}
\end{figure*}

\begin{figure*}
    \centering
    \includegraphics[width=\linewidth]{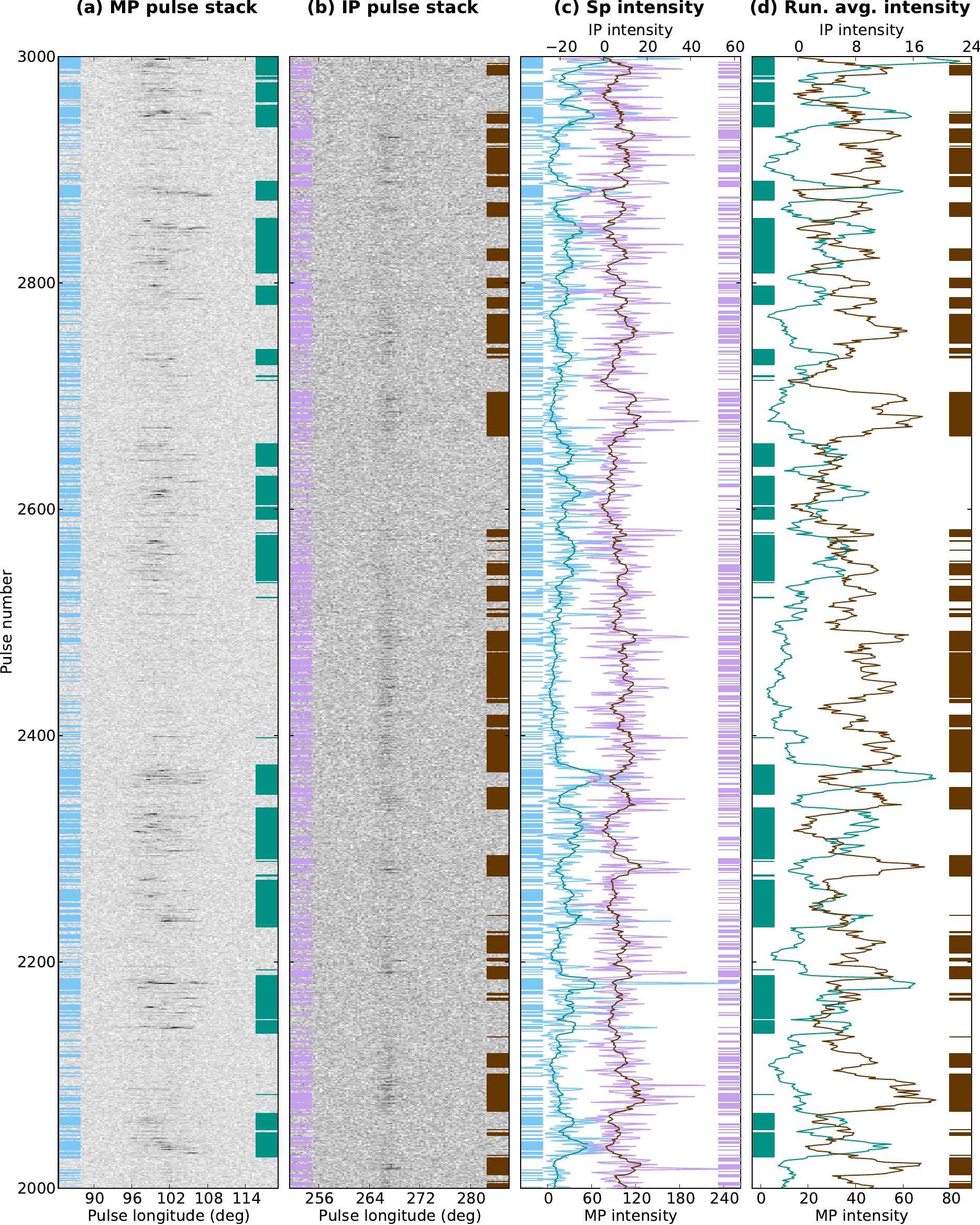}
    \caption{Pulse stacks of the MP (first panel) and the IP (second panel) for pulse number 2000 to 3000 for PSR~J1926+0737, and the corresponding single pulse (third panel) and running averaged intensities (fourth panel). See text for the explanation of the lines.}
    \label{fig:J1926sumall3}
\end{figure*}

\begin{figure*}
    \centering
    \includegraphics[width=\linewidth]{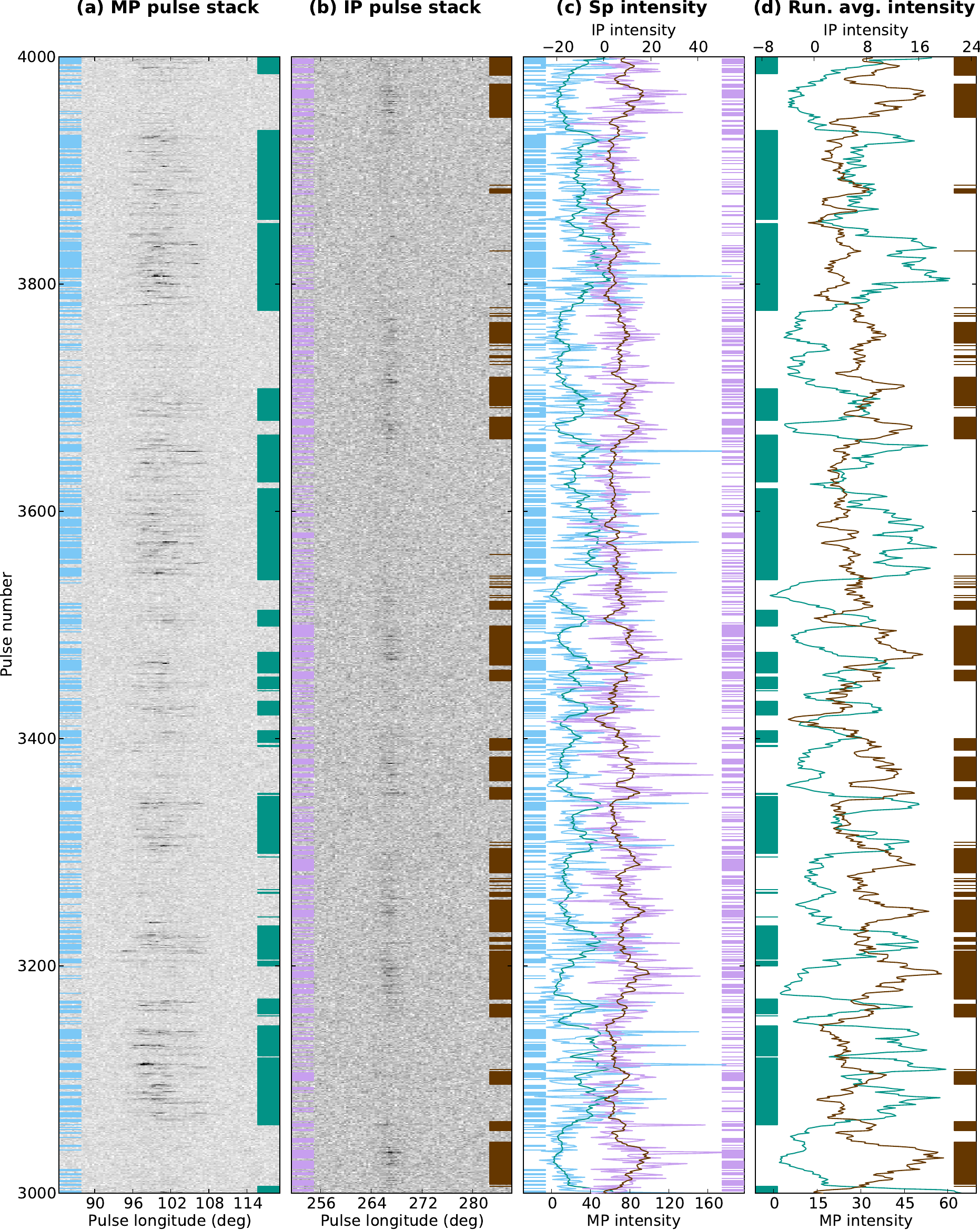}
    \caption{Pulse stacks of the MP (first panel) and the IP (second panel) for pulse number 3000 to 4000 for PSR~J1926+0737, and the corresponding single pulse (third panel) and running averaged intensities (fourth panel). See text for the explanation of the lines.}
    \label{fig:J1926sumall4}
\end{figure*}

\begin{figure*}
    \centering
    \includegraphics[width=\linewidth]{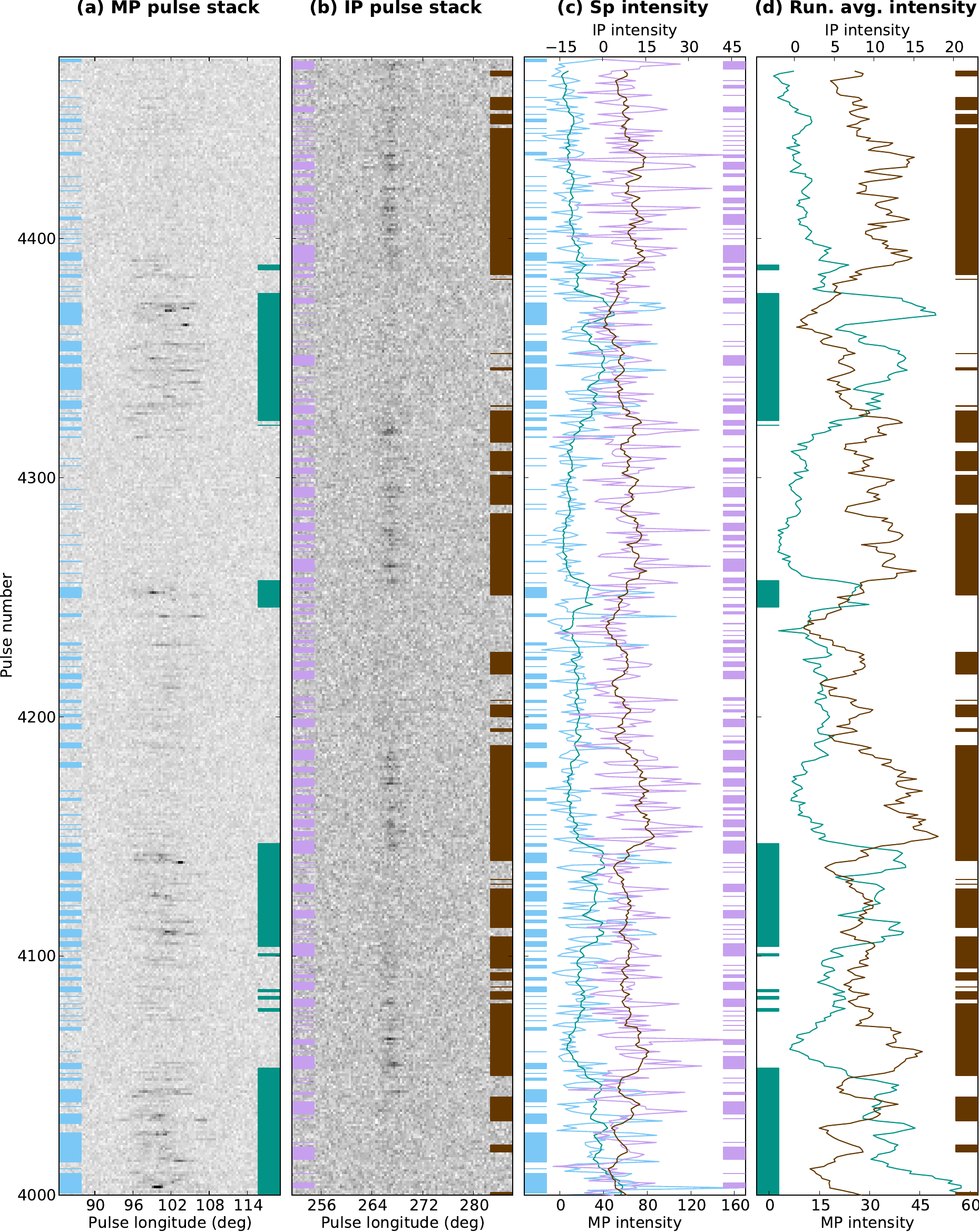}
    \caption{Pulse stacks of the MP (first panel) and the IP (second panel) for pulse number 4000 to 4476 for PSR~J1926+0737, and the corresponding single pulse (third panel) and running averaged intensities (fourth panel). See text for the explanation of the lines.}
    \label{fig:J1926sumall5}
\end{figure*}

\end{appendix}

\end{document}